\RequirePackage{ifpdf}
\documentclass[hyper,12pt,letterpaper]{JHEP3}
\usepackage{graphicx}
\usepackage{latexsym,amsmath,amsfonts,amssymb}
\usepackage{mathrsfs}
\usepackage[makeroom]{cancel}
\usepackage{bbm}
\usepackage{bm}
\usepackage{subfigure}
\usepackage{paralist}
\usepackage{url}

\newcommand{\spc}[2][c]{\begin{tabular}[#1]{@{}c@{}}#2\end{tabular}}

\newcommand{\vol}{\mathrm{vol}}

\newcommand{\eg}{\textit{e.g.}}

\newcommand{\ie}{\textit{i.e.}}

\numberwithin{equation}{section}

\newcommand{\nn}{\nonumber}
\newcommand{\mat}[1]{\begin{pmatrix} #1 \end{pmatrix}}

\newcommand{\be}{\begin{equation}} 
\newcommand{\ee}{\end{equation}}
\newcommand{\bea}{\begin{equation} \begin{aligned}} \newcommand{\eea}{\end{aligned} \end{equation}}

\newcommand{\bit}{\begin{itemize}} 
\newcommand{\eit}{\end{itemize}}

\newcommand{\cF}{\mathcal{F}}

\newcommand{\cH}{\mathcal{H}}

\newcommand{\cM}{\mathcal{M}}
\newcommand{\cN}{\mathcal{N}}

\newcommand{\cS}{\mathcal{S}}

\newcommand{\cW}{\mathcal{W}}

\newcommand{\cZ}{\mathcal{Z}}

\newcommand{\bZ}{\mathbb{Z}}

\newcommand{\Z}{\mathbb{Z}}
\newcommand{\C}{\mathbb{C}}
\newcommand{\R}{\mathbb{R}}

\renewcommand{\t}{\widetilde }
\renewcommand{\d}{\partial }

\renewcommand{\b}{\bar }

\newcommand{\half}{{1\over 2}}

\newcommand{\bz}{{\b z}}

\newcommand{\CA}{\mathcal{A}}
\newcommand{\CB}{\mathcal{B}}
\newcommand{\CC}{\mathcal{C}}
\newcommand{\CD}{\mathcal{D}}

\newcommand{\CF}{\mathcal{F}}

\newcommand{\CH}{\mathcal{H}}
\newcommand{\CI}{\mathcal{I}}
\newcommand{\CJ}{\mathcal{J}}
\newcommand{\CK}{\mathcal{K}}
\newcommand{\CL}{\mathcal{L}}
\newcommand{\CM}{\mathcal{M}}
\newcommand{\CN}{\mathcal{N}}
\newcommand{\CO}{\mathcal{O}}

\newcommand{\CQ}{\mathcal{Q}}
\newcommand{\CR}{\mathcal{R}}
\newcommand{\CS}{\mathcal{S}}
\newcommand{\CT}{\mathcal{T}}

\newcommand{\CV}{\mathcal{V}}
\newcommand{\CW}{\mathcal{W}}

\newcommand{\CZ}{\mathcal{Z}}

\newcommand{\FR}{\mathfrak{R}}
\newcommand{\Fg}{\mathfrak{g}}
\newcommand{\Fh}{\mathfrak{h}}

\newcommand{\GG}{\mathbf{G}}

\newcommand{\GH}{\mathbf{H}}

\newcommand{\rk}{{{\rm rk}(\GG)}}

\newcommand{\m}{\mathfrak{m}}
\newcommand{\n}{\mathfrak{n}}

\newcommand{\ba}{{\b a}}
\newcommand{\bb}{{\bb b}}

\newcommand{\h}{\hat}

\newcommand{\oneloop}{\text{1-loop}}

\newcommand{\Mgp}{\CM_{g,p}}
\newcommand{\pif}{{\Pi}}
\newcommand{\tCW}{{\t\CW}}

\DeclareMathOperator{\Tr}{Tr}
\DeclareMathOperator{\tr}{tr}

\DeclareMathOperator{\sign}{sign}

\newcommand{\SL}{{\mathscr L}}

\newcommand{\ov}{\over}

\newcommand{\fM}{\mathfrak{M}}

\newcommand{\setcond}[2]{\{\,#1\,:\,#2\,\}}

\newcommand{\dilog}{{\text{Li}_2}}

\title{Supersymmetric partition functions \\ and the three-dimensional A-twist }

\author{Cyril~Closset,$^\flat$ Heeyeon~Kim$^\sharp$ and Brian Willett$^{\natural}$\\

{}$^{\flat}$Theory Department, CERN\\
CH-1211, Geneva 23, Switzerland\\
{}$^{\sharp}$ Perimeter Institute for Theoretical Physics\\
31 Caroline Street North, Waterloo, N2L 2Y5, Ontario, Canada\\
{}$^{\natural}$ Kavli Institute for Theoretical Physics\\
 University of California, Santa Barbara, CA 93106
}

\preprint{CERN-TH-2017-006}
\keywords{Supersymmetry, Topological Field Theory}

\abstract{
We study three-dimensional $\CN=2$ supersymmetric  gauge  theories on $\Mgp$, an oriented circle bundle of degree $p$ over a closed Riemann surface, $\Sigma_g$.  We compute the $\Mgp$ supersymmetric partition function and correlation functions of supersymmetric loop operators. 
This uncovers interesting relations between observables on manifolds of different topologies. In particular, the familiar supersymmetric partition function on the round $S^3$ can be understood as the expectation value of a so-called ``fibering operator'' on $S^2 \times S^1$ with a topological twist. More generally, we show that the 3d $\CN=2$ supersymmetric partition functions (and supersymmetric Wilson loop correlation functions) on $\Mgp$ are fully determined by the two-dimensional A-twisted topological field theory obtained by compactifying the 3d theory on a circle. 
We give two complementary derivations of the result. We also discuss  applications to F-maximization and to three-dimensional supersymmetric dualities. }

\begin{document}

\tableofcontents

\section{Introduction}
Supersymmetric partition functions are useful to explore strongly-coupled theories with various amounts of supersymmetry---see {\it e.g.} \cite{Pestun:2007rz, Kapustin:2009kz, Hama:2010av, Jafferis:2010un, Benini:2012ui, Doroud:2012xw}. This is particularly true in three dimensions, where there are fewer non-perturbative tools available than in even dimensions. For instance,  in three-dimensional conformal field theories (CFT), the quantity
\be\label{def F}
F_{S^3}= - \log{|Z_{S^3}|}~,
\ee
where $Z_{S^3}$ is the partition function on the round three-sphere,~\footnote{The log of the partition function is UV divergent; $F$ is defined as its finite piece upon taking the UV cut-off to infinity.}
is a fundamental quantity analogous to the central charge $a$ in even dimensions \cite{Jafferis:2011zi, Klebanov:2011gs, Casini:2012ei}. In three-dimensional theories with $\CN=2$ supersymmetry, we can often compute \eqref{def F} exactly by supersymmetric localization in a gauge-theory UV completion of the CFT \cite{Kapustin:2009kz, Hama:2010av, Jafferis:2010un}.

In this work, we consider $\CN=2$ gauge theories with an $R$-symmetry $U(1)_R$. We study them on a three-manifold $\Mgp$, a $U(1)$ principal bundle over a Riemann surface:
\be\label{Mgp intro}
S^1 \longrightarrow \Mgp \stackrel{\pi}\longrightarrow \Sigma_g~.
\ee
This family of  geometries is indexed by two integers, $g\in \Z_{\geq 0}$, the genus of the Riemann surface $\Sigma_g$, and $p\in \Z$, the first Chern number of the principal bundle. It includes the round three-sphere and the product spaces $\Sigma_g \times S^1$:
\be
\CM_{0, 1}\cong S^3~, \qquad \qquad \CM_{g, 0}\cong \Sigma_g \times S^1~.
\ee
We derive general formulas for the supersymmetric partition functions $Z_{\Mgp}$ and for expectation values of supersymmetric Wilson loops (and other loop operators) wrapped on an $S^1$ fiber.
We heavily exploit the fact that the supersymmetric background on $\Mgp$ is a pull-back of the two-dimensional topological A-twist on $\Sigma_g$ \cite{Witten:1988xj, Closset:2012ru}.
Note that a very similar computation was performed in \cite{Ohta:2012ev} for theories with $\CN\geq 3$ supersymmetry.~\footnote{We differ from  \protect\cite{Ohta:2012ev} in our treatment of fermionic zero-modes and obtain different results for $g>0$.}

An interesting upshot of our analysis is that the $S^3$ partition function can be viewed as the expectation value of a particular loop operator $\CF$ wrapped on $S^1$ in the topologically-twisted theory on $S^2 \times S^1$---that is:
\be\label{S3 to S2S1}
Z_{S^3} = \big\langle  \CF\big\rangle_{S^2 \times S^1}~.
\ee
We call $\CF$ the {\it fibering operator}. Its insertion along an $S^1$ fiber in $\Mgp$ corresponds to shifting the Chern number of the total space, replacing $p$ by $p+1$.
 The relation \eqref{S3 to S2S1} is a generalization to $\CN=2$ theories of a similar relation in Chern-Simons theory \cite{Blau:2006gh}. We will compute the fibering operator explicitly for $\CN=2$ supersymmetric Yang-Mills-Chern-Simons-matter theories. 
Note that the relation \eqref{S3 to S2S1} only holds for theories with integer-quantized $R$-charges, because the $R$-charges are integer-quantized on $S^2 \times S^1$. However, the $S^3$ result can be analytically continued to real $R$-charges in a canonical fashion \cite{Jafferis:2010un, Closset:2014uda}.

In the rest of this introduction, we summarize our main results and discuss relations to previous works.

\subsection*{Seifert manifolds, three-dimensional $A$-twist and the Coulomb branch}
For 3d $\CN=2$ theories with an $R$-symmetry, the three-dimensional backgrounds that allow for some supersymmetry were classified in \cite{Closset:2012ru}. In order to preserve two supercharges of opposite $R$-charges, the three-manifold  $\CM_3$ must admit a nowhere-vanishing Killing vector $K^\mu$. There are two distinct possibilities: 
\begin{itemize}

\item $K^\mu$ is real. Then $\CM_3$ must be an (orientable) Seifert manifold---an $S^1$ bundle over a two-dimensional orbifold $\h \Sigma_g$.

\item $K^\mu$ is complex and generates two isometries. The only known example is the ``untwisted'' $S^2{\times}S^1$ background of  \cite{Imamura:2011su}.  The corresponding partition function computes the  so-called superconformal index  \cite{Kim:2009wb, Imamura:2011su}.
\end{itemize}
In this work, we focus on the simplest supersymmetric Seifert-manifold backgrounds satisfying two additional conditions: 1) the orbits of $K^\mu$ are the Seifert fibers; 2) the base space is a smooth closed Riemann surface $\Sigma_g$ (without orbifold points).

Condition 1) could be waived in the case of a base $\h\Sigma_g$ with genus $g=0$ or $1$. For $g=0$, this corresponds to turning on a `squashing' parameter---this is often denoted by $b\neq 1$  on lens space backgrounds---see {\it e.g.} \cite{Hama:2011ea, Alday:2013lba, Closset:2013vra}. It will be essential to our story that we do not allow any such squashing deformation. Condition 2) is not essential to our story but it is assumed for simplicity. We hope to report on the case of general Seifert manifolds in future work.

These two conditions imply that the supersymmetric background on $\CM_{g, p}$ is a {\it pull-back} of the ordinary A-twist for two-dimensional $\CN=(2,2)$ theories on the Riemann surface $\Sigma_g$.~\footnote{Strictly speaking, this is for a particular choice of the ``shift by $\kappa$'' in \protect\cite{Closset:2012ru}. See Appendix \protect\ref{app: 3d Atwist}.} In this language, localization on $\Mgp$ becomes a simple generalization of a recent localization computation on $\Sigma_g \times S^1$ \cite{Benini:2015noa, Benini:2016hjo, Closset:2016arn}.

As we will show, the supersymmetric partition functions on the A-twisted $\Mgp$  can be constructed in terms of the low-energy theory on the Coulomb branch of the flat-space theory on $\R^2 \times S^1$. We may view the three-dimensional theory on a circle of radius $\beta$ as a two-dimensional $\CN=(2,2)$ supersymmetric theory with an infinite number of fields. This theory has a classical Coulomb branch spanned by the coordinates $u = i \beta(\sigma + i a_0)$, where $\sigma$ is the real scalar in the 3d $\CN=2$ vector multiplet and $a_0$ is the holonomy of the gauge field on $S^1$ (along the Cartan of the gauge group). The low-energy dynamics on the Coulomb branch is governed by the effective twisted superpotential $\CW(u)$. We will discuss this function in depth, paying particular attention to the effect of the Chern-Simons terms. Schematically, it is given by:
\be\label{CW intro}
\CW(u)= {k\ov 2} u(u+1) + {1\ov 24} k_g+{1\ov (2 \pi i)^2} \sum_i \dilog(e^{2 \pi i Q_i(u)})~.
\ee
The first  term is the gauge CS term, the second term is a gravitational CS term, and the last term is the contribution of chiral fields $\Phi_i$ with gauge charges $Q_i$. Note that we assign a physical significance to the constant piece of $\CW$, which is identified with the gravitational CS level $k_g$ in three dimensions.

The coupling of this effective theory to curved space is governed by the so-called effective dilaton, which takes the schematic form:
\be\label{Om intro}
\Omega= k_{R} u +\half  k_{RR} -{1\ov 2 \pi i} \sum_i (r_i-1)\log(1- e^{2 \pi i Q_i(u)})- {1\ov 2 \pi i}\sum_\alpha \log(1- e^{2\pi i \alpha(u)})~.
\ee
Here $k_R$ and $k_{RR}$ are mixed $U(1)_R$-gauge and $U(1)_R$ CS levels, respectively. The other terms are the contribution of the chiral multiplets (with $R$-charges $r_i$) and  of  the W-bosons.  These two functions were derived in \cite{Witten:1993xi, Nekrasov:2014xaa}; in addition, we included the contribution of the $U(1)_R$ and gravitational supersymmetric CS terms of \cite{Closset:2012vp}.

From \eqref{CW intro} and \eqref{Om intro}, we construct two well-defined Coulomb-branch operators. The {\it handle-gluing operator} is given by \cite{Nekrasov:2014xaa}:
\be\label{H schematic}
\CH(u) = \exp\big({2 \pi i\, \Omega(u)}\big)\; \det_{ab}\left({\d^2 \CW(u)\ov
\d u_a \d u_b} \right)~.
\ee
It corresponds to adding a handle to the base $\Sigma_g$, shifting $g$ to $g+1$.  The {\it fibering operator} introduced above is a simple function of the twisted superpotential:
\be\label{F schematic}
\CF(u) = \exp\left({2 \pi i \left(\CW(u)- u_a {\d \CW(u)\ov\d u_a}\right)}\right)~.
\ee
This is our main result. The formula \eqref{F schematic} will be made more precise in the main text.

Consider a theory such that the ``Bethe vacua'' (the abelian Coulomb branch vacua, which are  the  gauge-inequivalent solutions to the Bethe equations \cite{Nekrasov:2009uh} of the theory) are distinct. This always happens, for instance, in theories with enough flavor symmetries and with generic fugacities.
In that case, the find the simple result:
\be\label{Zmgp intro}
Z_{\Mgp} = \sum_{\h u \in \CS_{\rm BE}} \CF(\h u)^p \, \CH(\h u)^{g-1}~,
\ee
with $\CS_{\rm BE}$ the set of Bethe vacua. For $p=0$, this is the so-called twisted index on $\CM_{g, 0}\cong \Sigma_g\times S^1$  \cite{Nekrasov:2014xaa, Benini:2016hjo, Closset:2016arn}.  In particular, the partition function on $\CM_{1,0}\cong T^3$ computes the 3d Witten index studied in \cite{Intriligator:2013lca}. For $p=1$, $g=0$, on the other hand, we have the familiar $S^3$ partition function, and the relation  \eqref{S3 to S2S1} directly follows.
 More generally, we have:
 \be\label{tft relations Mgp}
\big \langle W \big\rangle_{\Mgp}=\big \langle W \CF^p \big\rangle_{\Sigma_g\times S^1}= \big \langle W \CF^p \CH^{g}\big\rangle_{S^2\times S^1}
\ee
for the expectation value of a supersymmetric Wilson loop $W$ wrapped on the $S^1$ fiber.  We do not specify the insertion points on $\Sigma_g$ since the 2d theory is topological.  We may view $\CH$ and $\CF$ as particular defect loop operators in the 3d $\CN=2$ gauge theory, given by \eqref{H schematic} and \eqref{F schematic} on the Coulomb branch.

\subsection*{Localization formula}
Another formula for the partition function $Z_{\Mgp}$ can be obtained by supersymmetric localization in the UV. 
We follow the abelianization method of Blau and Thompson \cite{Blau:1993tv, Blau:2006gh} adapted to the supersymmetric context. From that point of view, the partition function for $p\neq 0$ can be written as:
\be\label{Zmgp JK intro}
Z_{\Mgp} = \sum_{\m \in \Z_p^\rk} \oint_{\CC^\eta} d^{\rk}u\;  \CI_{g, p, \m}(u)~. 
\ee
The sum is over flat torsion bundles for the Cartan subgroup of the gauge group $\GG$. The meromorphic integrand $\CI_{g, p, \m}(u)$ contains classical contributions (the Chern-Simons terms and FI parameters) and one-loop contributions from all the matter fields.  The contour integral is taken along a particular ``Jeffrey-Kirwan (JK) contour'' on a multiple cover of the Coulomb branch (with $u \in \C^\rk$). Using gauge invariance, one may also write \eqref{Zmgp JK intro} as:
\be\label{Zmgp JK2 intro}
Z_{\Mgp} = \sum_{\m \in \Z^\rk} \oint_{\CC^\eta_0} d^{\rk}u\;  \CI_{g, p, \m}(u)~. 
\ee
where the sum is over all GNO-quantized fluxes of $\GG$, while the $u$ variables are gauge-fixed to $0 \leq {\rm Re}(u)<1$. The expression \eqref{Zmgp JK2 intro} is also valid at $p=0$ \cite{Benini:2016hjo, Closset:2016arn}, in agreement with the relations \eqref{tft relations Mgp}. By resuming the fluxes in \eqref{Zmgp JK2 intro}, one can obtain the Bethe-vacua formula \eqref{Zmgp intro}.  Here we should note that we only rigorously derived the JK contour in \eqref{Zmgp JK intro} or \eqref{Zmgp JK2 intro} in the rank-one case. The higher-rank formula should be considered as a well-motivated conjecture. (It also follows in good part from earlier results relating supersymmetric localization \cite{Benini:2013xpa, Hori:2014tda, Benini:2015noa, Closset:2015rna} to JK residues \cite{JK1995,1999math......3178B}.)

For $p\neq 0$, the JK contour in \eqref{Zmgp JK intro} can be deformed to a simple ``$\sigma$-contour'' which lies along the imaginary $u$ axis---that is, we have an integral over real $\sigma$:
\be\label{Zmgp JK3 intro}
Z_{\Mgp} = i^\rk  \sum_{\m \in \Z_p^\rk}\int d^\rk \sigma\;  \CI_{g, p, \m}(\sigma)~,
\ee
in some appropriate region of parameter space. This is the familiar integral over real $\sigma$ on $S^3$ \cite{Kapustin:2009kz}.  For generic fugacities, the contour along $\sigma$ generally has to be deformed, so that it always ``separates'' the singularities of the integrand in the same way. For pure $\CN=2$ supersymmetric Chern-Simons theory, we may also rotate the contour to lie along the real axis in the $u$ plane; such a theory is equivalent to ordinary Chern-Simons theory (up to a shift of the CS level), and \eqref{Zmgp JK3 intro} indeed reproduces the known integral formula over the  holonomies  $a_0$ in that case \cite{Blau:2006gh}.

\subsection*{Parity anomalies, contact terms and Chern-Simons levels}
As is well known, a three-dimensional Dirac fermion coupled to gauge fields suffers from the so-called parity anomaly; one cannot quantize the fermion while preserving both gauge invariance and three-dimensional parity \cite{Redlich:1983dv,  Niemi:1983rq, AlvarezGaume:1984nf}. Throughout this work, we choose a gauge-invariant regularization of the 3d $\CN=2$ chiral multiplet. After integrating out the matter fields, the lack of parity invariance of the vector multiplet   effective action (both for dynamical gauge fields and background gauge fields for global symmetries) is encoded in certain parity-odd contact terms in two-points functions of the corresponding conserved currents. We denote these contact terms by $\kappa$. Unlike ordinary contact terms, which are generated by local terms in the effective action and are therefore ambiguous, the contact terms $\kappa$ correspond to Chern-Simons terms in the action, whose couplings $k$---the CS levels---are integer-quantized for compact gauge groups.  The contact terms $\kappa$, therefore,  are physically meaningful modulo integer shifts, $\kappa \rightarrow \kappa+k$  \cite{Closset:2012vp}. In this work, we are careful in distinguishing between $\kappa$ and $k$. Unless otherwise specified, the CS levels $k$ are always integer-quantized, while chiral multiplets contributes certain half-integers to $\kappa$. For instance, a single chiral multiplet coupled to a $U(1)$ (background) gauge field with charge $1$ will be quantized with a contact term $\kappa= -\half$ for that $U(1)$. This is sometimes referred to as a ``$U(1)_{-\half}$ quantization''.

This distinction is not only pedantic. It is crucial in order to compute partition functions, including all dynamical and background Chern-Simons terms, in a consistent manner. This resolves some confusions about ``sign ambiguities'' that appeared in  \cite{Benini:2015noa, Benini:2016hjo, Closset:2016arn}---there are no sign ambiguities except for the ones encoded in CS terms for global symmetries. Relatedly, we will correct some signs that arise from classical CS actions for abelian gauge groups. (See in particular Appendix \ref{app: spin CS}.)

\subsection*{Dualities and on-shell superpotential}
Many 3d $\CN=2$ supersymmetric gauge theories are related by infrared dualities. On general grounds, the partition function (and other $\CN=2$ supersymmetric observables) of two dual theories $\CT$ and $\CT_D$ should agree on any supersymmetric background:
\be\label{rel Z Zdual intro}
Z_{\Mgp}[\CT]=Z_{\Mgp}[\CT_D]~.
\ee
The Bethe-vacua formula \eqref{Zmgp intro} for $Z_{\Mgp}$ is particularly convenient to check these duality relations. The duality relations \eqref{rel Z Zdual intro}, and similar relations for loop operator insertions, can be rephrased as a statement about matching Bethe vacua in a one-to-one fashion. The duality statement is that the handle-gluing and fibering operators of the dual theories agree ``on-shell'', that is, when evaluated on a dual pair of Bethe vacua, $u= \h u$ and $u_D= \h u_D$. 

Equivalently, one can state the duality relations in terms of the effective twisted superpotential: The twisted superpotentials of the dual theories must agree on-shell:
\be\label{rel Wdual intro}
\CW^{\CT}(\h u)= \CW^{\CT_D}(\h u_D)~,
\ee
on any pair of dual vacua~\footnote{The twisted superpotential suffers from branch-cut ambiguities, and this relation holds for a particular choice of branches.} (and similarly for the so-called on-shell effective dilaton, that we will define later). Interestingly, the relations \eqref{rel Wdual intro} for gauge-theory dualities often follow from known dilogarithm identities \cite{Lewin1991}. We should emphasize that, even in the case of the $S^3$ partition functions, this provides a simpler derivation of the duality relation \eqref{rel Z Zdual intro} than previous investigations of complicated integral identities---see in particular \cite{vdbult_thesis, Willett:2011gp, Benini:2011mf}.~\footnote{On the other hand, those integral identities are valid on $S^3_b$ with non-zero squashing, $b\neq 1$, while we  only consider $b=1$.}

We will also discuss how we can use the on-shell twisted superpotential to gauge a flavor symmetry, independently of whether the original theory has a Lagrangian description.

\subsection*{Relation to previous works and outlook}
The three-dimensional A-twist vantage point relates the $S^3$ partition function  \cite{Kapustin:2009kz, Hama:2010av, Jafferis:2010un} with the $\Sigma_g \times S^1$ twisted indices \cite{Benini:2015noa, Benini:2016hjo, Closset:2016arn}. As already noted, this generalizes known results for pure CS theories  \cite{Blau:2006gh} to $\CN=2$ supersymmetric gauge theories with matter. This framework also explains the results of \cite{Kapustin:2013hpk} on the Wilson loop quantum algebra on $S^3$, which is encoded in the Bethe equations \cite{Closset:2016arn}. A similar relation between the twisted index and the $S^3$ partition function was also observed in large $N$ quiver gauge theories \cite{Hosseini:2016tor, Hosseini:2016ume}.

For generic values of $g, p$, the supersymmetric background $\Mgp$ only allows for quantized $R$-charges. When $g-1=0 \mod p$, including the case $\CM_{0,1}\cong S^3$, on the other hand, the $R$-charges can be varied continuously. We will explain how our formulas for $Z_{\Mgp}$ can account for any $R$-charge, in those cases. On $S^3$, this allows us to probe properties of the infrared CFT, where the $R$-charges are generally irrational.  Whenever the UV $R$-symmetry can mix with abelian flavor symmetries along the RG flow (and in the absence of accidental symmetries), the superconformal $R$-charge in the infrared can be determined by $F$-maximization \cite{Jafferis:2010un, Closset:2012vg}. That is, we need to maximize \eqref{def F}   over the possible trial $R$-charges. Our Bethe-vacua formula for $Z_{S^3}$ is well-suited for this computation, and the results compare well with previously-obtained results using the integral formula \eqref{Zmgp JK intro}.

Another important localization result available in the literature is the lens space $L(p,1)$ partition function  \cite{Benini:2011nc, Alday:2012au}. We should note that the supersymmetric background for the manifold:
\be\label{M0p intro}
\CM_{0,p} \cong L(p, p-1)~,
\ee
that we consider here, is {\it distinct} from the $L(p,1)$ background considered in  \cite{Benini:2011nc, Alday:2012au, Nieri:2015yia}, if $p> 2$. 
The main difference between the two supersymmetric backgrounds is that the $R$-symmetry line bundle present on \eqref{M0p intro} is topologically non-trivial, unlike the background of \cite{Benini:2011nc, Alday:2012au}.

In the A-twist language, the $L(p,1)$ background corresponds to a genus-zero Riemann surface $\h \Sigma_0 \cong S^2/\Z_p$ with two orbifold points. We hope to address this case in future work, along with generic Seifert manifolds. Pure Chern-Simons theory on a restricted class of Seifert manifolds was considered in \cite{Beasley:2005vf, Blau:2013oha}.

The formula \eqref{Zmgp intro} for the supersymmetric partition functions is reminiscent of the surgery prescription for pure CS theory \cite{Witten:1988hf}. Here we have a potentially richer quasi-topological structure that depends holomorphically on various parameters. It would be very interesting to explore that point of view further.

Another construction of supersymmetric partition functions is in terms of holomorphic blocks \cite{Beem:2012mb}. They are partition functions on $D^2 \times S^1$, with $D^2$ a disk, which are in one-to-one correspondence with the Bethe vacua. Despite the similarities, that approach seems somewhat orthogonal to the one of the present paper, especially since the squashing parameter (or $\Omega$-deformation on $D^2$) plays such an important role in \cite{Beem:2012mb}, while we set it to zero throughout. Nonetheless, it would be very interesting to understand better the relation between the two approaches. Relatedly, our results should be of interest in the context of the 3d/3d correspondence \cite{Dimofte:2011ju, Dimofte:2011py, Dimofte:2014zga}. In particular, one might ask what kind of topological field theory can be obtained by compactifying M5-branes on the supersymmetric background $\Mgp$; progress on understanding such systems has been made recently in \cite{Gukov:2015sna,Gukov:2016gkn}.

Finally,  let us mention that results completely analogous to the ones of this paper can be obtained for four-dimensional $\CN=1$ theories on $\Mgp\times S^1$ \cite{toappearCKW}.

%%%%%
This paper is organized as follows.  In Section \ref{sec: BE formula}, we explore the two-dimensional A-model point of view and we derive the Bethe-vacua formula \eqref{Zmgp intro}. In Section \ref{sec: 3}, we summarize important aspects of curved-space supersymmetry on $\Mgp$. In Section \ref{sec: Loc}, we discuss supersymmetric localization and we obtain the localization formula \eqref{Zmgp JK intro}.
In Section \ref{sec: S3 and Fmax}, we compute the $S^3$ partition function with the Bethe-vacua formula, and we present some non-trivial examples of $F$-maximization. In Section \ref{sec:duality}, we study the matching of supersymmetric partition function across gauge-theory dualities. Additional material is contained in various appendices.

%%%%%%%%%%%%%%%%%%%%%%%%%%%%%%%%%%%%%%%%
%%%%%%%%%%%%%%%%%%%%%%%%%%%%%%%%%%%%%%%%
\section{The partition function as a sum over Bethe vacua}\label{sec: BE formula}
In this section, we start by reviewing some relevant results about two-dimensional $\CN=(2,2)$ gauge theories. We then consider three-dimensional $\CN=2$ gauge theories on a circle as a two-dimensional $\CN=(2,2)$ theory and discuss in detail the low-energy theory on the Coulomb branch. We argue that the partition function on $\Mgp$ can be obtained as a sum over ``Coulomb branch vacua'' (Bethe vacua) by a simple modification of the formula for the $\Sigma_g \times S^1$ twisted indices discussed in \cite{Nekrasov:2014xaa,Closset:2016arn,Benini:2016hjo}.  We will  give a microscopic derivation of this result in Section \ref{sec: Loc}.

\subsection{The Bethe-vacua formula in two dimensions}\label{subsec: 2d W etc}
As a preliminary, consider a two-dimensional $\CN=(2,2)$ gauge theory with gauge group $\GG$ and chiral multiplets $\Phi_i$ in representations $\FR_i$ of $\Fg= {\rm Lie}(\GG)$.  From the vector multiplet $\CV$, one can build a $\Fg$-valued twisted chiral multiplet $\Sigma = -i D_- \t D_+ \CV$ with components:
\be\label{Sigma components}
\Sigma=\; \left(\sigma~, \, \Lambda_1~, \, - \t\Lambda_{\b 1}~, \,   -4 f_{1\b 1} \right)~.
\ee
Here we follow the A-twist conventions of \cite{Closset:2015rna}. In particular, the gauginos $\Lambda_1$, $\t\Lambda_{\b1}$ are $(1,0)$- and $(0,1)$-forms after the twist, respectively.~\footnote{To avoid any possible confusion, let us recall that there are two distinct but standard usages of the term ``twist'' in two dimensions.  The terms ``twisted chiral multiplet'' and ``twisted mass'' refer to representations of $\CN=(2,2)$ supersymmetry, while the ``A-twist'' is a topological twist of the theory.} See also Appendix \ref{app: 3d Atwist}.

Let us denote by $\GG_F$ the {\it flavor} symmetry group (the non-R global symmetry group) of the theory. It is natural to couple the flavor currents to a background vector multiplet $\CV_F$. The so-called twisted masses corresponds to constant expectations values $\sigma_F= m_F$  for its complex scalar component. A particular chiral multiplet $\Phi_i$ has twisted mass $m_i = \omega_i(m_F)$, where $\omega_i$ is a weight of the flavor representation.

At a generic point on the classical Coulomb branch, the gauge group is broken to the Cartan subgroup $\GH\subset \GG$, and the massive chiral multiplets and W-bosons can be integrated out. The low-energy dynamics on the Coulomb branch is governed by the effective twisted superpotential \cite{Witten:1993yc, Witten:1993xi,Nekrasov:2009uh}:
\be\label{CW eff gen 2d}
 \CW= \tau(\sigma)-{1\ov 2 \pi i}\sum_i \sum_{\rho_i \in \FR_i} (\rho_i(\sigma)+ m_i)\big(\log (\rho_i(\sigma)+ m_i)-1\big)- \half \sum_{\alpha\in \Fg_+} \alpha(\sigma)~.
 \ee
The first term in  \eqref{CW eff gen 2d}  is the contribution from the two-dimensional complexified Fayet-Iliopoulos (FI) parameters, with $\tau$ a projection on the free abelian subgroup $\prod_I U(1)_I\subset \GG$. The second term in \eqref{CW eff gen 2d} is the contribution from the chiral multiplets $\Phi_i$, with $\rho_i$ the weights of the representation $\FR_i$. The last term in \eqref{CW eff gen 2d}  is the contribution from the W-bosons and their superpartners, with a sum over the positive roots of $\Fg$.

In the following, it will be useful to pick a basis $e^a$ of the Cartan $\GH$ of $\GG$, and a basis $e_F^\alpha$ of the Cartan of $\GG_F$, such that:
\be
\sigma = \sigma_a e^a~,\qquad \qquad m_F= m_\alpha e_F^\alpha~.
\ee
We choose a basis $\{e^a\}$ that generates the coweight lattice $\Lambda_{\rm cw}$, so that $\rho(e^a)\equiv \rho^a \in \Z$ for all weights $\rho \in \Lambda_{\rm w}$, and similarly for the flavor group. 

We view the low energy theory on the Coulomb branch as an A-twisted Landau-Ginzburg (LG) model  \cite{Vafa:1990mu}   with twisted superpotential $\CW$ for the twisted chiral multiplets $\Sigma_a$. However, we see from \eqref{Sigma components} that the highest component of $\Sigma_a$ is an abelian field strength. We may treat $f_{1\b 1}$ as the fundamental variable if we also impose flux quantization by hand \cite{Nekrasov:2009uh}:
\be\label{flux quantization}
{1\ov 2 \pi } \int d^2 \sqrt{g} \; (-2 i f_{1\b 1})_a \in \m_a \in \Z~,
\ee
on any compact space. Relatedly, the twisted superpotential \eqref{CW eff gen 2d} suffers from branch cut ambiguities due to the logarithms:
\be \label{W branch cut}
\CW(\sigma_a, m_\alpha)\; \rightarrow \; \CW(\sigma_a, m_\alpha) + n^a \sigma_a + n^\alpha m_\alpha~, \qquad n^a, n^\alpha \in \Z~.
\ee
The quantization condition \eqref{flux quantization} ensures that \eqref{W branch cut} only shifts the effective action by an integer multiple of $2 \pi i$, so that the path integral remains well-defined. 
When looking for the vacua of the theory, we have to take the ambiguity \eqref{W branch cut} into account. This leads to the so-called {\it Bethe equations}  \cite{Nekrasov:2009uh}:
\be\label{BE0}
\exp\left(2 \pi i {\d \CW\ov \d\sigma_a}\right)=1~, \qquad a=1, \cdots, \rk~.
\ee
Note the left-hand side is independent of the branch-cut ambiguity---in fact, it is a rational function of $\sigma_a$ and $m_i$.

If $\GG$ is abelian, the solutions to \eqref{BE0} correspond directly to the vacua of the theory.  In a non-abelian theory, we must divide by the action of the Weyl group of $\GG$, $W_\GG$.  In addition, solutions which are not acted on freely by the Weyl symmetry correspond to putative vacua with unbroken non-abelian gauge symmetry, wherein the derivation of \eqref{CW eff gen 2d} is unreliable. Following \cite{Hori:2006dk}, we will exclude these solutions, which are believed not to correspond to physical vacua. (See also \cite{Aharony:2016jki} for a related recent discussion.) Thus the set of vacua  of the Coulomb branch theory is given by:~\footnote{For simplicity in this paper, we consider compact connected gauge groups, such that the Weyl group of $\GG$ and the Weyl group of its Lie algebra $\Fg$ coincide. Then the condition $w. \h \sigma \neq \h\sigma$, $\forall w\in W_\GG$ is  equivalent to $\alpha(\h\sigma)\neq 0$, $\forall \alpha \in \Fg$.}
\be\label{S BE 2d}
\cS_{BE}= \left\{ \;\h \sigma_a \; \bigg| \; \exp{\left(2 \pi i {\d\CW\ov \d\sigma_a}(\h \sigma)\right)} = 1~, \;\; \forall  a~, \quad w \cdot \h\sigma \neq \h\sigma, \;\; \forall w \in W_\GG \; \right\} /W_\GG~.
\ee
We refer to the solutions $\h\sigma= (\h\sigma_a)$ (modulo the Weyl symmetry) as the ``Bethe vacua''.

\subsubsection{Coulomb branch correlation functions}
The Coulomb branch operators are the twisted chiral ring operators given by gauge-invariant polynomials $P(\sigma)$ in the scalar field $\sigma \subset \CV$. On the classical Coulomb branch, they correspond to Weyl-invariant polynomials in the variables $\sigma_a$. The effective twisted superpotential provides us with twisted chiral ring quantum relations. 

Let us consider the $\CN=(2,2)$ theory on a closed orientable Riemann surface $\Sigma_g$ with the topological A-twist. The low energy topological field theory for the twisted chiral multiplets $\Sigma_a$ has an effective action:
\be\label{S TFT}
S_{\rm TFT}= \int_{\Sigma_g} d^2 x \sqrt{g} \left(-2  {f_{1 \b1}}_a {\d \CW(\sigma)\ov \d \sigma_a} + \t\Lambda^a_{\b1}\Lambda_1^b {\d^2 \CW(\sigma)\ov \d \sigma_a\d\sigma_b}   \right)+ {i\ov 2}  \int_{\Sigma_g} d^2 x \sqrt{g} \,  \Omega(\sigma)\, R~,
\ee
up to $Q$-exact terms. The first term in \eqref{S TFT} depends on the effective twisted superpotential $\CW$, and it is explicitly topological (since $f_{1\b1}$ and $\Lambda_{\b1}\Lambda_1$ are naturally 2-forms). The second term involves $R$ the Ricci scalar, and it is topological for the constant modes of $\sigma_a$ due to the Gauss-Bonnet theorem. It corresponds to the ``improvement'' Lagrangian of \cite{Closset:2014pda}.  The holomorphic function $\Omega(\sigma)$ is the {\it effective dilaton} which governs the coupling of the theory to the A-twist background. In our two-dimensional $\CN=(2,2)$ gauge theory, it is given by \cite{Witten:1993xi, Nekrasov:2014xaa}:
\be
\Omega(\sigma)=- {1\ov 2 \pi i}\sum_i \sum_{\rho_i \in \FR_i} (r_i-1)\log (\rho_i(\sigma)+ m_i)
- {1\ov 2 \pi i} \sum_{\alpha \in \Fg}  \log{\alpha(\sigma)}~,
\ee
up to an arbitrary constant. Here $r_i \in \Z$ denote the $R$-charges of the chiral multiplets $\Phi_i$, which should be integers so that the theory can be defined on any $\Sigma_g$.

The correlation functions of Coulomb branch operators can be computed as a sum over the Bethe vacua, by a direct generalization of Vafa's formula for ordinary topological LG models (LG)  \cite{Vafa:1990mu}. One finds \cite{Nekrasov:2014xaa}:
\be\label{corr 2d}
\langle P(\sigma) \rangle_{\Sigma_g} = \sum_{\h \sigma \in \cS_{BE} }  \CH(\h\sigma)^{g-1} \, P(\h \sigma)~,
\ee
with
\be\label{DEA in 2d}
 \CH(\sigma) = e^{2 \pi i \, \Omega(\sigma)} \, \det_{ab}\left(-2 \pi i \;{\d^2 \CW(\sigma)\ov  \d{\sigma_a}\d{\sigma_b}}\right) 
\ee
the so-called handle-gluing operator. The first factor in \eqref{DEA in 2d} comes from the last term in \eqref{S TFT} evaluated on the Coulomb branch, and the Hessian determinant of the superpotential arises because of the gaugino zero-modes on $\Sigma_g$. One can also obtain \eqref{corr 2d}  by supersymmetric localization in the UV \cite{Benini:2016hjo}.

There is an important caveat to this discussion: we have assumed that the Bethe vacua are {\it isolated}. This generally happens in theories with enough flavor symmetries and with generic twisted masses. Many important two-dimensional theories do not satisfy this condition, however---for instance, any GLSM that flows to a Calabi-Yau NLSM in the IR has a degenerate $\CW$; on the other hand, such theories can still be studied by localization methods, at least at genus $g=0$  \cite{Morrison:1994fr, Closset:2015rna}. Isomorphic comments apply in three dimensions.

\subsubsection{Flux operators}
In the presence of a flavor symmetry group $\GG_F$, it is natural to turn on supersymmetric {\it background fluxes} for the gauge field in $\CV_F$,
\be
{1\ov 2 \pi} \int_{\Sigma_g} d^2 x \sqrt{g}\, (-2 i f_{1\b1})_\alpha= \n_\alpha \in \Z~,
\ee
in addition to the twisted masses $\sigma_\alpha= m_\alpha$. This adds a term:
\be
S_{\rm flux} =  \int_{\Sigma_g} d^2 x \sqrt{g} \left(-2  {f_{1 \b1}}_\alpha {\d \CW(\sigma, m)\ov \d m_\alpha} \right)
\ee
to the topological effective action \eqref{S TFT}. We are free to choose the background gauge field at will. In particular, we may consider the addition of a $\delta$-function flux at a point $x_0$ on $\Sigma_g$:
\be
(-2i  f_{1 \b1})_\alpha =2\pi \n_\alpha \, \delta^2(x-x_0)
\ee
for each $U(1)_\alpha \subset \GG_F$. In this case, we have:
\be
e^{-S_{\rm flux}} = \prod_\alpha \pif_\alpha(\sigma, m)^{\n_\alpha}~, \qquad \quad 
\pif_\alpha(\sigma, m)\equiv \exp{\left(2\pi i {\d\CW(\sigma, m)\ov \d m_\alpha}\right)}~.
\ee
Therefore, the insertion of a unit of $U(1)_\alpha$ background flux on $\Sigma_g$ can be viewed as the insertion of a {\it local operator} $\pif_\alpha$ at $x=x_0$. We will call such operators the {\it flux operators}. 

Incidentally, the handle-gluing operator \eqref{DEA in 2d} can itself be thought of as a flux operator for the vector-like $R$-symmetry. On the A-twist background, the $R$-symmetry background flux is:
\be
{1\ov 2 \pi}\int dA^{(R)}= g-1~,
\ee
in order to preserve supersymmetry.
Therefore, adding a handle has the same effect as adding one unit of $U(1)_R$ flux.

%%%%%%%%

\subsection{Three-dimensional $\CN=2$ gauge theories on a circle}\label{subsec: 3d on S1}
Let us now consider a three-dimensional $\CN=2$ supersymmetric gauge theory compactified on a circle $S^1_\beta$ of radius $\beta$. We view this theory as a two-dimensional $\CN=(2,2)$ theory with an infinite number of fields, corresponding to the Kaluza-Klein (KK) modes of each three-dimensional field. 

At finite $\beta$, the complex scalar in any $U(1)$ vector multiplet is cylinder-valued due to large gauge transformations. We introduce the notation:
\be\label{def u nu}
u_a = i \beta(\sigma_a + i a_{0\, a})~, \qquad \qquad \qquad
\nu_\alpha = i \beta(m_\alpha + i a^F_{0\, \alpha})
\ee
for the scalar fields in the Cartan of $\GG\times \GG_F$. Here $\sigma$ and $m$ are {\it real} scalars in 3d $\CN=2$ vector multiplets, and $a_0$ denotes the holonomy along $S^1$. We have the identifications $u_a \sim u_a +1$ and $\nu_\alpha \sim \nu_\alpha +1$ under large gauge transformations. The dimensionless quantity $u$ in \eqref{def u nu} is related to the two-dimensional complex scalar of Section \ref{subsec: 2d W etc} by $u = \beta \sigma_{(2d)}$. It is often convenient to work with the single-valued fugacities:
\be
x_a= e^{2 \pi i u_a}~, \qquad \qquad \qquad  y_\alpha = e^{2 \pi i \nu_\alpha}~.
\ee
The low energy theory on the Coulomb branch (with coordinates $u_a$) is still governed by the topological effective action \eqref{S TFT}, but the twisted superpotential $\CW(u)$ and the effective dilaton $\Omega(u)$ have  new features intimately related to three-dimensional physics.
In the following, it will be convenient to rescale $\CW$ according to $\CW_{\rm 3d}= \beta \CW_{\rm 2d}$, so that both $\CW$ and $\Omega$ are dimensionless quantities.

%%%%%%%%
\subsubsection{The three-dimensional twisted superpotential}\label{subsec: W3d}
The classical part of the twisted superpotential is related to Chern-Simons interactions in three dimensions.  Consider any $U(1)_a$ vector multiplet. A Chern-Simons interaction with level $k_{aa}\in \Z$ contributes to the twisted superpotential as:
\be\label{WCS aa}
\CW_{{\rm CS}, aa} =  k_{aa}\, \half u_a(u_a+1)~.
\ee
This can be derived by direct evaluation of the Chern-Simons functional on $\Sigma_g \times S^1$, for instance, as we will explain in Section \ref{sec: Loc}.  The quadratic piece is essentially a mass term, corresponding to the well-known fact that the CS interaction lifts the three-dimensional Coulomb branch classically.
The linear piece in \eqref{WCS aa} is related to the subtle signs alluded to in the introduction (see also Section \ref{sec: Loc} and Appendix \ref{app: spin CS}). Although this is not single-valued, it may only shift by terms of the form \eqref{W branch cut}, which do not affect the path-integral.
For future reference, we may rewrite \eqref{WCS aa} as a function of $x_a$:
\be\label{WCS aa bis}
\CW_{{\rm CS}, aa} =     {k_{aa}\ov (2 \pi i)^2} \half \left(\log^2(-x_a)+ \pi^2\right)~,
\ee
with a branch cut along the positive real axis $x_a\in [0, \infty)$. (Here the $\log$ is on its principal branch, so that $\log(-x)= \log{x}+ \pi i$.)

Similarly, a mixed CS term between $U(1)_a$ and $U(1)_b$ contributes:
\be\label{WCS ab}
\CW_{{\rm CS}, ab} =  k_{ab}\,u_a u_b= {k_{ab}\ov (2 \pi i)^2} \log{x_a}\log{x_b}~.
\ee
In addition, we claim that the supersymmetric gravitational Chern-Simons term \cite{Closset:2012vp} contributes a constant term:
\be\label{W CS g}
\CW_{{\rm CS}, g}= {k_g\ov 24}~,
\ee
with $k_g\in \Z$. We will give several justifications for this claim below. In total, the contribution of all the gauge, flavor and gravitational CS terms to the twisted superpotential read:
\bea\label{WCS gen}
&\CW_{{\rm CS}}(u, \nu)&=&\; \sum_a {k_{aa}\ov 2} u_a (u_a+1)+ \sum_{a> b} k_{ab} u_a u_b 
+ \sum_\alpha {k_{\alpha\alpha}\ov 2} \nu_\alpha (\nu_\alpha+1)\cr
&&& + \sum_{\alpha> \beta} k_{\alpha\beta} \nu_\alpha \nu_\beta + \sum_{a, \alpha} k_{a\alpha} u_a \nu_\beta + {k_g\ov 24}~,
\eea
where all the levels are integer-quantized. For any simple group $\GG_\gamma \subset \GG$, we have $k_{a' b'}= h_{a' b'} k_\gamma$ with $h_{a' b'}$ the Killing form of $\GG_\gamma$ (and $a', b'$ running over its Cartan subgroup). 

Consider next the one-loop contribution of the three-dimensional chiral multiplets. 
A chiral multiplet $\Phi$ with charge $1$ under some $U(1)$ symmetry contributes:
\be\label{Wphi}
\CW_\Phi =- {1\ov 2 \pi i} \sum_{n\in \Z} (u+ n) \left(\log{(u+n)}-1\right)= {1\ov (2 \pi i)^2} \dilog{(x)}~,
\ee
with $u$ the effective twisted mass of $\Phi$ and $x=e^{2\pi i u}$; the $U(1)$ symmetry could be dynamical, flavor or a combination of both.
The first equality in \eqref{Wphi} gives $\CW_\Phi$ as a formal sum over KK modes. Upon regulating that expression, we obtain the dilogarithm of $x$. As we will explain in Section \ref{sec: Loc}, we have implicitly chosen a regularization scheme that preserves gauge invariance at the expense of ``parity''. This is often stated as a ``$U(1)_{-\half}$ quantization'' of the chiral multiplet, wherein we turn on a ``half-integer CS level to cancel the parity anomaly''.  In this work, we never consider ``half-integer'' CS levels since they are not well-defined. The quantization of the chiral multiplet implicit in \eqref{Wphi} is gauge-invariant and includes a contact term $\kappa=-\half$ for the $U(1)$ current two-point function \cite{Closset:2012vp}. We also have a gravitational contact term $\kappa_g=-1$.
The only scheme ambiguity is in shifting $\kappa$ by an integer CS level $k$ (and $\kappa_g$ by an integer $k_g$), corresponding to
\be
\CW=  {1\ov (2 \pi i)^2} \left(\dilog{(x)} +{k\ov 2}\left(\log^2(-x)+ \pi^2\right) -  {\pi^2\ov 6} k_g\right)~,
\ee
which would correspond to a ``$U(1)_{-\half +k}$ quantization''. An important consistency check of the twisted superpotential  \eqref{Wphi} is that it reproduces the correct decoupling limits at large value of the three-dimensional real mass $\sigma$. We have:
\be\label{W limit}
\lim_{\sigma \rightarrow +\infty}\CW_\Phi = 0~, \qquad\qquad \lim_{\sigma \rightarrow -\infty}\CW_\Phi = -{1\ov 12} + {1\ov 8 \pi^2} \left(\log^2(-x)+ \pi^2\right)~,
\ee
which corresponds to the expected shift of the contact terms:
\be
\delta \kappa = \half \sign(\sigma)~, \qquad\qquad \delta \kappa_g = \sign(\sigma)~.
\ee
For large positive $\sigma$, we obtain an empty theory and the twisted superpotential vanishes, while at large negative $\sigma$ we are left with the background $U(1)$ and gravitational CS levels $k=-1$ and $k_g=-2$, as we can see by comparing \eqref{W limit} to \eqref{WCS aa bis} and \eqref{W CS g}. This gives a first consistency check of  the detailed form of \eqref{WCS gen}. We can easily generalize this consistency check to chiral multiplets coupled to arbitrary background gauge fields. We refer to Section \ref{subsec: regulated Zphi} for additional discussions of our treatment of the chiral multiplets. 

As another consistency check, let us consider a pair of two chiral multiplets $\Phi_1$, $\Phi_2$ of $U(1)$ charges $\pm 1$. Since this allows for a superpotential mass term $W= \Phi_1\Phi_2$, the low energy theory should be empty. More precisely, it is empty if we consider two multiplets with opposite contact terms, which amount to adding CS level $k=1$ and $k_g=2$, with our choice of quantization. We then have:
\be
\CW_{\Phi_1\Phi_2}= {1\ov (2 \pi i)^2}\left(\dilog(x)+ \dilog(x^{-1})+ \half \left(\log^2(-x)+ \pi^2\right)\right)+ {1\ov 12}=0~.
\ee
Here we have used the dilogarithm identity:
\be\label{dilog id 1}
\dilog(x)+ \dilog(x^{-1})  +{\pi^2\ov 6}+\half\log^2(-x)=0~.
\ee
In Section \ref{sec:duality}, we will relate other dilograrithm identities to non-trivial dualities between different gauge theories.

Finally, we should consider the effect of the W-bosons and their superpartners on the Coulomb branch, which contribute like chiral multiplets $W_\alpha$ of gauge charges $\alpha$ and $R$-charge $2$. For every pair of roots $\alpha, -\alpha$, we choose the ``symmetric'' quantization, with opposite contact terms for $W_{\alpha}$ and $W_{-\alpha}$.  Therefore, due to the identity \eqref{dilog id 1}, the W-bosons do not contribute at all to the effective twisted superpotential in three dimensions.

For general $\CN=2$ Chern-Simons-Yang-Mills matter theories with gauge group $\GG$ and chiral multiplets $\Phi_i$ in representations $\FR_i$ of $\GG$, we have the twisted superpotential:
\bea\label{CW 3d full}
\CW(u, \nu)= \CW_{\rm CS}(u, \nu) + {1\ov (2\pi i)^2}\sum_i \sum_{\rho_i \in\FR_i} \dilog(x^\rho y_i)
\eea
where the classical contribution $\CW_{\rm CS}$ is given by \eqref{WCS gen}.  Here we introduced the short-hand notation:
\be
\nu_i = \omega_i(\nu)~, \qquad\qquad y_i = y^{\omega_i}= e^{2 \pi i \nu_i}~,
\ee
where $\omega_i$ is the flavor charge of $\Phi_i$ (that is, a weight of the flavor group). Note that this twisted superpotential is only defined modulo the branch-cut ambiguities:
\be\label{ambiguity W3d}
\CW\rightarrow \CW + n^a u_a +  n^\alpha \nu_\alpha + n^0~, \qquad\qquad n^a, n^\alpha, n^0 \in \Z~.
\ee
However, all the physical observables that we will define are free from such ambiguities.

\subsubsection{Flux operators and Bethe equations}
As in two dimensions, we may define the {\it flux operators}:
\be\label{flux op 3d}
\pif_a(u, \nu) \equiv \exp{\left(2\pi i {\d\CW(u, \nu)\ov \d u_a}\right)}~, \qquad \qquad
\pif_\alpha(u, \nu) \equiv \exp{\left(2\pi i {\d\CW(u, \nu)\ov \d \nu_\alpha}\right)}~, 
\ee
for the gauge and flavor symmetries, respectively. This is obviously invariant under \eqref{ambiguity W3d}.
One can check that these operators are rational functions of the fugacities $x_a$ and $y_\alpha$. The three-dimensional flux operators are loop operators supported along the $S^1$ direction. They can be identified with the vortex loops discussed in \cite{Drukker:2012sr,Kapustin:2012iw}. The Bethe vacua are given by:
\be\label{S BE 3d}
\cS_{BE} =\left\{ \;\h u_a \; \bigg| \; \Pi_a(\h u, \nu) = 1~, \;\; \forall  a~, \quad w \cdot \h u \neq \h u, \;\; \forall w \in W_G \; \right\} /W_\GG~.
\ee
In particular, they are rational equations for the single-valued variables $x_a$.

\subsubsection{The effective dilaton and the handle-gluing operator}
If we couple the 3d $\CN=2$ theory to a $\Sigma_g\times S^1$ background with the A-twist along $\Sigma_g$, the effective dilaton $\Omega$ can be computed like in two dimensions \cite{Nekrasov:2014xaa}. As we will further discuss in Section \ref{sec: Loc}, the classical Chern-Simons terms for the $U(1)_R$ background gauge field \cite{Closset:2012vp} contributes:
\be\label{OmegaCS}
\Omega_{\rm CS}(u, \nu)=  \sum_a k_{aR} u_a + \sum_\alpha k_{\alpha R}\nu_\alpha + \half k_{RR}~.
\ee
Here $k_{aR}$, $k_{\alpha R}$ denote mixed $R$-gauge and $R$-flavor CS levels, and $k_{RR}$ is the $U(1)_R$ CS level. All these levels are integer-quantized. 
A chiral multiplet $\Phi$ of $U(1)$ gauge charge $1$ and $R$-charge $r\in \Z$ contributes:
\be
\Omega_\Phi = -{1\ov 2 \pi i }(r-1) \log{\left(1-x\right)}~.
\ee
This corresponds to the same ``$U(1)_{-\half}$ quantization'' discussed above, which includes the contact terms $\kappa_{R}= -\half(r-1)$ and $\kappa_{RR}= - \half (r-1)^2$
for the gauge-$R$ and $R$-$R$ conserved-current two-point functions, respectively. The limits
\be\label{lim Omega 3d}
\lim_{\sigma \rightarrow +\infty}\Omega_\Phi = 0~, \qquad\qquad \lim_{\sigma \rightarrow -\infty}\Omega_\Phi = -{1\ov 2 \pi i} (r-1)\log{x}- \half(r-1)~,
\ee
reproduce the correct shifts of the  $U(1)_R$ CS terms upon integrating out a chiral multiplet.~\footnote{That is, taking into account that $\Omega$ is only defined modulo an integer. The second limit in \protect\eqref{lim Omega 3d} corresponds to CS levels $k_{R}=-(r-1)$ and $k_{RR}= - (r-1)^2$.}  The W-bosons contributes similarly like chiral multiplets of $R$-charge $r=2$. Due to our choice of ``symmetric quantization'' mentioned above, we also have a shift of $k_{RR}$ by $ \half{\rm dim}(\Fg/\Fh)$.

In total, the effective dilaton of our 3d $\CN=2$ supersymmetric gauge theory compactified on $S^1$ reads:
\bea\label{Omega full}
&\Omega(u, \nu) &=&\; \sum_a k_{aR} u_a + \sum_\alpha k_{\alpha R}\nu_\alpha + \half \Big(k_{RR}+ \half{\rm dim}(\Fg/\Fh)\Big)\cr
&&&\; - {1\ov 2 \pi i}  \sum_i (r_i-1)\sum_{\rho_i\in \FR_i}  \log(1- x^{\rho_i} y_i)   - {1\ov 2 \pi i}\sum_{\alpha\in \Fg} \log(1-x^\alpha)~,
\eea
with $r_i\in \Z$ the $R$-charge of $\Phi_i$. The three-dimensional handle-gluing operator is given by:
\be\label{HGO 3d}
\CH(u, \nu) = \exp\Big(2 \pi i \, \Omega(u, \nu)\Big) \, \det_{ab}\left({\d^2 \CW(u, \nu)\ov  \d{u_a}\d{u_b}}\right)~.
\ee
This directly leads to an expression for the $\Sigma_g\times S^1$ twisted index as a sum over Bethe vacua   \cite{Nekrasov:2014xaa, Closset:2016arn, Benini:2015noa}. Note that we accounted for the effect of the CS level $k_{RR}$ in \eqref{Omega full}. This leads to a subtle sign $(-1)^{k_{RR}}$ in \eqref{HGO 3d}, which was previously overlooked.

\subsection{Induced charges of monopole operators}
For future reference, let us consider the induced charges of the bare monopole operators $T_{a\pm}$. These operators are associated with the limit $\sigma_a \rightarrow \mp \infty$ on the classical Coulomb branch. We define their induced charges by:
\bea\label{induced charges T}
&{Q_{a\pm}}^b &\equiv&\; Q^b[T_{a\pm}] &=&\; \pm \lim_{u_a \rightarrow \mp i \infty} \d_{u_a}\d_{u_b} \CW~, \cr
&{Q^F_{a\pm}}^\alpha &\equiv&\; Q^\alpha[T_{a\pm}] &=&\; \pm \lim_{\nu_\alpha \rightarrow \mp i \infty} \d_{u_a}\d_{u_b} \CW~, \cr
&{r_{a\pm}} &\equiv&\; R[T_{a\pm}] &=&\; \pm \lim_{u_a \rightarrow \mp i \infty} \d_{u_a}\Omega~,
\eea
for their gauge, flavor and $R$-charges, respectively. One can easily check that these formula reproduce the standard one-loop formula for the induces charges; see {\it e.g.} \cite{Closset:2016arn}. By construction, the charges \eqref{induced charges T} are always integers.

\subsection{The fibering operator in three dimensions}\label{subsec:fibering op}
In addition to the three-dimensional flavor symmetry group $\GG_F$, the effective two-dimensional theory has a $U(1)_{KK}$ symmetry whose charge is the KK momentum. We may turn on a supersymmetric background vector multiplet $\CV_{KK}$ for $U(1)_{KK}$. It originates from the three-dimensional $\CN=2$ ``new-minimal'' supergravity multiplet---see {\it e.g.} \cite{Closset:2012vp, Kuzenko:2013uya}---which decomposes into a supergravity and a vector multiplet upon KK reduction to two dimensions.  The twisted mass associated to $\CV_{KK}$ is $m_{KK}= {1\ov \beta}$. Indeed, the twisted masses for the KK tower of any 3d chiral multiplet takes the form $\sigma_{\rm 2d}+ {n\ov \beta}$, with $n\in \Z$ the KK momenta.

In any three-dimensional $\CN=2$ theory, there must exist a distinguished flux operator for $U(1)_{KK}$, which we denote by $\CF$. The insertion of $\CF$ at a point on $\Sigma_g$ has the effect of introducing one unit of flux for $U(1)_{KK}$, which is nothing but a shift of the first Chern class of the $U(1)$ principal bundle over $\Sigma_g$.  In particular, the partition function of $\Mgp$ can be written in terms of $p$ insertions of $\CF$ on $\Sigma_g \times S^1$:
\be
Z_{\Mgp}= \langle \CF^p \rangle_{\Sigma_g \times S^1}~.
\ee
Since $\CF$ introduces a non-trivial fibration of the circle over $\Sigma_g$, we call it the {\it fibering operator}.
Reinstating dimensions, we have:
\be
\CF(u, \nu) = \exp{\left(2 \pi i\, {\d   \ov \d m_{KK}}  \Big(m_{KK} \CW(u, \nu)\Big) \right)}~,\qquad u= {\sigma_{\rm 2d}\ov m_{KK}}~, \quad \nu = {m_{\rm 2d}\ov m_{KK}}~,
\ee
with the dimensionless $\CW(u, \nu)$ given by \eqref{CW 3d full}. This gives us the explicit form of the fibering operator for any 3d $\CN=2$ gauge theory:
\be\label{CF explicit 3d}
\CF(u, \nu)  = \exp{\left(2\pi i \left(\CW(u, \nu)- u_a {\d \CW(u, \nu)\ov \d u_a} - \nu_\alpha {\d \CW(u, \nu)\ov \d \nu_\alpha}  \right) \right)}~.
\ee
We immediately see that \eqref{CF explicit 3d} is insensitive to the branch-cut ambiguities \eqref{ambiguity W3d} of the twisted superpotential. On the other hand, it transforms non-trivially under large gauge transformations $u\sim u+1$ or $\nu\sim \nu+1$ (for either the gauge or flavor group). We find:
\be\label{prop F shift}
\CF(u_a- \m_a, \nu_\alpha - \n_\alpha)=\CF(u, \nu)\, \prod_a \pif_a(u, \nu)^{\m_a} \,   \prod_\alpha \pif_\alpha(u, \nu)^{\n_\alpha}~, \quad \forall \m_a, \n_\alpha \in \Z~,
\ee
where $\pif_a, \pif_\alpha$ are the flux operators defined in \eqref{flux op 3d}.

\subsubsection{The Chern-Simons and chiral multiplet fibering operator}
For future reference, we note that the effect of the classical CS terms \eqref{WCS gen} on the fibering operator is:
\be
\CF_{\rm CS}= \exp{\left(- \pi i \sum_{a,b}k_{ab} u_a u_b - 2 \pi i \sum_{a, \alpha} k_{a\alpha}u_a\nu_\alpha - \pi i \sum_{\alpha, \beta} k_{\alpha,\beta} \nu_\alpha\nu_\beta + {\pi i \ov 12 } k_g \right)}~.
\ee
Similarly, a chiral multiplet of charge $1$ under some $U(1)$ contributes:
\be\label{CFPhi def}
\CF_\Phi(u) = \exp{\left({1\ov 2 \pi i} \dilog\left(e^{2\pi i u}\right)+ u \log\left(1-e^{2\pi i u}\right)\right)}~.
\ee
This defines a meromorphic function of $u$ on the complex plane, as the branch cuts of the dilogarithm and logarithm cancel each other. The function \eqref{CFPhi def} has poles of order $n$ at $u=-n$, $n\in \Z_{>0}$ and zeros of order $n$ at $u=n$, $n\in \Z_{>0}$. (This is proven {\it e.g.} by proposition 5.1 of \cite{1999math......7061F}.) It is closely related to the chiral multiplet one-loop determinant on $S^3$, as we discuss further in Section \ref{subsec: S3 free chiral}.

We note that the Chern-Simons and chiral fibering operators satisfy:
\be \label{parity transformatin} u \rightarrow -u, \;\;\; \nu \rightarrow -\nu, \;\;\; ``k" \rightarrow -``k"\qquad\Rightarrow \qquad \CF \rightarrow \CF^{-1} \ee
where ``$k$'' denotes all Chern-Simons levels and contact terms in the theory, including the gravitational Chern-Simons level and the contact terms appearing in the quantization of the chiral multiplet.  This operation thus has the same effect as taking $p \rightarrow -p$.  As discussed further in Section \ref{subsec: regulated Zphi}, this reflects the fact that the $\Mgp$ and $\CM_{g,-p}$ backgrounds are related by a parity transformation.

\subsection{Partition function and loop-operator correlation functions}
Combining all the ingredients introduced so far, we can write the supersymmetric partition function on $\Mgp$ as:
\be\label{ZMgp main}
Z_{\Mgp}(\nu)= \sum_{\h u \in \CS_{\rm BE}} \CF(\h u,\nu)^p\, \CH(\h u,\nu)^{g-1} \,\prod_{\alpha} \pif_\alpha(\h u,\nu)^{\n_\alpha}
\ee
Here we introduced generic background fluxes $\n_\alpha$ for the flavor symmetry. As we discussed, we can also view these background fluxes as inserting flux operators $\Pi_\alpha$ at points on $\Sigma_g$. (A constant background flux is then viewed as a ``smeared'' flux operator.)
Note that, in the presence of any abelian flavor symmetry $U(1)_F$, we may shift the $R$-symmetry by $R\rightarrow R+ t F$, where $t$ is quantized to preserve the Dirac quantization of the $R$-charge. The net effect on the partition function is to shift the background flux $\n_F \rightarrow \n_F + (g-1)t$. This amounts to a shift:
\be
\Omega \rightarrow \Omega + t {\d \CW\ov \d \nu_F}
\ee
in the topological effective action \eqref{S TFT}. The partition function \eqref{ZMgp main} is unaffected if we shift the $R$-symmetry current by any abelian gauge current.

We are also interested in supersymmetric Wilson loop operators along the $S^1$ fiber.  
Any such Wilson loop correspond to a Weyl-invariant Laurent polynomial in the fugacities $x_a$,
\be
W(x) \in \C[x_a, x_a^{-1}]^{W_\GG}~.
\ee
For a Wilson loop in a representation $\FR$ of $\GG$, we have:
\be
W_{\FR} = \Tr_{\FR} {\rm Pexp}{\left( -i \int_{S^1} \left(a_\mu dx^\mu - i \beta \sigma d\psi\right)\right)}= \Tr_{\FR}\left(x\right)= \sum_{\rho\in \FR} x^\rho~,
\ee
where $\psi$ the fiber coordinate. We then have the expectation value:
\be\label{W corr 3d}
\langle W(x) \rangle_{\Mgp} = \sum_{\h u \in \CS_{\rm BE}} W(\h x)\, \CF(\h u,\nu)^p\, \CH(\h u,\nu)^{g-1} \,\prod_{\alpha} \pif_\alpha(\h u,\nu)^{\n_\alpha}~.
\ee
From this formula, we can read off the quantum algebra of Wilson loops, which is an $S^1$ uplift of the 2d $\CN=(2,2)$ twisted chiral ring  \cite{Kapustin:2013hpk, Closset:2016arn}. The quantum relations are the relations satisfied by solutions to the Bethe equations \eqref{S BE 3d}.

Let us briefly comment on the defect operators $\Pi_\alpha$. They enter in \eqref{W corr 3d} in the same way as the Wilson loops, in agreement with their interpretation as operators supported along $S^1$ at a particular point on the base $\Sigma_g$. These line operators can be identified with the vortex loop operators discussed in \cite{Drukker:2012sr,Kapustin:2012iw}.  In principle, one can  insert fractional flux at points on $\Sigma_g$ as long as the total flux is integer. The effect of such operators is to impose that matter fields charged under the flavor symmetry induce a non-trivial holonomy as they wind around the vortex loop. The Bethe equations imply relations satisfied by flux operators, just like for Wilson loops.

%%%%%
\subsection{Gauging flavor symmetries and the on-shell twisted superpotential}\label{sec:onshell}
Given a flavor symmetry, it is natural to gauge it, by promoting background vector multiplets to dynamical ones. 
 This is an important operation for producing new theories from old ones, and we would like to perform it at the level of the partition function \eqref{ZMgp main}.  This can be done most conveniently by working with the ``on-shell'' effective twisted superpotentials and effective dilatons,
 \bea\label{Wonshell}
 &\CW^l(\nu)\equiv \CW(\h u^l(\nu), \nu)~,\cr
& \Omega^l(\nu)\equiv \Omega(\h u^l(\nu)~, \nu)+  \log\left(\det_{a,b} {\d\CW\ov \d u_a \d u_b}\right)\Big|_{u=\h u^l(\nu)}~, 
 \eea
 which are evaluated at solutions $\h u^l$ to the Bethe equations. 
The functions \eqref{Wonshell} are particularly useful because we can use them to construct all of the ingredients in the partition function \eqref{ZMgp main}, even if one does not have access to a Lagrangian description of the theory.

As described above, the supersymmetric vacua of the theory are determined by solutions to the Bethe equation \eqref{S BE 3d}.  For generic-enough mass parameters $\nu_\alpha$, this has a finite number of solutions, 
\be \label{Bethe solutions}
 \h u_a^l(\nu)~, \qquad  l =1,...,|\CS_{\rm BE}|
\ee
It is important to stress that the functions $\h u_a^l(\nu)$ generically have branch points, where two or more solutions become equal, and branch cuts, where the solutions are permuted.  Thus it is more natural to think of $\h u_a^l(\nu)$ as functions on an $|\CS_{\rm BE}|$-fold branched cover of the space of  the $\nu_\alpha$'s.

To any  Bethe vacua, we may associate the  ``on-shell'' effective twisted superpotential and effective dilaton \eqref{Wonshell}, which we consider as a function on the $|\CS_{\rm BE}|$-fold branched cover of the parameter space.  Nonetheless, the twisted superpotential $\CW^l$ is not yet well-defined due to branch cut ambiguities of $\CW$ itself.  To partially fix this ambiguity, we impose a ``physical branch'' condition:\footnote{Namely, the Bethe equation, \protect\eqref{S BE 3d}, only imposes that the RHS is an integer, however, by ``changing the branch'' by adding appropriate integer multiples of $u_a$ to $\CW$, we may arrange that the RHS is precisely zero.}
\be \label{physical branch condition}
\frac{\partial \CW}{\partial u_a}(u_a^l(\nu),\nu) = 0~, \qquad a=1,...,\rk~.
\ee
This function will still have branch cut ambiguities associated to the background gauge multiplets, \ie, it is defined only up to shifts $\CW^l \rightarrow \CW^l + m^\alpha \nu_\alpha + m_0$, $m^\alpha,m_0\in \Z$,  but it will  not have any branch cuts associated to shifts by the dynamical gauge field.  This must be the case, as the dynamical gauge field should play no role in the low energy effective theory.  Up to these shifts, the on-shell effective twisted superpotential is a physically-meaningful observable of the low-energy theory. In particular, it should match across dualities.  Similar statements hold for the on-shell effective dilaton.

If one has access to the on-shell effective twisted superpotentials of a theory, one may construct the on-shell flux and fibering operators, even if the theory lacks a known Lagrangian description. They are given by:
\be \label{on shell fiber and flux}
\pif_\alpha^l(\nu) = \exp\left( 2 \pi i\frac{\partial \CW^l(\nu)}{\partial \nu_\alpha} \right)~,\quad
 \CF^l(\nu) = \exp\left( 2 \pi  i \left( \CW^l(\nu) - \nu_\alpha \frac{\partial \CW^l(\nu)}{\partial \nu_\alpha} \right)\right)~.
 \ee
 We can easily see that this agrees with the gauge-theory definitions \eqref{flux op 3d} and \eqref{CF explicit 3d} upon using \eqref{physical branch condition}.  Similarly, the on-shell handle-gluing operator is simply defined by:
 \be
 \CH^l(\nu) =  \exp\Big(2 \pi i \, \Omega^l(\nu)\Big)~,
 \ee
  which obviously agrees with \eqref{HGO 3d}.
  
Using the on-shell twisted superpotential, it is straightforward to gauge a flavor symmetry.  For instance, suppose we want to gauge a subgroup of the flavor group $\GG_F$, with parameters $\{\nu_a \} \subset\{\nu_\alpha\}$.    We simply write the Bethe equation for $\nu_a$  in terms of $\CW^l$, namely:
\be
\exp\left(2 \pi i {\d \CW^l(\nu)\ov \d \nu_a}\right)= 1~.
 \ee
These equations should be solved for each $l$, and may have zero, one, or several solutions for each $l$.  The vacua of the new gauge theory is the union of these solutions for all $l$, and the resulting on-shell twisted superpotential can be used to construct the $\Mgp$ partition function of the new theory.  This procedure is described in more detail in Appendix \ref{app: gauging}. We will see an example of this procedure in Section \ref{sec:duality}.

%%%%%%%%%%%%

%%%%%%%%%%%%
\section{$A$-twisted supersymmetric theories on $\Mgp$}\label{sec: 3}
In this section, we  study curved-space rigid supersymmetry on $\Mgp$. We introduce a particular three-dimensional supergravity background which realizes the ``three-dimensional A-twist'' in a precise sense.
We also discuss curved-space supermultiplets and Lagrangians on this background, 
following the general results of \cite{Closset:2012ru, Closset:2013vra}.

\subsection{Supersymmetric background on $\Mgp$}
Consider the three-manifold $\Mgp$, a principal $U(1)$ bundle of first Chern number $p\in \Z$ over a closed oriented Riemann surface $\Sigma_g$ of genus $g$:
\be\label{Mgp as bundle}
S^1 \longrightarrow \Mgp \stackrel{\pi}\longrightarrow \Sigma_g~.
\ee
This is a simple example of a Seifert fibration. The topology of $\Mgp$ is fully specified by the two integer $p\in \Z$ and $g\in \Z_{\geq 0}$. 
In particular, if $p\neq 0$ we have the second cohomology:
\be\label{H2 with Zp}
H^2(\Mgp, \Z)  \cong \Z^{2g}\oplus \Z_p~,
\ee
which includes the torsion subgroup $\Z_p$.
A more detailed account of the topology and geometry of $\Mgp$ is provided in Appendix \ref{app: geom}. 
Let us consider the metric
\be\label{metric Mgp}
ds^2(\Mgp)= \beta^2 \big(d\psi +  \CC(z,\bz) \big)^2 + 2 g_{z\bz}(z, \bz) dz d\bz~,
\ee
with $\psi \in [0, 2 \pi)$ the fiber coordinate, and $z$ a complex coordinate on the base $\Sigma_g$ (in a given patch). 
The principal bundle connection $\CC$ has field strength:
\be
\d_z \CC_\bz- \d_\bz \CC_z = {2 \pi i \, p\ov {\rm vol}(\Sigma_g)} g_{z\bz}~.
\ee
We normalize the volume of the base to ${\rm vol}(\Sigma_g)= \pi$, so that $\CC$ has flux 
\be
{1\ov 2\pi}\int_{\Sigma_g} d\CC = p
\ee
on $\Sigma_g$.
The metric \eqref{metric Mgp} admits a Killing vector $K$ whose orbits are the $S^1$ fibers. The dual one-form $\eta$ determines a transversely holomorphic foliation (THF) 
of $\Mgp$. We define:
\be
K\equiv K^\mu \d_\mu = {1\ov \beta} \d_\psi~,\qquad\qquad
\eta \equiv K_\mu dx^\mu =  \beta\left(d\psi + p \CA\right)~.
\ee
Note that $K^\mu \eta_\mu=1$. We also define the tensor:
\be
{\Phi_\mu}^\nu= - {\epsilon_\mu}^{\nu\rho}\,\eta_\mu~,
\ee
which acts as a three-dimensional `complex structure'. (We summarize important aspects of this geometric structure in Appendix \ref{app: geom}.)
The complex coordinates $\psi, z, \bz$ introduced above are coordinates adapted to the THF.

In order to preserve half of the flat-space supersymmetry on $\Mgp$, we turn on additional background fields in the three-dimensional `new minimal'  $\CN=2$ supergravity multiplet \cite{Festuccia:2011ws, Klare:2012gn, Closset:2012ru}. This includes a scalar $H$ and a $U(1)_R$ gauge field $\CA_\mu^{(R)}$:
\be\label{H and CAR}
H= i p \beta~, \qquad \qquad
\CA^{(R)}_\mu= {1\ov 8} {\Phi_\mu}^\nu \d_\nu \log{g} + \d_\mu s
\ee
with $g$ the metric determinant. The complete supergravity background is spelled out in Appendix  \ref{app: 3d Atwist}.  The expression for  $\CA^{(R)}_\mu$ in \eqref{H and CAR} is only valid in the  adapted coordinates $\psi, z, \bz$. Let us also define the adapted frame:
\be\label{canonical frame}
e^0= \beta\left(d\psi + p \CA\right)~, \qquad\quad  e^1=\sqrt{2 g_{z\bz}}dz~, \qquad \quad e^{\b1}=\sqrt{2 g_{z\bz}} d\bz~.
\ee
Any one-form $\alpha$ can be decomposed into `vertical', `holomorphic' and `anti-holomorphic' components:
\be\label{alpha decomp}
\alpha= \alpha_0 \eta + \alpha_z dz+ \alpha_\bz d\bz~,
\ee
and similarly for any tensor. (In the following, we will mostly use the frame basis.)
The holomorphic component $\alpha_z$ in \eqref{alpha decomp} transforms as a section of a ``canonical line bundle'' on $\Mgp$, denoted by $\CK$,  which is the pull-back of the canonical line bundle on the Riemann surface $\Sigma_g$ through the projection $\pi$ in \eqref{Mgp as bundle}. Its first Chern class is given by:
\be\label{c1K}
c_1(\CK)= 2g-2 \in \Z_p~,
\ee
where $\Z_p$ is the torsion subgroup in \eqref{H2 with Zp}. It is very natural to introduce a modified Levi-Civita connection $\hat\nabla_\mu$ that preserves the decomposition \eqref{alpha decomp}.
Following \cite{Closset:2012ru}, we define the modified spin connection:
\be\label{def omega hat}
\h \omega_{\mu\nu\rho}= \omega_{\mu\nu\rho} - i H \left(\eta_\nu \Phi_{\mu\rho}-\eta_\rho \Phi_{\mu\nu}+\eta_\mu \Phi_{\nu\rho}\right)~,
\ee
with $\omega_{\mu\nu\rho}$ the standard spin connection.
 In particular, we have:
 \be\label{compatibility condition}
 \h\nabla_\mu g_{\nu\rho} = 0~, \qquad \qquad   \h\nabla_\mu \eta_\nu = 0~.
 \ee
 The price to pay is that the modified connection has torsion, with the torsion tensor ${ T^{\mu}}_{\nu\rho}=2 i H \, \eta^\mu\, \Phi_{\nu\rho}$  proportional to $H$.

The supergravity background \eqref{metric Mgp}-\eqref{H and CAR} preserves two (generalized) Killing spinors $\zeta$ and $\t\zeta$, of $R$-charge $1$ and $-1$, respectively, which satisfy: 
\be\label{KSE twist}
\big(\h \nabla_\mu - i \CA^{(R)}_\mu\big)\zeta=0~, \qquad \qquad
\big(\h \nabla_\mu + i \CA^{(R)}_\mu\big)\t\zeta=0~,
\ee
with $\CA^{(R)}_\mu$ given above. The holonomy of  the modified connection $\h\nabla_\mu$ is contained in $U(1)$, therefore it can be ``twisted'' away by a compensating $U(1)_R$ transformation. The Killing spinors are then essentially constant in the adapted frame:
\be\label{KS explicit}
\zeta=e^{i s}\mat{0\cr 1}~, \qquad\qquad \t\zeta=e^{-i s}\mat{1\cr 0}
\ee
 This is the three-dimensional version of the A-twist. Geometrically, it corresponds to choosing the $U(1)_R$ line bundle $L^{(R)}$ such that:
\be\label{A twist as bundles}
L^{(R)}\cong \sqrt\CK~.
\ee
This is a torsion line bundle with first Chern class:
\be\label{c1R}
c_1(L^{(R)})= g-1 \in \Z_p~,
\ee
with the connection  $\CA_\mu^{(R)}$ given by \eqref{H and CAR}.
It follows that the $R$-charges must be integers in general. More precisely, we have the Dirac quantization condition:
\be\label{dqc R}
r (g-1) \in \Z~,
\ee
with $r$ the $R$-charge of any field. 
Note that the $U(1)_R$ bundle is topologically trivial if and only if $g-1=0 \mod p$. For instance, this is the case for the three-sphere $\CM_{0,1}\cong S^3$.

The function $s$ in \eqref{H and CAR} and \eqref{KS explicit} corresponds to  a $U(1)_R$ gauge transformation. The Killing spinors \eqref{KS explicit} are globally well-defined if we choose $s=0$. We may call this choice the ``A-twist gauge''. More generally, we can choose a gauge $s=- n \psi$, where $n$ is any integer; we will come back to this point below.

 Note also that the Killing vector $K$ and the covector $\eta$ are built out of the Killing spinors \eqref{KS explicit} according to:
 \be\label{K as spinor}
K^\mu = \zeta\gamma^\mu \t \zeta~, \qquad\qquad 
\eta_\mu=-{ \zeta^\dagger \gamma_\mu \zeta�\over |\zeta|^2} = {\t \zeta^\dagger \gamma_\mu\t \zeta�\over |\t\zeta|^2} ~, 
 \ee
with $\eta_\mu= K_\mu$ in our background. All the background fields are invariant under the isometry generated by $K$. The compatibility condition \eqref{compatibility condition} directly follows from \eqref{KSE twist} and \eqref{K as spinor}.

\subsubsection{Background vector multiplets}
In addition to the background supergravity fields  \eqref{metric Mgp}-\eqref{H and CAR}, we may also turn on background vector multiplets:
\be
\CV^{(F)}= \big(a_\mu^{(F)}~, \quad \sigma^{(F)}~, \quad D^{(F)}\big),
\ee
for any flavor symmetry of the theory.  To preserve the same supersymmetry as the geometric background, we take:
\be
\sigma^{(F)}= m^{(F)}~,
\ee
a constant, which is the {\it real mass} associated to the flavor symmetry, and:
\be\label{susy cond backg F}
f^{(F)}_{01}=f^{(F)}_{0\b1}=0~, \qquad \qquad   D^{(F)} = 2 i f^{(F)}_{1\b 1} +\sigma^{(F)} H~,
\ee
with $f_{\mu\nu}^{(F)}$ the field strength of $a_\mu^{(F)}$ and  $H$ given in \eqref{H and CAR}.
This implies that $a_\mu^{(F)}$ is the connection of a {\it holomorphic} vector bundle over $\Mgp$ \cite{Closset:2013vra}. In particular, let us choose a holomorphic line bundle $L^{(F)}$ associated to a $U(1)_F$ flavor symmetry. Its first Chern class has to lie in the torsion subgroup of the second cohomology \eqref{H2 with Zp}:
\be\label{c1F}
c_1(L^{(F)})= \n_F \in \Z_p~,
\ee
assuming $p\neq 0$. (See \cite{Closset:2016arn} for the $p=0$ case.)  Let us also define:
\be\label{def nuF}
\nu^{(F)} = i \beta \big(m^{(F)} + i a_0^{(F)}\big)~,
\ee
where $a_0^{(F)}= \eta^\mu a_\mu^{(F)}$ is taken to be constant. The quantity \eqref{def nuF} has a nice geometric interpretation as a complex modulus of the holomorphic line bundle $L^{(F)}$   \cite{Closset:2013vra}. Under a large gauge transformation along the circle fiber, the parameters $\nu^{(F)}$ and $\n^{(F)}$ transform as:
\be
\left(\nu^{(F)}~,\; \n^{(F)} \right)\sim \left(\nu^{(F)}+1~,\; \n^{(F)}+p \right)~.
\ee
This must be an invariance of any physical observable.

\subsubsection{$U(1)_R$ vector multiplet}
From the supergravity multiplet, one can also construct an abelian vector multiplet for the $R$-symmetry \cite{Closset:2014uda}. In terms of the supersymmetric background \eqref{metric Mgp}-\eqref{H and CAR}, it is given by:
\be\label{VR def}
\CV^{(R)}\equiv  \big(a_\mu^{(R)}~, \; \sigma^{(R)}~, \;  D^{(R)}\big) =\left(\CA_\mu^{(R)}+ i H  \eta_\mu~, \; H~, \; {1\ov 4} (R-6 H^2)\right)~,
\ee
where $R$ is the Ricci scalar of $g_{\mu\nu}$.
In particular, one can check that the supersymmetry conditions \eqref{susy cond backg F} are satisfied:
\be
f^{(R)}_{01}=f^{(R)}_{0\b 1}=0~, \qquad\qquad   D^{(R)} = 2 i f^{(R)}_{1\b 1} + H^2~.
\ee
It follows that $L^{(R)}$ is a holomorphic line bundle, which is determined by its torsion flux \eqref{c1R} and by the modulus:
\be
\nu^{(R)}= i\beta\big({\sigma^{(R)}+ i a_0^{(R)}}\big)=  - \d_\psi s= n \in \Z~. 
\ee
Interestingly, $\nu^{(R)}$ is fully determined by the supergravity background. A large $U(1)_R$ gauge transformation  along the circle fiber corresponds to:
\be\label{R large gauge shift}
\left(\nu^{(R)}~,\; g-1 \right)\sim \left(\nu^{(R)}+1~,\; g-1+p \right)~.
\ee
Note that we can set $\nu^{(R)}=0$, but it is sometimes useful to keep track of $\nu^{(R)}$ as a formal parameter, together with the $U(1)_R$ gauge redundancy \eqref{R large gauge shift}.

\subsubsection{Parameter dependence and $R$-charge dependence}\label{subsec: continuous R}
Supersymmetric observables on $\Mgp$ depend explicitly on the discrete parameters $p$ and $g$ as well as on the torsion fluxes $\n^{(F)}$ for flavor symmetries.  They are also {\it  locally holomorphic} functions of the complex parameters $\nu^{(F)}$  \cite{Closset:2013vra, Closset:2014uda}.
Note that a line bundle $L^{(F)}$ generally has additional moduli, corresponding to flat connections along the one-cycles from $\Sigma_g$. In our two-supercharge background, however,  these additional moduli couple to $Q$-exact operators and supersymmetric observables are completely independent of them \cite{Closset:2013vra}. 

We can similarly understand the dependence of supersymmetric observables on the choice of $R$-symmetry \cite{Closset:2014uda}. 
In a theory with abelian flavor symmetries, the $R$-symmetry current can mix with flavor currents. Let us consider:
\be\label{shift jR}
j_\mu^{(R)} \rightarrow  j_\mu^{(R)}+ t\,  j_\mu^{(F)}~,
\ee
for some parameter $t$,  which shifts the $R$-charge by the $U(1)_F$ charge according to $R\rightarrow R+ t \,F$. (The flavor charge $F$ is integer quantized by assumption.) This is equivalent to a shift of the $U(1)_F$ vector multiplet by the $U(1)_R$ vector multiplet:
\be\label{shift Vf}
\CV^{(F)} \rightarrow  \CV^{(F)}+ t\, \CV^{(R)}~.
\ee
On our geometric background, the shift \eqref{shift jR} is only allowed if it preserves the Dirac quantization condition \eqref{dqc R}. This implies that $t \in \Z$ in general.
On the other hand, if $L^{(R)}$ is topologically trivial (that is, if $g-1=0 \mod p$), there is no restriction on the $R$-charge and we can take $t\in \R$. 
Geometrically, the shift  \eqref{shift Vf} is a tensor product of line bundles:
\be
L^{(F)}\rightarrow  L^{(F)} \otimes \left(L^{(R)}\right)^{\otimes t}~,
\ee
with $t$ integer or real, respectively. This corresponds to a shift of parameters:
\be
\nu^{(F)}\rightarrow  \nu^{(F)}+ t\, \nu^{(R)}~, \qquad \qquad
\n^{(F)}\rightarrow  \n^{(F)}+ t\, (g-1)~.
\ee
The partition function (or any supersymmetric observable) shifts accordingly.  Note that the complex modulus  $\nu^{(F)}$ stays invariant in the ``$A$-twist gauge'' $\nu^{(R)}=0$. This is the gauge that we used implicitly in Section \ref{sec: BE formula}.
When  $L^{(R)}$ is topologically trivial, another particularly interesting gauge is:
\be\label{global gauge}
\nu^{(R)} = {1-g\ov p}~, \qquad \qquad {\rm if} \quad g-1=0\mod p
\ee
In such a case, the dependence of supersymmetric observables on the $R$-charge is entirely through the combination:
\be\label{nF plus i r}
\nu^{(F)}+ t \, {1-g\ov p}~,
\ee
with $t\in \R$.  Let us note that, in the case $M_{0, 1}\cong S^3$, the supersmmetric background considered in \cite{Kapustin:2009kz, Hama:2010av, Jafferis:2010un} has $\nu^{(F)}= i \sigma$ (setting $\beta=1$ for simplicity) and therefore the dependence on the $R$-charge is holomorphic in the parameter $\sigma - i t$  \cite{Jafferis:2010un, Closset:2014uda}. As we can see from \eqref{nF plus i r}, that property generalizes to any $\Mgp$ background admitting continuous $R$-charges.

\subsubsection{Comparison with three-sphere and lens space backgrounds}\label{subsec:compare to lens}
It is interesting to compare our family of curved-space backgrounds to the ones previously studied in the literature. 
The genus zero case, $g=0$,  corresponds to the lens space $\CM_{0,p}\cong S^3/\Z_p$. For instance, we can consider the metric:
\be\label{round S3 metric}
ds^2 = \beta^2 \left(d\psi +{p\ov 2}(1-\cos\theta) d\phi\right)^2+ {1\ov 4}(d\theta^2 + \sin^2 \theta d\phi^2)~,
\ee 
with the angular coordinates $\theta\in [0, \pi]$ and 
\be\label{hopf angle periods}
(\phi, \psi) \sim (\phi, \psi + 2\pi)\sim (\phi + 2 \pi, \psi + p \pi)~.
\ee
This is the total space of a degree $p$ $U(1)$ bundle over the round $S^2$, written down on the ``northern patch'' $\theta \in [0, 2 \pi)$.~\footnote{The usual Hopf coordinates are $\theta, \phi$ and $\h\psi = {2\psi\ov p}$, with $\h\psi \in [0,{4 \pi \ov p})$.}  If we choose $\beta= {1\ov p}$, we obtain the round metric on the  $S^3/\Z_p$ quotient, and the remaining  supergravity fields (see Appendix \ref{app: 3d Atwist}) are:
\be\label{background round S3}
H= i~, \qquad V_\mu=-2 \eta_\mu~, \qquad A_\mu^{(R)}dx^\mu ={1\ov p} d\psi~.
\ee
For $p=1$, we can set $A_\mu^{(R)}=0$ by a large gauge transformation. 
This background is related to the three-sphere background of \cite{Kapustin:2009kz, Hama:2010av, Jafferis:2010un} by a so-called ``$\kappa$ ambiguity'' shift \cite{Closset:2012ru, Closset:2013vra} which we briefly discuss in Appendix \ref{app: 3d Atwist}. While one can preserve four supercharges on $S^3$, our background only preserves two of them. 

For $p > 1$, we have a non-trivial holonomy of the $R$-symmetry gauge field $A_\mu^{(R)}$ along the Hopf fiber, corresponding to the fact that $c_1(L^{(R)})=-1 \mod p$. This is in contrast with the supersymmetric backgrounds considered  in \cite{Benini:2011nc, Alday:2012au, Nieri:2015yia}, which studied the same geometry \eqref{round S3 metric}  with a topologically trivial $L^{(R)}$. The reason is that there exists two {\it distinct} supersymmetric backgrounds on the same topological space, corresponding to topologically distinct THFs. To explain this point, let us consider the lens space $L(p,q)$ defined as the quotient of the three-sphere 
\be
\{|z_1|^2 +|z_1|^2= 1\} \subset \C^2
\ee
by the freely-acting $\Z_p$ action:
\be\label{Zp quotient}
(z_1\,,\; z_2)\sim (e^{2 \pi i q\ov p}z_1\, , \; e^{-{2 \pi i \ov p}} z_2)~,
\ee
with $p$ and $q$ two non-zero integers. The Hopf fibration considered above is given by the map:
\be
\pi: (z_1, z_2) \mapsto z= {z_2 \ov z_1}
\ee
to the two-sphere, where $z$ is the complex coordinate on $\C\mathbb{P}^1$ on the northern patch ($z\neq \infty$), related to the angular coordinates above by $z= \tan{\theta\ov 2} e^{i\phi}$.
 The quotient \eqref{Zp quotient} acts on the base as:
\be
z\sim e^{-{2 \pi i (q+1)\ov p}}z~,
\ee
leaving it invariant if and only if $q=p-1$ (mod $p$). It follows that:
\be
\CM_{0,p}\cong L(p, p-1)~,
\ee 
as Seifert manifolds equipped with a particular THF.
In contrast, the previous literature dealing with $\CN=2$ theories on lens spaces \cite{Benini:2011nc, Alday:2012au, Nieri:2015yia}  considered $L(p,1)$ instead. While $L(p,1)$ and $L(p,p-1)$ are homeomorphic, the THFs induced on them by the quotient \eqref{Zp quotient} are distinct (if $p>2$).~\footnote{The THFs are inherited from the complex structure on $\C^2$. A closely related statement is that there exists two distinct families of complex structures on the Hopf surface $L(p,q) \times S^1$ if $p>2$ \protect\cite{nakagawa1995}, as discussed in \protect\cite{Nishioka:2014zpa}.} 
We should note that the methods of this paper do not apply directly to $L(p,1)$ or other lens spaces, because they would correspond to circle fibrations over the sphere with orbifold points. (These are examples of general Seifert fibrations, as mentioned in the introduction.) We also note that \cite{Ohta:2012ev} studied gauge theories on the $L(p, p-1)$ supersymmetric background.

\subsection{Supersymmetric multiplets and Lagrangians}
Given the supersymmetric background above, the $\CN=2$ supersymmetric multiplets and Lagrangians directly follow from the general results of \cite{Closset:2012ru}. In this subsection, we spell out those multiplets and Lagrangian in ``A-twisted variables''---see Appendix \ref{app: twisted fields} and {\it e.g.}  \cite{Closset:2015rna,Closset:2016arn}---in order to emphasize the relation to the A-twist on $\Sigma_g$. 

In the following, we write all the fields in the canonical frame basis. In that case, the  holomorphic line bundle $\CK$ on  $\Mgp$ is really a $U(1)$ bundle,~\footnote{We are being slightly cavalier in our notation since  $\CK$ may denote either a holomorphic line bundle or the associated $U(1)$ bundle:  $\alpha_z$ is a section of the holomorphic line bundle $\CK$ and $\alpha_1 = e_1^z \alpha_z$ is a section of the  associated $U(1)$ bundle.}  and $\b \CK\cong \CK^{-1}$.
The corresponding $U(1)$ charge is the ``two-dimensional spin'' of a field---in other words, a field of integer two-dimensional spin $s_0\in \Z$  is a section of $(\CK)^{s_0}$, and similarly for $s_0$ half-integer for some choice of square root.  The three-dimensional A-twist \eqref{A twist as bundles} corresponds to a ``twist'' of the two-dimensional spin by the $R$-symmetry according to:
\be\label{def twisted spin}
s= s_0 +{r\ov 2}~,
\ee
with $r$ the $R$-charge. By definition, the A-twisted variables have vanishing $R$-charge and definite twisted spins. 
Note that  $2 s\in \Z_p$  since $\CK$ is a torsion bundle.  The real connection on $\CK$ is given by:
\be\label{def AmuK}
A^{(\CK)} = - {i\ov 4} \d_z \log g\; dz+  {i\ov 4} \d_\bz \log g\; d\bz+  2 ds= 2 \CA^{(R)}~,
\ee
with $\CA^{(R)}$ defined in \eqref{H and CAR}.
Let us also define the covariant derivative
\be\label{def Dmu torsion}
D_\mu = \hat\nabla_\mu - i s A_\mu^{(\CK)}~,
\ee
acting on tensors valued in $(\CK)^s$, with $s\in \half \Z$ and  $\h\nabla_\mu$ the connection defined by \eqref{def omega hat}. 

\subsubsection{Supersymmetry algebra}
The two Killing spinors \eqref{KS explicit}  correspond to two supersymmetry transformations:
\be
\delta= \zeta \CQ~, \qquad\qquad\qquad \t\delta =\t\zeta \t\CQ~, 
\ee
which satisfy the supersymmetry algebra:
\be\label{susy algebra 3d}
\delta^2= 0~, \qquad\qquad \t\delta^2=0~,\qquad\qquad \{\delta~,\, \t\delta \} = -2i \left(Z+ \CL_K\right)~.
\ee
Here $Z$ is the real central charge of the $\CN=2$ superalgebra in flat space, and  $\CL_K$ is the  $\CK$-covariant Lie derivative along the Killing vector $K$. 
For a vector multiplet $\CV$ in Wess-Zumino (WZ) gauge, the real scalar component $\sigma$ also enters \eqref{susy algebra 3d} as  $Z= Z_0- \sigma$, where $Z_0$ is the actual central charge and  $\sigma$ is valued in the appropriate gauge representation. We should note that the Lie derivative and the covariant derivative $\h\nabla_\mu$ coincide along $K^\mu$, which means that:
\be
\CL_K = K^\mu D_\mu~.
\ee
Note that we traded the $R$-symmetry gauge field for $A_\mu^{(\CK)}$ in \eqref{susy algebra 3d} since we are considering A-twisted fields, which are $R$-neutral by definition.

\subsubsection{Vector multiplet}

Let $\GG$ and $\Fg={\rm Lie}(\GG)$ denote a compact Lie group and its Lie algebra, respectively.
In  WZ gauge, a $\Fg$-valued vector multiplet $\CV$ has components:
\be\label{V components}
\CV= \left(a_\mu~,\, \sigma~, \,\Lambda_\mu~, \, \t\Lambda_\mu~,\,  D\right)~.
\ee
The $A$-twisted fermions $\Lambda_\mu$ decompose as:
\bea\label{gauginiA} 
&\Lambda_\mu dx^\mu =  \Lambda_0 e^0 + \Lambda_1 e^1~, \quad\qquad
&\t\Lambda_\mu dx^\mu =  \t \Lambda_0 e^0 + \t \Lambda_{\b 1} e^{\b 1}~,
\eea
where the vertical components $\Lambda_0$  $\t\Lambda_0$ are scalar fields and the horizontal components $\Lambda_1$, $\t\Lambda_{\b1}$ are sections of $\CK$ and $\b\CK$, respectively.
Let us define the field strength
\be
f_{\mu\nu} = \d_\mu a_\nu - \d_\nu a_\mu -i [a_\mu, a_\nu]~,
\ee
and denote by $D_\mu$ the covariant and gauge-covariant derivative. 
The supersymmetry transformations of \eqref{V components} are%:
\bea\label{susyVector twisted}
&\delta a_\mu = i \t\Lambda_\mu~,   
\quad &&  \t\delta a_\mu = - i \Lambda_\mu\cr 
&\delta\sigma =\t\Lambda_0~,
\quad &&  \t\delta \sigma = -\Lambda_0~,\cr  
& \delta \Lambda_0 =i \left(D -\sigma H - 2 i f_{1\b 1}\right)+ i D_0 \sigma~,
\quad  && \t\delta \Lambda_0 =0~,\cr
& \delta \Lambda_1 =2 f_{01} + 2 i D_1 \sigma~,
\quad &&   \t\delta \Lambda_1 =0~,  \cr
& \delta \t\Lambda_0 =0~,
\quad  && \t\delta \t\Lambda_0 =i \left(D -\sigma H - 2 i f_{1\b 1}\right)- i D_0 \sigma~,\cr
&\delta \t\Lambda_{\b1} =0~,
\quad &&\t\delta \t\Lambda_{\b1}= -2 f_{0\b 1} - 2 i D_{\b1}\sigma~, \cr
&\delta D= -D_0\t\Lambda_0 - 2 D_1 \t\Lambda_{\b 1}\quad
 && \t\delta D= -D_0\Lambda_0- 2D_{\b 1} \Lambda_1 \cr
 &\quad \quad -H \t\Lambda_0+ [\sigma, \t\Lambda_0]~,\quad 
 &&\quad\qquad +H \Lambda_0 +[\sigma, \Lambda_0] 
\eea
The dependence of \eqref{susyVector twisted} on the geometric background is mostly implicit, through the covariant derivatives written in the frame basis. To check the supersymmetry algebra, it is important to note that:
\be
f_{01}= D_0 a_1- D_1 a_0~, \qquad
f_{0\b1}= D_0 a_{\b1}- D_{\b1} a_0~, \qquad
f_{1{\b1}}= D_1 a_{\b1}- D_{\b1} a_1 + H a_0~,
\ee
where $H$ appears due to the non-zero torsion of the covariant derivative.

\subsubsection{Chiral multiplet}
Consider a chiral multiplet $\Phi$ of (integer) $R$-charge $r$, transforming in a representation $\FR$ of $\Fg$.
 In $A$-twisted notation \cite{Closset:2015rna}, we denote the components of $\Phi$ by
\be\label{compo Phi}
\Phi= \left(\CA~,\, \CB~,\, \CC~,\, \CF\right)~.
\ee
Similarly, the charge-conjugate antichiral multiplet $\t\Phi$ of $R$-charge $-r$ in the representation  $\b\FR$ has components
\be\label{compo tPhi}
\t\Phi= \left(\t\CA~,\, \t\CB~,\, \t\CC~,\, \t\CF\right)~,
\ee
The fields are valued in the canonical line bundle to the appropriate power. We have:
\bea
&\CA~,\, \CB \,\in \Gamma(\CK^{r\over 2}\otimes V_{\FR})~,\qquad && \CC~,\,\CF\, \in \Gamma(\CK^{r\over 2}\otimes \b\CK \otimes V_{\FR})~,\cr
&\t\CA~,\, \t\CB \,\in \Gamma(\b\CK^{r\over 2}\otimes \b V_{\b\FR})~,\qquad && \t\CC~,\,\t \CF\, \in \Gamma(\b\CK^{r\over 2}\otimes \CK\otimes \b V_{\b\FR})
\eea
where $V_{\FR}$, $\b V_{\b \FR}$ are the gauge vector bundles. In particular, $\CA, \CB$ have two-dimensional  spin ${r\ov 2}$, while $\CC, \CF$ have two-dimensional spin ${r-2\ov 2}$.
The supersymmetry transformations of the chiral multiplet read:
\bea\label{susytranfoPhitwistBis}
& \delta \CA = \CB~, \qquad\qquad  && \t\delta \CA=0~,\cr
& \delta \CB=0~, \qquad \qquad && \t\delta \CB= -2i\big(- \sigma +D_0\big)\CA~,\cr
& \delta \CC=\CF~, \qquad \qquad && \t\delta \CC= 2i D_{\b1}\CA~,\cr
& \delta \CF=0~, \qquad \qquad && \t\delta \CF=- 2i \big(-\sigma +D_0\big)\CC -2i D_{\b1}\CB -2i \t\Lambda_{\b1}\CA~,
\eea
where $D_\mu$ is appropriately gauge-covariant and $\sigma$ and $\t\Lambda_{\b1}$ act in the representation $\FR$. We have:
\be\label{def Dmu chiral}
D_\mu \CA =\left( \d_\mu - i a_\mu - i {r\ov 2} A_\mu^{(\CK)}\right)\CA~, \qquad
D_\mu \CC =\left( \d_\mu - i a_\mu - i {r-2\ov 2} A_\mu^{(\CK)}\right)\CC~,
\ee
with $A^{(\CK)}_\mu$ defined in \eqref{def AmuK}.
For the antichiral multiplet, we similarly have:
\bea\label{susytranfotPhitwistBis}
& \delta \t\CA = 0~, \qquad\qquad\qquad\qquad\qquad\qquad\qquad
& \t\delta \t\CA=\t\CB~,\cr
& \delta \t\CB= -2i \big(\sigma +D_0 \big)\t\CA~, \qquad\qquad\qquad \qquad\quad\quad &\t\delta \t\CB=0~,\cr
& \delta \t\CC= -2i D_1 \t\CA~,
\qquad\qquad\qquad \qquad\quad 
&\t\delta \t\CC=\t\CF~,\cr
& \delta \t\CF= -2i \big(\sigma +D_0 \big)\t\CC +2i D_1  \t\CB + 2i \Lambda_1\t\CA~,
& \t\delta \t\CF=0~.
\eea
Using \eqref{susyVector twisted}, one can check that \eqref{susytranfoPhitwistBis} and \eqref{susytranfotPhitwistBis} realize the supersymmetry algebra:
\be\label{susy with gauge field i}
\delta^2=0~, \qquad\qquad \t\delta^2=0~, \qquad \qquad
\{\delta,\t\delta\}=- 2i  \left(- \sigma+ \CL_K^{(a)} \right)~,
\ee
where $\CL_K^{(a)}$ is the gauge-covariant Lie derivative, and $\sigma$ acts in the appropriate representation of the gauge group.

\subsubsection{Supersymmetric Lagrangians}
To conclude this section, let us write down the most important  supersymmetric Lagrangians for our three-dimensional $\CN=2$ gauge theories \cite{Closset:2012ru}.

\paragraph{Vector multiplet.}
The curved-space super-Yang-Mills (SYM) Lagrangian reads:
\bea\label{S YM full}
&\SL_{YM}&=&\;{1\ov e^2} \Big({1\ov 4} f_{\mu\nu}f^{\mu\nu} + \half D_\mu \sigma D^\mu \sigma - \half (D+ \sigma H)^2 + 4 i H \sigma f_{1\b 1} + 2 H^2 \sigma^2  \cr
&&&\qquad+ i \t\Lambda_0 D_0 \Lambda_0 + 2 i \t \Lambda_{\b1} D_1 \Lambda_0
 +2 i \t\Lambda_{0} D_{\b1} \Lambda_1 - i \t\Lambda_{\b1} D_0 \Lambda_1  \cr
 &&&\qquad - i \t\Lambda_0 [\sigma, \Lambda_0] - i \t\Lambda_{\b1} [\sigma, \Lambda_1] \Big)~. 
\eea
Here and below, the trace over gauge indices is left implicit.
The Lagrangian \eqref{S YM full} is $\delta$-exact, like any well-defined $D$-term. One can check that:
\be
\SL_{YM}={1\ov e^2} \delta \t\delta \left(\half \t\Lambda_0 \Lambda_0 - \half \t\Lambda_{\b1}\Lambda_1 + 2 \sigma f_{1\b1} - 2 i H \sigma^2 \right)~.
\ee
Another important Lagrangian is the Chern-Simons (CS) term. For any  gauge group $\GG$, we have
\be\label{classical CS Lag}
\SL_{\rm CS} ={k \over 4 \pi} \left(i \epsilon^{\mu\nu\rho} \left(a_\mu \d_\nu a_\rho- {2i\ov 3}a_\mu a_\nu a_\rho\right) - 2 D \sigma + 2 i \t \Lambda_0 \Lambda_0 + 2 i  \t\Lambda_{\b1}\Lambda_1 \right)~,
\ee
with $k\in \Z$ the CS level.~\footnote{In general, we have a distinct CS level for each simple factor and for each $U(1)$ factor in $\GG$.}  
 In the presence of an abelian sector, we can  also have mixed CS terms between $U(1)_I$ and $U(1)_J$, with $I\neq J$:
\be\label{mixed CS lag}
\SL_{{\rm CS}, IJ} = {k_{IJ} \over 2 \pi} \left(i \epsilon^{\mu\nu\rho} a^{(I)}_\mu \d_\nu a_\rho^{(J)} -  D^{(I)} \sigma^{(J)}- D^{(J)}\sigma^{(I)} +  i \t\lambda^{(I)}\lambda^{(J)}+ i \t\lambda^{(J)}\lambda^{(I)} \right)~,
\ee
with $\t\lambda^{(I)}\lambda^{(J)}=   \t \Lambda_0^{(I)} \Lambda_0^{(J)} +   \t\Lambda_{\b1}^{(I)}\Lambda_1^{(J)}$.  
For each $U(1)_I$  factor, we may also turn on the Fayet-Iliopoulos parameter:
\be\label{2Q FI}
\SL_{\rm FI} = -{\xi_I\over 2\pi} \tr_I(D-  (\sigma+ 2 i a_0)H)~,
\ee
where we normalized $\xi_I$  like in \cite{Closset:2016arn}. The FI term is a special case of a mixed CS term between the $U(1)_I$ vector multiplet $\CV_I$ and the background vector multiplet $\CV_{T_I}$ (with real mass  $\sigma_{T_I}= \xi_I$ and vanishing flux $\n_{T_I}=0$) for the associated topological symmetry $U(1)_{T_I}$, with  level $k_{I T_I}=1$.

\paragraph{Chiral multiplet.}
The standard kinetic term for a chiral multiplet coupled to a vector multiplet (in WZ gauge) reads:
\bea\label{kin chiral}
&\SL_{\t\Phi\Phi} &=&\; \t\CA \left(-D_0 D_0 - 4 D_1 D_{\b1} + \sigma^2 + D- \sigma H - 2 if_{1\b1}  \right)\CA - \t\CF \CF \cr
&&&  -{i\ov 2} \t\CB (\sigma+ D_0) \CB + 2 i \t\CC (\sigma-D_0)\CC + 2 i \t \CB D_1 \CC - 2 i \t\CC D_{\b1} \CB\cr
&&& - i \t\CB \t\Lambda_0 \CA + i \t\CA \Lambda_0 \CB - 2 i \t\CA \Lambda_1 \CC + 2 i \t\CC \t\Lambda_{\b1} \CA~.
\eea
This Lagrangian is $\delta$-exact:
\be
\SL_{\t\Phi\Phi}  = \delta\t\delta \left({i\ov 2} \t\CA (\sigma+ D_0)\CA - \t\CC \CC\right)~.
\ee
Finally, we may write down superpotential interactions in terms of a superpotential $W= W(\Phi)$ of $R$-charge $2$. Those interaction terms are $Q$-exact and do not play any crucial role in the following. The only way the superpotential appears in the localization computation  is by the constraints it imposes on the flavor symmetry and $R$-charges.

\paragraph{$U(1)_R$ and gravitational Chern-Simons terms.}
Three additional supersymmetric Chern-Simons Lagrangians are available in curved space \cite{Closset:2012vp, Closset:2012ru}. Let us consider them on our $\Mgp$ background.

The first Lagrangian is simply a mixed CS term between a $U(1)_I$ vector multiplet and the $U(1)_R$ vector multiplet \eqref{VR def}. It reads:
\be\label{CS RI}
\SL_{{\rm CS}, RI} =  {k_{RI} \over 2 \pi} \Big(i \epsilon^{\mu\nu\rho} (\CA^{(R)}_\mu+ i H \eta_\mu) \d_\nu a_\rho^{(I)} -  H D^{(I)} - {1\ov 4}(R-6 H^2)\sigma^{(I)} \Big)~,
\ee
in terms of the supergravity background fields defined above.
The second Lagrangian is a supersymmetric CS term for the $U(1)_R$ vector multiplet:~\footnote{The full non-linear expression for $\SL_{{\rm CS}, zz}$ has not appeared explicitly in the literature, but it is easily obtained by realizing that this supergravity Lagrangian only depends on the $U(1)_R$ vector multiplet, instead of the full supergravity multiplet.} 
\be\label{CS zz}
\SL_{{\rm CS}, zz} =  {k_{zz} \over 4 \pi} \Big(i \epsilon^{\mu\nu\rho} (\CA^{(R)}_\mu+ i H \eta_\mu) \d_\nu (\CA^{(R)}_\rho+ i H \eta_\rho) -    {1\ov 2}H R + 3 H^3 \Big)~.
\ee
The third CS Lagrangian is the $\CN=2$ supersymmetric completion of the gravitational CS terms:
\bea\label{CS grav}
&\SL_{{\rm CS}, g} &=&  {k_g \over 192 \pi} \Big( i\epsilon^{\mu\nu\rho} \Tr \big(\omega_\mu \d_\nu \omega_\rho + {2 \over 3} \omega_\mu \omega_\nu \omega_\rho\big)\cr
&&&\qquad+ 4 i \epsilon^{\mu\nu\rho}  (\CA^{(R)}_\mu- i H \eta_\mu) \d_\nu (\CA^{(R)}_\rho- i H \eta_\rho)\Big)~.
\eea
We need $k_g \in \Z$ for the non-supersymmetric gravitational CS term to be well-defined by itself.
On the other hand, the coefficient of the $R$-symmetry CS term $\CA^{(R)} d\CA^{(R)}$ is:
\be
k_{RR} \equiv k_{zz}+ {1\ov 12} k_g~.
\ee
This ``RR CS level'' must be integer, $k_{RR} \in \Z$, whenever the $U(1)_R$ line bundle is topologically non-trivial. The level $k_{zz}$ itself does not need to be quantized because it is a CS level for the gauge field coupling to the central charge  \cite{Closset:2012vp}, which is never quantized in our family of backgrounds.

The mixed CS term \eqref{CS RI} can involve either a dynamical or background $U(1)_I$ vector multiplet. The two other terms \eqref{CS zz} and \eqref{CS grav} only depend on the geometric background. The CS levels $k_{RR}$ and $k_g$ correspond to contact terms in two point functions of the $R$-symmetry current and energy-momentum tensor, respectively \cite{Closset:2012vp}.

%%%%%%%%%
\section{Localization on the Coulomb branch}\label{sec: Loc}
In this section, we sketch the Coulomb branch localization argument, which gives an independent derivation of the results of Section \ref{sec: BE formula}.

\subsection{Vector multiplet localization}\label{sub: Vector localization}
Let us first consider the supersymmetry equations for the vector multiplet $\CV$. It follows from \eqref{susyVector twisted} that the gaugino variations vanish if and only if:
\be\label{susy vec 01}
D_0 \sigma=0~, \qquad f_{01}+ i D_1 \sigma=0~, \qquad
 f_{0\b1}+ i D_{\b1} \sigma=0~, \qquad
 D= 2 i f_{1\b 1} + \sigma H~.
\ee
In addition, we consider the partial gauge-fixing condition:
\be
\eta^\mu (\CL_K a_\mu)=0~,
\ee
which is simply $D_0 a_0=0$. To understand the supersymmetry equations, it is useful to define the complexified gauge field:
\be
\CA_\mu = a_\mu - i \sigma \eta_\mu~,
\ee
with field strength $\CF_{\mu\nu}$,
in terms of which the equations \eqref{susy vec 01} read:
\be\label{susy vec 02}
D_0 \sigma=0~, \qquad \CF_{01}=0~, \qquad
 \CF_{0\b1}=0~, \qquad
 D+\sigma H= 2 i \CF_{1\b 1} ~.
\ee
These conditions imply that $\CA_\mu$ is the connection of a holomorphic vector bundle \cite{Closset:2013vra}, together with the gauge-fixing condition $D_0 \CA_0=0$. 
Let us define the quantities:
\be\label{u ub def}
u = i \beta (\sigma + i a_0)~, \qquad\qquad \t u =-i \beta (\sigma - i a_0)~,
\ee 
for the constant modes of $\sigma$ and $a_0$. We also define:
\be\label{def x holo}
x= e^{-i \int_\gamma \CA} = e^{2 \pi i u}~,
\ee
the holonomy of $\CA_\mu$ along the $S^1$ fiber.

We would like to localize the path integral onto the constant modes \eqref{u ub def}. 
The bosonic part of the SYM action  \eqref{S YM full} can be written as:
\be\label{bosonic_ym}
\SL_{YM}\Big|_{\rm bos}= {1\ov e^2} \Big(2f_{01}f_{0\b1} + \half D_\mu \sigma D^\mu \sigma  +\half \left(2 i \CF_{1\b1}\right)^2 - \half (D+ \sigma H)^2\Big)~.
\ee
Since the action \eqref{bosonic_ym} is the bosonic part of $Q$-exact action, we can localize the path integral by taking the limit $e\rightarrow 0$. We choose a standard reality condition for the dynamical fields $a_\mu$ and $\sigma$, which are taken to be real, while we remain agnostic about the reality condition for $D$.~\footnote{Note that the $Q$-exact action \protect\eqref{bosonic_ym} is not positive definite in general. This makes it harder to argue for the validity  of the localization argument. We leave a clearer understanding of this point for future work.}
Then the BPS configurations \eqref{susy vec 01} simplify to:
\be\label{susy eq 3}
D_\mu\sigma=0~, \qquad f_{01}=f_{0\b 1} = 0~, \qquad  D= 2if_{1\bar 1}+\sigma H~.
\ee
If, in addition, we take $D$ to be purely imaginary, we have $f_{1\b1}=0$ and we localize onto flat connections. This is slightly too strong, however, and in the following we will also allow for constant modes of $f_{1\b1}$ that satisfy \eqref{susy eq 3}.

We may use the residual two-dimensional gauge freedom to diagonalize $a_0$: \be
a_0 = {\rm diag}(a_{0,a})~,\qquad \qquad a=1, \cdots, \rk,
\ee
breaking the gauge group $\GG$ to the Cartan subgroup 
\be
\GH \cong \prod_{a=1}^\rk U(1)_a~.
\ee
From $D_\mu \sigma=0$ and the reality condition, $\sigma$ is also localized onto the constant diagonal modes $\sigma = {\rm diag}(\sigma_a)$. 
The constant modes $u_a = i\beta(\sigma_a+ia_{0,a})$ will be identified with the Coulomb branch parameters of Section~\ref{sec: BE formula}.
In the diagonal gauge, we should sum over $\GH$-bundles over $\Mgp$ which are pull-backs of $\GH$-bundles on $\Sigma_g$  \cite{Blau:1994rk,Blau:2006gh}. All such bundles are torsion bundles \cite{Blau:2006gh}. Here we assume that $p\neq 0$. (We briefly review the $p=0$ case below.)
The torsion flux $\m$ takes value in the finite group:
\be\label{Gammap}
\Gamma_{\mathbf{G}^\vee}^{(p)} = \setcond{\m}{\rho(\m) \in \bZ ~~ \forall \rho \in \Gamma_\mathbf{G}~,\, \m\in \Z_p^\rk}\cong \Z_p^\rk~,
\ee
which is a $\Z_p$ reduction of the ordinary magnetic flux lattice \cite{Englert:1976ng,Kapustin:2005py}.
Here  $\Gamma_\mathbf{G}\subseteq i\mathfrak{h}^*$ is the weight lattice of electric charges of $\GG$.

In a given topological sector $\m$, the non-trivial connection can be chosen to be flat. We take:
\be
a_\mu = \h a_0 \eta_\mu + a_\mu^{(\rm flat)}~, \qquad   \h a_0\in \R~.
\ee
Note that $ \h a_0$ is the coefficient of a well-defined one-form, therefore it cannot affect the topological properties of the gauge field. Some basic properties of flat connections are reviewed in Appendix \ref{app: geom}. Importantly, we have the holonomy:
\be
e^{- i \int_\gamma a^{(\rm flat)}}= e^{2 \pi i{ \m \ov p}}~,
\ee
along the fiber. Note that we have:
\be\label{def u with m}
u= i\beta\left(\sigma+i \h a_0\right)+ {\m\ov p}
\ee
in a given topological sector. Under a $U(1)_a$ large gauge transformations, the parameters $\m_a$ and $u_a$ transform as:
\be\label{u large gauge transfo}
(u_a,\, \m_a) \sim (u_a+1,\, \m_a+p )~.
\ee
In addition to these parameters, the $U(1)_a$ line bundles are also characterized by flat connections along $\Sigma_g$, corresponding to elements of the cohomology group $H^1(\Mgp, \R)\cong \R^{2 g}$. We can parametrize these flat connections by:
\be
\sum_{I=1}^g \left(\alpha_I \omega^I_z dz + \t\alpha_I \t\omega^I_{\bz} d\bz\right)~, \qquad \qquad [\omega^I]\in H^1(\Mgp, \R)
\ee
The $U(1)_a$ holonomies $\alpha_I$, $\t\alpha_I$ live in a compact domain.

Importantly, the kinetic terms for the gaugino appearing in the localizing action \eqref{S YM full} admit fermionic zero-modes, which satisfy:
\be\label{zeromode equations}
D_0 \Lambda_0= D_{1} \Lambda_0=0~,\qquad \qquad
D_0 \Lambda_1= D_{\b1}\Lambda_1=0~,
\ee
and similarly for the charge-conjugate fermions $\t\Lambda_0$, $\t\Lambda_{\b1}$. These zero-modes are directly related to the more familiar zero modes of the $A$-twisted Dirac operator on $\Sigma_g$. We have the constant mode of $\Lambda_0$, and $g$ one-form zero-modes for $\Lambda_1$:
\be
\Lambda_0= {\rm constant}~, \qquad\qquad \Lambda_1 = \sum_{I=1}^g \Lambda_I \omega^I_1~.
\ee
The cohomology classes
$[\omega^I_z dz]\in H^1(\Mgp, \R)\cong \R^{2 g}$
are the pull-back of the holomorphic one-forms on the Riemann surface $\Sigma_g$. Note that the torsion of the covariant derivative $D_\mu= \h\nabla_\mu$ plays a crucial role here, since it is such that  the equations  \eqref{zeromode equations} are independent of $p$. Therefore, the localization of the  path integral can be performed in a manner identical to the $p=0$ case studied in \cite{Closset:2016arn}. The vector multiplet localizes to an integral over the zero-mode supermultiplets:
\be\label{V0 VI def}
\CV_0 = (\sigma~,\, a_0~,\, \Lambda_0~,\, \t\Lambda_0~,\, \h D)~, \qquad\quad \CV_I = (\alpha_I~,\, \t \alpha_I~,\, \Lambda_I~,\, \t\Lambda_I)~, \quad I=1, \cdots, g~,
\ee
where the constant mode $\h D$ is defined by
\be
 D= 2 i f_{1\b 1} + \sigma H + i \h D\ . 
\ee
We have turned on a non-BPS constant mode $\h D$ as a regulator. In order to have a positive definite localizing  action, the contour for $\h D$ is chosen to be $\h D = \mathbb{R}-2f_{1\b 1}$, which allows the constant modes for $f_{1\b 1}$ \cite{Benini:2015noa, Benini:2016hjo,Closset:2016arn}. Then we deform the $\h D$-contour to be along the real axis. When we deform the contour, we pick up the residues of the pole in the region $0<\h D < -2f_{1\bar 1}$, but the residues of these poles are exponentially suppressed as we take the limit $e\rightarrow 0$ \cite{Benini:2015noa}. 
Schematically, we obtain the partition function:
\be\label{Z inter}
Z =\sum_{\m\in \Gamma_{\mathbf{G}^\vee}^{(p)}} \int d\CV_0 \int \prod_I d\CV_I  \; e^{-S_0}\; \CZ^\oneloop_{\m}(\CV_0, \CV_I)~.
\ee
where the sum is over all topological sectors,  $S_0$ is the classical action evaluated on the supersymmetric locus, including the fermionic zero-modes, and $\CZ^\oneloop_{\m}$ is the one-loop determinant in a given topological sector. The integrand of \eqref{Z inter} enjoys a residual supersymmetry, which follows from \eqref{susyVector twisted} restricted to the zero-modes. Following \cite{Closset:2016arn}, we can  argue that \eqref{Z inter} reduces to a certain multi-dimensional contour integral on $u_a$-space with a meromorphic integrand. 

While the integrand of that contour integral can be straightforwardly computed, the precise form of the contour is more complicated to derive. We will give a complete derivation of the contour in the rank-one case in Appendix \ref{app: loc}, and we will present the higher-rank generalization as a conjecture.

\subsection{Classical action contribution: CS terms}\label{subsec:classical contrib CS}
Let us first consider the classical action evaluated on the supersymmetric locus. For the vector multiplet, this corresponds to the parameters $u, \t u$ and $\m$. The only non-vanishing contributions come from the Chern-Simons terms (including the FI terms). On general ground, the result should be holomorphic in $u$. We provide a summary of some subtle properties of the CS functional in Appendix \ref{app: spin CS}.

\paragraph{Ordinary CS term.}
For simplicity, let us first consider a $U(1)$ vector multiplet as described above, with parameters $(u, \m)$. The Chern-Simons term \eqref{classical CS Lag} can be decomposed as:
\be
S_{\rm CS} =S_{\rm CS}^{(1)}+ S_{\rm CS}^{(2)}
\ee
with
\bea\label{SCS1and2}
& S_{\rm CS}^{(1)}= i{k\ov 4 \pi}\int a^{(\rm flat)}\wedge da^{(\rm flat)}~, \cr
& S_{\rm CS}^{(2)}=  i{k\ov 4 \pi}\int (\h a_0)^2 \eta \wedge d\eta+  {k\ov 4 \pi}\int d^3x \sqrt{g}\left(-2 \sigma^2 H -4 i \sigma f_{1\b 1}\right)~.
\eea
The expression for $S_{\rm CS}^{(1)}$ is formal since it involves a non-trivial gauge connection. We claim that the exponentiated CS functional for the flat connection of a torsion $U(1)$ bundle, of first Chern class $\m\in \Z_p$, is given by:
\be\label{CS torsion and sign}
e^{-S_{\rm CS}^{(1)}}= (-1)^{k\m} e^{\pi i k {\m^2\ov p}}~.
\ee
See {\it e.g.} \cite{2002math......9403H, Alday:2012au}  in the case $g=0$. We conjecture that \eqref{CS torsion and sign} also holds on $\Mgp$ with $g>0$. A proper computation should be done by using the four-dimensional definition of the CS functional, as explained in Appendix \ref{app: spin CS}.
Note that \eqref{CS torsion and sign} is invariant under the large gauge transformations $\m\sim \m+p$ for any $\m\in \Z_p$, if and only if $k\in \Z$, as it should be.

The integrand of $ S_{\rm CS}^{(2)}$ in \eqref{SCS1and2}, on the other hand, is well-defined, and the action can be evaluated straightforwardly. We find:
\be
e^{-S_{\rm CS}^{(2)}}= e^{\pi i k p \beta^2 \left(\sigma+ i \h a_0\right)^2}~,
\ee
which is holomorphic in $\sigma + i \h a_0$, as expected.
The total contribution of the supersymmetric CS action takes the simple form:
\be\label{SCS full}
e^{-S_{\rm CS}} = \exp{\big(- \pi i p \, k\, u^2+ 2 \pi i k \big(u+\half\big) \m\big)} 
=e^{- \pi i p \, k\, u^2}\, (-x)^{k\m}
\ee
when written in terms of $u$ as defined in \eqref{def u with m}, with $x$ defined in \eqref{def x holo}.  For a more general gauge group $\GG$, we similarly obtain:
\be
e^{-S_{\rm CS}} = \prod_{a=1}^\rk e^{- \pi i p\, k\, u_a^2}\,  (-x_a)^{k \m_a}
\ee
after diagonalization. 

\paragraph{Mixed CS term.} Consider two $U(1)$ vector multiplets with parameters $(u_I, \m_I)$ and $(u_J, \m_J)$. We claim that the mixed Chern-Simons term \eqref{mixed CS lag} has a contribution from the flat connections:
\be
e^{-S_{{\rm CS}, IJ}^{(1)} }=\exp{\Big(- i {k_{IJ}\ov2 \pi} \int a_I^{(\rm flat)} d a_J^{(\rm flat)}\Big)}= e^{2 \pi i k_{IJ} {\m_I \m_J\ov p}}~,
\ee
similarly to \eqref{CS torsion and sign}.
This is invariant under large gauge transformations for $k_{IJ}\in \Z$.
The remaining terms are well-defined and give:
\be
e^{-S_{{\rm CS}, IJ}^{(1)} }=\exp{\big(2 \pi i \,p\,  k_{IJ}\,  \beta^2(\sigma_I + i \h a_{I0})(\sigma_J+ i \h a_{J0})\big)}~.
\ee
 The full supersymmetric action  \eqref{mixed CS lag} can be written as:
\be\label{mixed CS on susy loc}
e^{-S_{{\rm CS}, IJ} } 
= e^{-2 \pi i \, p\, k_{IJ} u_I u_J} (x_J)^{k_{IJ} \m_I}(x_I)^{k_{IJ} \m_J}~,
\ee
with $x_I= e^{2\pi i u_I}$ and $x_J= e^{2\pi i u_J}$. Note that this includes the (generalized) FI parameter for a $U(1)_I$ gauge group, which is given by mixed CS term between $U(1)_I$ and the topological symmetry $U(1)_{T_I}$,  at level $k_{I T_I}=1$, with fugacity:
\be
x_{T_I} \equiv q_I = e^{2�\pi i \tau_I}~,
\ee
and background flux $\n_{T_I}$.

\paragraph{$U(1)_R$ and gravitational CS terms.}
By direct computation, one can check that the mixed $U(1)_R$-$U(1)_I$ CS term \eqref{CS RI} evaluates to:
\be\label{CS RI susy}
e^{-S_{{\rm CS}, RI} }  = e^{2\pi i k_{RI} (g-1) u_I }=(x_I)^{k_{RI} (g-1)}~.
\ee
This simply corresponds to \eqref{mixed CS on susy loc} with the $U(1)_R$ vector multiplet parameters plugged in.  We wrote down \eqref{CS RI susy}  in the ``A-twist gauge'' $\nu^{(R)}=0$. (More generally, we have $\nu^{(R)}\in \Z$ and therefore $x_R=e^{2\pi i \nu^{(R)}}=1$.)

In the $A$-twist gauge, the $U(1)_R$ and gravitational CS terms \eqref{CS zz} and \eqref{CS grav} give a subtle contribution:
\be\label{S RR and g}
e^{-S_{{\rm CS}, zz}-S_{{\rm CS}, g}}=  (-1)^{k_{RR} (g-1)}\, e^{\pi i p {k_{g}\ov 12}}~.
\ee
The $k_{RR}$ term can be inferred by replacing $u$ and $\m$ by  $\nu^{(R)}=1$ and $\m_R=g-1$ in \eqref{SCS full}. The $k_g$ term is a further conjecture. We do not provide a complete proof of \eqref{S RR and g}, but it passes a number of consistency checks. For instance, these CS classical terms can be generated from the chiral multiplet effective action on $\Mgp$ in the appropriate decoupling limits. 

In the language of Section \ref{sec: BE formula}, all these supersymmetric Chern-Simons terms correspond to the classical twisted superpotential \eqref{WCS gen} and effective dilaton \eqref{OmegaCS}, that is:
\be\label{CW Omega CS expl}
\CW = \half k u (u+1)+ k_{IJ} u_I u_J +{1\ov 24} k_g~, \qquad\qquad
\Omega = k_{IR} u_I + \half k_{RR}~.
\ee
Note again that this only makes sense for $k_{RI}$, $k_{RR}$ integer-quantized. Whenever $U(1)_R$ can be taken non-compact, the general result for a theory with continuous $R$-charges  can be obtained by starting with integer-quantized $R$-charges and deforming the fugacities in the way explained in Section \ref{subsec: continuous R}.

\subsection{One-loop determinants}
Next, we discuss the one-loop determinant contributions to the localized path integral. 

%%%%%%%%%%%
\subsubsection{Chiral multiplet contribution}
Consider a chiral multiplet $\Phi$ coupled to a $U(1)_I$ vector multiplet $\CV$ with charge $Q=1$, and coupled to our geometric background with $R$-charge $r\in \Z$. 
We contribution of $\Phi$ in the supersymmetric background $(u, \m)$ for $\CV$ can be computed with the $Q$-exact action \eqref{kin chiral}. The Gaussian integral
\be
Z^\Phi = \int [d\Phi d\t\Phi] e^{-S_{\Phi\t\Phi}}
\ee
only receives non-trivial contributions from the zero-modes of the operator $D_{\b1}$, with $D_\mu$ defined in \eqref{def Dmu chiral}. By a standard argument,~\footnote{See {\it e.g.} the discussion in Appendix C of \protect\cite{Closset:2015rna} which easily generalizes to our case.} we find:
\be\label{Zphi gen}
Z^\Phi = {{\rm det}_{{\rm coker}D_{\b1}}(-\sigma+ D_0)  \ov  {\rm det}_{{\rm ker}D_{\b1}}(-\sigma+ D_0) }~.
\ee
All other modes cancel out by supersymmetry.  
Note that the modified covariant derivative $D_\mu$ is the pull-back of the ordinary covariant derivative on $\Sigma_g$.
 We can then expand any 3d field along the $S^1$ fiber:
\be
\varphi = \sum_{n\in \Z} \varphi_n \;e^{i n \psi}~,
\ee
with the modes $\varphi_n$ living on $\Sigma_g$. 
 In particular, the zero-modes $\CA_n$ that contribute to the denominator in \eqref{Zphi gen},  satisfy:
\bea
& D_0 \CA_n = i{ n- a_\psi\ov \beta} \CA_n~,\qquad\qquad
& (D_\bz - i \CC_\bz)\CA_n=0~.
\eea
In other words, the modes $\CA_n$ correspond to holomorphic sections of the line bundle:
\be\label{line bundle chiral}
\CO(pn+ \m)\otimes \CK^{r\ov 2}
\ee
 on $\Sigma_g$, where $\CO(n)$ denotes a line bundle of first Chern number $n$. These bundles pull-back to torsion bundles on $\CM_{g, p}$ \cite{Blau:2006gh}.  Similar considerations hold for the fermionic zero-modes $\CC$ that satisfy $D_1\CC=0$, corresponding to the numerator of \eqref{Zphi gen}.
In this way, we find that \eqref{Zphi gen} �is given by the formal expression:
\be\label{Zphi explicit}
Z^\Phi_{g, p, \m}(u) = \prod_{n\in \Z} \left(1\ov n+ u\right)^{p n + \m + (g-1)(r-1)}~,
\ee
with $u$ defined in \eqref{def u with m}. This infinite product has to be regulated carefully, but it is clear that it possesses the expected properties. Firstly, it is formally invariant under the large gauge transformation $(u, \,\m) \sim (u+1, \, \m +p)$. Secondly, it takes the form:
\be
Z^\Phi_{g, p, \m}(u)=\CF^\Phi(u)^p \, Z_{g, 0, \m}^\Phi(u)~,
\ee
 where  $Z_{g, 0, \m}^\Phi(u)$ is the result of \cite{Benini:2015noa} for a chiral multiplet on $\Sigma_g\times S^1$, in the presence of $\m$ units of flux on $\Sigma_g$. The function:
 \be\label{CF for phi}
 \CF_\Phi(u)\equiv \prod_{n\in \Z} \left(1\ov n+ u\right)^n
 \ee
gives the contribution of a chiral multiplet to the fibering operator introduced in Section \ref{subsec:fibering op}. A similar one-loop determinant was first obtained in \cite{Ohta:2012ev}.

\subsubsection{Regulated chiral multiplet one-loop determinant}\label{subsec: regulated Zphi}
The formal product \eqref{Zphi explicit} is invariant under large gauge transformations. It is also invariant under a ``parity'' transformation which acts on  \eqref{Zphi explicit} as:
\be
{\rm P}: \qquad u \rightarrow -u~, \qquad p \rightarrow -p~,
\ee
leaving all other parameters fixed. This reflects the fact that the kinetic Lagrangian \eqref{kin chiral} is both gauge invariant and parity invariant.~\footnote{On a fixed background, the coupling to curved space breaks parity explicitly; in particular, the background supergravity field $H$ is parity odd.  Here we are considering a family of supersymmetric backgrounds $\CM_{g,p}$ on which parity acts naturally as $\CM_{g,p}\rightarrow \CM_{g,-p}$.} 
The quantum theory, however, has a ``parity anomaly'' \cite{Redlich:1983dv, Niemi:1983rq, AlvarezGaume:1984nf}. This is the statement that we cannot quantize a three-dimensional Dirac fermion coupled to a background gauge field (and a background metric) while preserving both gauge invariance (and diffeomorphism invariance) and parity. In the present case, the parity anomaly shows up upon regulating the formal product \eqref{Zphi explicit}. We naturally choose to preserve gauge invariance.

The parity anomaly is sometimes loosely stated as the fact that one should ``add a CS term with level $\half$'' to compensate for the lack of gauge invariance of the fermion effective action. This is misleading since there is no such thing as a Chern-Simons action with half-integer level. Instead, the gauge-invariant effective action necessarily breaks parity. (See \cite{Seiberg:2016rsg} for a recent discussion of this point.)  In particular, a Dirac fermion coupled to $U(1)$ gauge fields contributes half-integer contact terms $\kappa$ to two-point functions of $U(1)$ currents (and similarly for the coupling to the metric). These contact terms can be shifted by integers (by adding CS terms at levels $k$ for the gauge fields in the effective action) but the non-integer parts of $\kappa$ are physical \cite{Closset:2012vp} and violate parity.

In order to identify the correct gauge-invariant regularization for the chiral multiplet one-loop determinant, we recall that integrating out a chiral multiplet  $\Phi$ by scaling the real mass $\sigma \rightarrow \pm \infty$ leads to a shift of the relevant contact terms by:
\bea\label{shift CS}
& \delta\kappa_{II}=  \half Q^2 \sign{(Q\sigma)} ~, \qquad\qquad &&\delta\kappa_{RI}=  \half Q(r-1)  \sign{(Q\sigma)} ~, \cr
&\delta\kappa_{RR}=\half (r-1)^2 \sign{(Q\sigma)} ~,\qquad \qquad && \delta\kappa_g =  \sign{(Q\sigma)} ~.
\eea
Here we reintroduced the $U(1)_I$ gauge charge  $Q$, which we had set to $1$ before.
We would like to identify a ``$U(1)_{-\half}$ regularization'', corresponding to contact terms:
\be\label{contact term ourreg}
\kappa_{II}= - \half Q^2~, \qquad \kappa_{RI}= - \half Q(r-1)~, \qquad \kappa_{RR}=- \half (r-1)^2~, \qquad \kappa_g = -1~,
\ee
for a free chiral multiplet coupled to background fields.
This is such that:
\be\label{limit Z constraint}
\lim_{Q \sigma \rightarrow +\infty} Z^\Phi_{g, p, \m}(Qu) =1~,
\ee
since the IR theory with large positive real mass $Q\sigma$ is then an empty theory with vanishing background Chern-Simons levels.
Let us first consider the $p=0$ contribution to \eqref{Zphi explicit} (with $Q=1$):
\be
Z_{g, 0, \m}^\Phi(u)= \pif_\Phi(u)^{\m + (g-1)(r-1)}~, \qquad \qquad \pif_\Phi(u)\equiv \prod_{n\in \Z}{1\ov u+n}~.
\ee
The infinite product can be regularized in various ways, but there is a unique gauge-invariant answer that satisfy \eqref{limit Z constraint}. It is given by:
\be\label{def pif gauge}
\pif_\Phi(u) = {1 \ov 1- x}~,
\ee
with $x= e^{2 \pi i u}$. Similarly, the ``fibering operator'' contribution \eqref{CF for phi} gives:
\be\label{def FPhi of u}
 \CF_\Phi(u)= \exp{\left( {1\ov 2 \pi i}\dilog(x) + u \log(1-x)\right)}~,
\ee
in agreement with \eqref{CFPhi def}.
 As mentioned before, $\CF_\Phi(u)$ is a meromorphic function of $u$ with poles at $u= -n$, $n\in \Z_{>0}$. It is also the contribution of a chiral multiplet of $R$-charge $r=1$ to the $S^3$ partition function \cite{Jafferis:2011zi,Closset:2012vg}, as we will discuss in Section \ref{sec: S3 and Fmax}. The full one-loop determinant on $\Mgp$ is given by:
\be\label{ZPhi full}
  Z^\Phi_{g, p, \m}(u) = \CF_\Phi(u)^p \; \pif_\Phi(u)^{\m + (g-1)(r-1)}~.
\ee
Note that $\CF_\Phi(u)$ satisfies the difference equation:
\be\label{F monodromy}
 \CF_\Phi(u+1)= \CF_\Phi(u) \; \pif_\Phi(u)^{-1}~,
\ee
which implies that \eqref{ZPhi full} is invariant under the large gauge transformations \eqref{u large gauge transfo}. We may also view $\CF_\Phi$ as a function of $x$:
\be\label{CF of x}
 \CF_\Phi(x)= \exp{\left( {1\ov 2 \pi i}\Big(\dilog(x) + \log{x} \log(1-x)\Big)\right)}~,
\ee
 in which case  \eqref{F monodromy} corresponds to a monodromy around $x=0$.~\footnote{Note that $x=0$ is the only branch point in \protect\eqref{CF of x}. The branch cut of $\dilog(x)$ at $x\in [1,\infty)$ is cancelled by the second term in the exponent.}
In the limit of large negative real mass, one can show that:
\be
\lim_{Q \sigma \rightarrow -\infty} Z^\Phi_{g, p, Q\m}(Qu) = e^{\pi i p Q^2 u^2} e^{-{\pi i p \ov 6}}  (-x^{-Q})^{Q\m + (g-1)(r-1)}~.
\ee
Comparing to the classical Chern-Simons contributions discussed in Section \ref{subsec:classical contrib CS}, we see that this limit reproduces the classical supersymmetric Chern-Simons action with integer levels:
\be
k_{II}= -  Q^2~, \qquad k_{RI}= - Q(r-1)~, \qquad k_{RR}=- (r-1)^2~, \qquad k_g = -2~,
\ee
provided that $r\in \Z$.
This agrees with the expected shift \eqref{shift CS} of the bare contact terms \eqref{contact term ourreg}.

%%%%%%
\subsubsection{Vector multiplet contribution}
The W-boson and their superpartners also give a non-trivial contribution on the Coulomb branch. They contribute like chiral multiplets of gauge charges $\alpha$, with $\alpha$ the roots of $\Fg$, and $R$-charge $2$ \cite{Closset:2015rna, Benini:2015noa}. The W-bosons come in pairs of charges $\alpha$ and $-\alpha$.  As was already mentioned in Section \ref{sec: BE formula}, we choose a symmetric quantization, such that there is no shift of any contact term. This implies that the W-bosons do not contribute to the effective twisted superpotential, while they do contribute to the effective dilaton. We have:
\be\label{oneloop vec}
Z^{\rm vec}_{g}(u)  =(-1)^{(g-1)\half {\rm dim}(\Fg/\Fh)}\, \prod_{\alpha\in \Fg} (1- x^\alpha)^{1-g}= \prod_{\alpha\in \Fg_+} (1-x^\alpha)^{2-2g}~,
\ee
where $\Fg_+$ denotes the positive roots.
The one-loop determinant  \eqref{oneloop vec} is independent of $p$ and of the topological sector $\m$. It naturally agrees with previous results for $S^3$ \cite{Kapustin:2009kz} and $\Sigma_g \times S^1$ \cite{Benini:2015noa}.

%%%%%%%
\subsection{A comment on the $\CM_{g,0}\cong \Sigma_g \times S^1$ case}
The case $p=0$  was studied in \cite{Benini:2015noa, Benini:2016hjo, Closset:2016arn}. Let us emphasize the presence of some subtle signs that were previously overlooked. When $p=0$,  the sum over topological sectors is over all GNO-quantized fluxes on $\Sigma_g$:
\be
{1\ov 2 \pi} \int_{\Sigma_g} da = \m~.
\ee
The Coulomb branch parameters $u_a$ are cylinder-valued, $u_a\sim u_a+1$, corresponding to complexified flat connections along $S^1$. 

The classical and one-loop contributions can be obtained by setting $p=0$ in the results above. In particular, a $U(1)$ CS term at level $k$ contributes:
\be
e^{-S_{\rm CS}} = (-x)^{k \m}
\ee
in the presence of a flux $\m\in \Z$.
We see that, even in the absence of flat connection along $S^1$ (that is, if $u=0$), we have a contribution $(-1)^{k\m}$. This is because of the choice of spin structure dictated by supersymmetry, with periodic boundary conditions for fermions around $S^1$. This explicit dependence on the spin structure for $k$ an odd integer \cite{Dijkgraaf:1989pz} was discussed recently in \cite{Mikhaylov:2015nsa, Seiberg:2016gmd}, and we review some relevant material in Appendix \ref{app: spin CS}. The $U(1)_R$ CS term contributes a sign $(-1)^{(g-1)k_{RR}}$ for the same reason. Note that {\it mixed} CS terms do not introduce any additional signs, because they are independent of the spin structure.

%%%%%%%%
\subsection{The contour-integral formula}
Combining the classical and one-loop contribution, and integrating over the fermionic zero modes, the path integral \eqref{Z inter} can be written as a particular contour integral on $\{u_a\} \cong  \Fh_\C$. (This is proven in Appendix \ref{app: loc}  in the rank-one case, and it is a well-motivated conjecture in general.) On $\Mgp$ with background fluxes $\n_\alpha$ for the flavor symmetry, we have:
\be\label{Z inter1}
Z_{\Mgp}=  {1 \ov |W_\GG|}  \sum_{\m\in \Gamma_{\mathbf{G}^\vee}^{(p)}} \int_\CC  \CI_{g,p,\m}(u)   \, d u_1 \wedge \cdots \wedge d u_{\bf r}~,
\ee
 with ${\bf r}=\rk$ and the integrand:
\be\label{density}
\CI_{g,p,\m} (u) =(-1)^{\bf r} \; Z_{\m, g, p}^\text{classical}(u)\; Z_{\m, g, p}^\oneloop(u) \; \big(\det_{ab}\d_{u_a}\d_{u_b}\CW(u)\big)^g~.
\ee
The term $Z_{\m, g, p}^\text{classical}(u)$ is the classical contribution due to the Chern-Simons terms discussed in Section \ref{subsec:classical contrib CS}. The one-loop determinants contribute as:
\be
Z_{\m, g, p}^\oneloop(u)= \prod_{\alpha\in \Fg_+} (1-x^\alpha)^{2-2g} \; \prod_i \prod_{\rho_i \in \FR_i} \CF_\Phi(x^{\rho_i} y_i)^p \, \pif_\Phi(x^{\rho_i} y_i)^{\rho_i(\m) + \n_i + (g-1)(r_i-1)}~,
\ee
with $\CF_\Phi(x)$ and $\pif_\Phi(x)$ defined as in \eqref{CF of x} and \eqref{def pif gauge}, respectively.
The last term in \eqref{density} originates from the integration over the gaugino zero-modes $\Lambda_1$, $\t\Lambda_{\b1}$.

The integrand \eqref{density} may be conveniently written in terms of the effective twisted superpotential and  effective dilaton of Section \ref{subsec: 3d on S1}:
\bea\label{def CJ}
&\CI_{g,p,\m} (u) &=&\;    \CJ(u) \prod_{a=1}^{\bf r}\pif_a(u)^{\m_a}~,\cr
& \CJ(u)  &\equiv &\; (-1)^{\bf r}\,  \CF(u)^p \, e^{2 \pi i(g-1)\Omega(u)} \big(\det_{ab}\d_{u_a}\d_{u_b}\CW(u)\big)^g\,  \prod_\alpha \pif_\alpha(u)^{\n_\alpha}~,
\eea
with $\Omega$,  $\CF$ and $\pif_a, \pif_\alpha$ the effective dilaton \eqref{Omega full}, the  fibering operator \eqref{CF explicit 3d},  and the flux operators \eqref{flux op 3d}, respectively. (We suppressed the dependence on the flavor parameters $\nu_\alpha$ to avoid clutter.)

Note that the integrand \eqref{density}  is invariant under the large gauge transformations $(u_a, \m_a)\sim (u_a+1, \m_a+p)$. In particular, when $p=0$, the integrand is periodic, $u_a \sim u_a+1$, in each topological sector, and the integration contour lies on the classical Coulomb branch $\t \fM$ \cite{Closset:2016arn}. For $p\neq 0$, it is useful to decompose the (as yet unspecified)  real codimension-${\bf r}$ integration contour $\CC \subset \C^{\bf r}$ as:
\be
\CC\cong \cup_{n\in \Z^{\bf r}}  \CC_n~, \qquad\qquad  \CC_n \subset \{ u \;|\; n_a \leq {\rm Re}(u_a)\leq n_a+1\}~,
\ee 
where $\CC_n$ is a contour that lies in the vertical strip $n \leq {\rm Re(u)}<n+1$, as indicated.  We then have the formal identities:
\bea\label{identities integrals}
& \sum_{\m\in \Z_p^{\bf r}} \sum_{n\in \Z^{\bf r}} \int_{\CC_{n}} d^{\bf r}u\, \CJ(u) \prod_a \pif_a(u)^{\m_a}
=\sum_{\m\in \Z_p^{\bf r}} \sum_{n\in \Z^{\bf r}} \int_{\CC_0} d^{\bf r}u\, \CJ(u+n) \prod_a \pif_a(u)^{\m_a}\cr
&= \sum_{\m\in \Z_p^{\bf r}} \sum_{n\in \Z^{\bf r}} \int_{\CC_0} d^{\bf r}u\, \CJ(u) \prod_a \pif_a(u)^{\m_a-p n_a}= \sum_{\m\in \Z^{\bf r}} \int_{\CC_0} d^{\bf r}u\, \CJ(u) \prod_a \pif_a(u)^{\m_a}~,
\eea
where we used the property \eqref{prop F shift} in the second equality, and we relabelled the fluxes  $\m- p n$ as $\m$ in the last one.  Therefore, the partition function \eqref{Z inter1} can be written as a sum over the whole flux lattice of $\GG$, like in the $p=0$ case:
\be\label{Z inter1b}
Z_{\Mgp}= {1 \ov |W_\GG|}  \sum_{\m\in \Gamma_{\mathbf{G}^\vee}} \int_{\CC_0}  \CJ(u) \prod_{a=1}^{\bf r}  \pif_a(u)^{\m_a} \, d u_1 \wedge \cdots \wedge d u_{\bf r}~.
\ee
This formula realizes the relation \eqref{tft relations Mgp} at a formal level, since \eqref{Z inter1b} looks like an explicit insertion of the operator $\CF^p$ in the Coulomb-branch localization formula on $\Sigma_g \times S^1$. This is only formal, however, because the localization argument must be adapted to accommodate for the insertion of the fibering operator. This generally results in a different contour prescription for $p\neq 0$, consistent with the fact that the integrand  $\CJ(u)$ is no longer invariant under $u\sim u+1$ in this case.

\subsubsection{Singularities of the integrand}
Before discussing the integration contour, let us summarize the structure of the integrand singularities. We have four distinct types of singularities:

\paragraph{Matter field singularities.} First of all, we have potential singularities along the hyperplanes:
\be\label{hyperplane 1}
H_{\rho_i, n} = \{ u \in \Fh_\C \; |\; \rho_i(u) + \nu_i + n=0~, \;\; n\in \Z~\}~.
\ee
They correspond to the poles at $x^{\rho_i}y_i=1$ in the one-loop determinant of the chiral multiplet $\Phi_i$, corresponding to points in the moduli space where the chiral multiplet develops a bosonic zero mode. There is a pole along the hyperplane $H_{\rho_i, n}$ if and only if:
\be\label{N pole}
N_{\rho_i, n}\equiv p n + \rho(\m) + \n_i + (g-1)(r_i-1)>0~,
\ee
as is evident {\it e.g.} from \eqref{Zphi explicit}. Note that $N_{\rho_i, n}$ is the order of the pole. For $N_{\rho_i, n}<0$, on the other hand, we have a zero of order $|N_{\rho_i, n}|$ along the hyperplane.

\paragraph{Large  ${\rm Im}(u)$ region (monopole singularities).} The second type of singularities originate from the large imaginary $u$ region. We define the ``hyperplanes'':
\be\label{hyperplane 2}
H_{a\pm} = \{ u \in \Fh_\C \; |\; {\rm Im}(u_a) = \mp \infty\}~.
\ee
That is $x_a= \infty$ and $x_a=0$, respectively, in the $x_a$ variables. 
The integrand has potential singularities of the form: 
\bea\label{singularity monopole}
&\CI_\m(u) \sim \cr
&\quad e^{\mp 2\pi i p \left( \half{Q_{+a}}^a u_a^2 + \sum_{b\neq a} {Q_{a+}}^b u_b u_a +\sum_{\alpha} {Q^F_{a+}}^\alpha \nu_\alpha u_a  \right) } x_a^{\pm\left(Q_{a\pm}(\m)+ Q_{a\pm}^F(\n) + (g-1) r_{a \pm}\right)}~,
\eea
in the limit $\sigma_a \rightarrow \mp \infty$, where $Q_{a\pm}$, $Q^F_{a\pm}$ and $r_{a\pm}$ are  the monopole charges \eqref{induced charges T}. 
We refer loosely to the singularity at ${\rm Im}(u) \rightarrow \mp \infty$ as a ``pole at infinity''. More precisely, we have an actual pole at $x=\infty$ or $x= 0$, respectively, when $p=0$. For $p\neq 0$, it is more natural to use the variable $u$. The  regions in the $u$-plane where \eqref{singularity monopole} diverges generally contribute non-trivially to the partition function.

\paragraph{Large ${\rm Re}(u)$ regions.} When $p \neq 0$, the integrand may diverge at ${\rm Re}(u) \rightarrow \pm \infty$.  Using the property $\CF(u+N)= \CF(u) \pif(u)^{-N}$ with $N$ a large integer, we can understand that divergence as follows. Suppose we have a part of the integration contour that probes the large ${\rm Re}(u)$ region.  For $p>0$, the integrand diverges as ${\rm Re}(u) \rightarrow - \infty$ along portions of that contour such that $|\pif(u)|>1$; similarly, it diverges  as ${\rm Re}(u) \rightarrow  \infty$ when $|\pif(u)|<1$.

%%%%%
\paragraph{W-boson singularities.} In addition, at higher genus $g>1$ and for a non-abelian gauge group, we also have potential singularities at:
\be\label{hyperplane 3}
H_{\alpha, n} =   \{ u \in \Fh_\C \; |\; \alpha(u) = n~, \;\; n\in \Z~\}~,
\ee
for any simple root $\alpha\in \Fg$.  These hyperplanes correspond to the walls of the Weyl chambers, where part of the non-abelian symmetry is restored. Following previous works, our prescription will be to {\it exclude} the contribution from any singularity that includes $H_{\alpha, n}$ \cite{Blau:2006gh, Benini:2016hjo, Closset:2016arn}.

\subsection*{Example: $U(1)_{-1/2}$ with one chiral}
To illustrate some of these general features, we will consider a simple example, the $U(1)$ theory with a charge one chiral and an effective CS level $\kappa=-\frac{1}{2}$.  For simplicity let us consider the case $g=0,p=1$, \ie, $\Mgp=S^3$, and an $R$-charge such that $\Omega=0$.  Then the partition function is given by:
\bea 
Z_{S^3}(\nu,\tau)  &= -\int_{\CC} du \;e^{-2 \pi i \tau u}  \cF_\Phi(u+\nu) \\
&  =- \sum_{n \in \Z} \int_{\CC_{n}} du \; e^{-2 \pi i \tau u} \cF_\Phi(u+\nu )  \\
&  =- \sum_{\m \in \Z} \int_{\CC_0} du \; e^{-2 \pi i \tau u} \cF_\Phi(u+\nu )  \Pi(u)^{\m}~,
\eea
where:
\be
 \Pi(u) = \frac{e^{2\pi i \tau}}{1-e^{2 \pi i (u+ \nu)}}~.
\ee
Here we have included a mass parameter $\nu$ for the chiral multiplet, as well as an FI parameter $\tau$.~\footnote{We may eliminate $\nu$ by a shift of $u$, but it will be instructive to include it.}

\begin{figure}[t]
\begin{center}
\includegraphics[width=8cm]{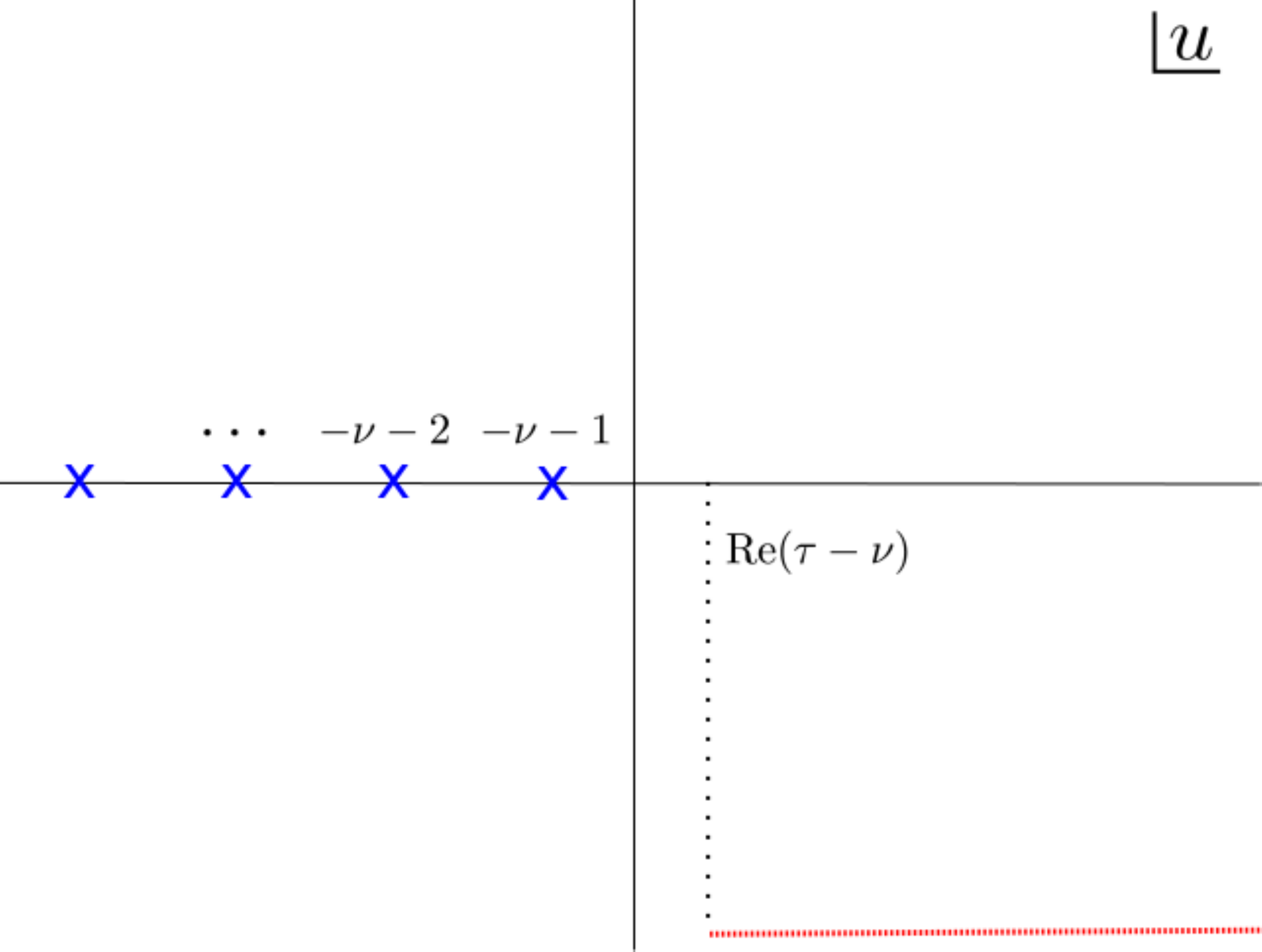}
\caption{Poles of integrand for $U(1)_{-1/2}$ with one chiral, for $g=0,\;p=1$.  Poles due to the positively charged chiral multiplet are shown in blue, and the ``poles at infinity'' from the negatively charged monopole, $T_+$, are denoted by the red line.}\label{fig:ex1poles}
\end{center}
\end{figure}

From \eqref{N pole}, the integrand has a pole of order $n$ at $u=-\nu-n$, $n=1,2,\cdots$.  For large $|u|$ the integrand behaves as:
\be \label{example large u} 
e^{-2 \pi i \tau u}  \cF_\Phi(u+\nu) \underset{|u|\rightarrow \infty} \longrightarrow \left\{ \begin{array}{cc}
e^{-2 \pi i \tau u} \;\; & \;\; {\rm if} \quad \text{Im}(u) >0~, \\
e^{\pi i (u+\nu)^2 -2 \pi i \tau u- \frac{\pi i}{6}} \;\; & \;\; {\rm if} \quad  \text{Im}(u) <0~. \end{array} \right. 
\ee
The behavior of the integrand is shown in Figure \ref{fig:ex1poles}.  There are poles due to the charged chiral multiplet, as well as the charged monopole $T_+$, at $\text{Im}(u) \rightarrow -\infty$, however the monopole $T_-$ is uncharged.  
%%%%%%%
%%%%%%
We will revisit this example below as we discuss more properties of the $\Mgp$ partition function.

%%%%%%%%
\subsubsection{Jeffrey-Kirwan contour: the rank-one case}
\label{sec:jkrankone}

\begin{figure}[t]
\begin{center}
\includegraphics[width=15cm]{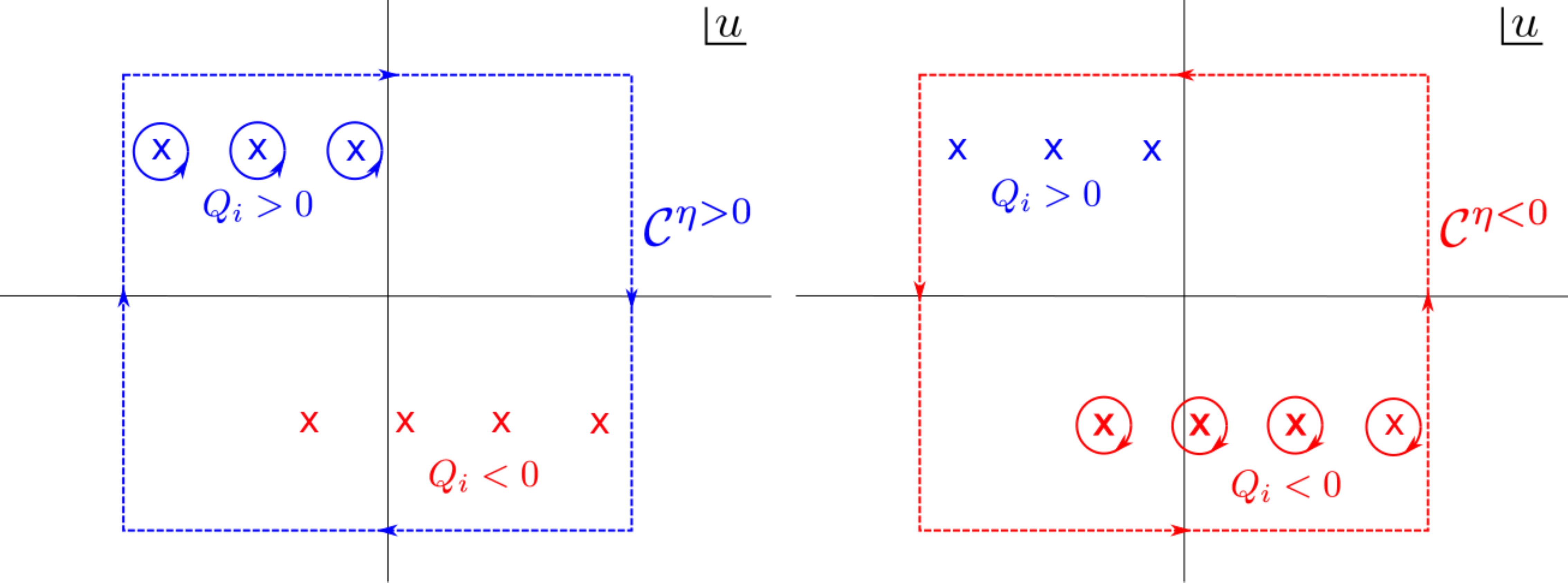}
\caption{The $p \neq 0$ JK contour, for $\eta >0$ and $\eta<0$, respectively.  For $\eta>0$ the contour surrounds the poles due to positively charged chirals in an anti-clockwise manner, and for $\eta<0$ it surrounds the poles due to negatively charged chirals in a clockwise manner.  Only the part of the contour at infinity that satisfies the condition $ \sign\left(\text{Im}(\d_u \CW)\right) =- \sign(\eta) $ should be included in the respective contours. }\label{fig:JKcontours}
\end{center}
\end{figure}

In the rank-one case (${\bf r}=1$), we can derive a precise contour on the $u$-plane, as we explain in Appendix \ref{app: loc}. The supersymmetric partition is given by:
\be \label{JK formula rank one}
Z_{\Mgp} ={-1 \ov |W_\GG|}  \sum_{\m \in \Z_p} \int_{\CC^\eta} du \; \CF(u)^p\, \pif(u)^\m\, \,\pif_\alpha(u)^{\n_\alpha}   \,e^{2 \pi i (g-1) \Omega(u)}\, H(u)^g~,
\ee
where $H(u)\equiv \d^2_u \CW(u)$, and $\pif(u)$, $\pif_\alpha(u)$  are the gauge and flavor flux operators, respectively.
The contour $\CC^\eta$, with $\eta$ a non-zero real number, is defined as follows.  We excise an $\epsilon$-neighborhood of all singularities in the integrand, as well as a box of size $R$, which we take very large, leaving a compact region, $\hat{\frak{M}}$ in which the integrand is regular.  We then define:
\be 
\CC^\eta = \big\{ u \in \partial \hat{\frak{M}} \; | \; \sign\left(\text{Im}(\d_u \CW)\right) =- \sign(\eta) \big\}~.
 \ee
For $\eta>0$, this includes those portions of $\partial \hat{\fM}$ encircling the poles due to positively-charged chiral multiplets, as well as the positively-charged monopole singularities at ${\rm Im}(u)\rightarrow \pm \infty$. Similarly, the contour for $\eta<0$ picks the contributions from  the negatively charged singularities.  The orientation of $\CC^\eta$ is positive or negative for $\eta>0$ or $\eta<0$, respectively. This is such that the residues from the positively charged chiral multiplets are counted with a plus sign (respectively, the residues from the negatively charged fields are counted with a minus sign). A corresponding orientation is assigned to the boundary components,  as shown in Figure \ref{fig:JKcontours}.  Since this contour integral is a slight modification of the Jeffrey-Kirwan residue prescription at rank one, we will call $\CC^\eta$ the ``JK contour''.
%%%%
%%%%

For some purposes, it will be useful to rewrite \eqref{JK formula rank one} as:
\be \label{JK formula rank one sum over fluxes}
Z_{\Mgp} ={-1 \ov |W_\GG|} \sum_{\m \in \Z} \int_{\CC_0^\eta} du \;  \CF(u)^p\, \pif(u)^\m\, \,\pif_\alpha(u)^{\n_\alpha}   \,e^{2 \pi i (g-1) \Omega(u)}\, H(u)^g~,
\ee
by using the identities \eqref{identities integrals}.
 Let $\hat{\frak{M}}_0$ be the restriction of $\hat{\frak{M}}$ to the vertical strip $0 \leq \text{Re}(u) \leq 1$ on the $u$-plane.
 The contour $\CC_0^\eta$ is defined as:
\be
\CC_0^\eta = \big\{ u \in \partial \hat{\frak{M}}_0 \; | \; \sign\left(\text{Im}(\d_u \cW)\right) = -\sign(\eta) \big\}~,
 \ee
 with the orientation depending on $\sign(\eta)$ as before.
Note that this contour generally includes vertical lines along ${\rm Re}(u)=0$ and  ${\rm Re}(u)=1$, 
as we will see in explicit examples below. For $p=0$, we have the same formula  \eqref{JK formula rank one sum over fluxes} with a periodic integrand, and the contributions of those vertical lines  cancel out. For $p \neq 0$, on the other hand, they are an important part of the JK contour $\CC_0^\eta$ in the quasi-periodic representation \eqref{JK formula rank one sum over fluxes}.   More generally, we may define $\hat{\frak{M}}_n$ to be the restriction of $\hat{\frak{M}}$ to the vertical strip $n \leq \text{Re}(u) \leq n+1$, and define the contour $\CC_n^\eta$ analogously.

We emphasize that, for each $\m$, the integral in \eqref{JK formula rank one sum over fluxes} is independent of the choice of $\eta$.\footnote{One way to see this is to note that $\CC_0^{\eta>0} - \CC_0^{\eta<0}$ encloses the region $\hat{\frak{M}}_0$, inside of which the integrand has no poles, and so the integral over this difference of the contours vanishes.}  Due to the non-periodicity of the integrand under $u \rightarrow u+1$ for $p \neq 0$, this property would not hold if we did not inlcude the segments of the vertical lines along $\text{Re}(u)=0$ and $1$.

\begin{figure}[t]
\begin{center}
\includegraphics[width=9cm]{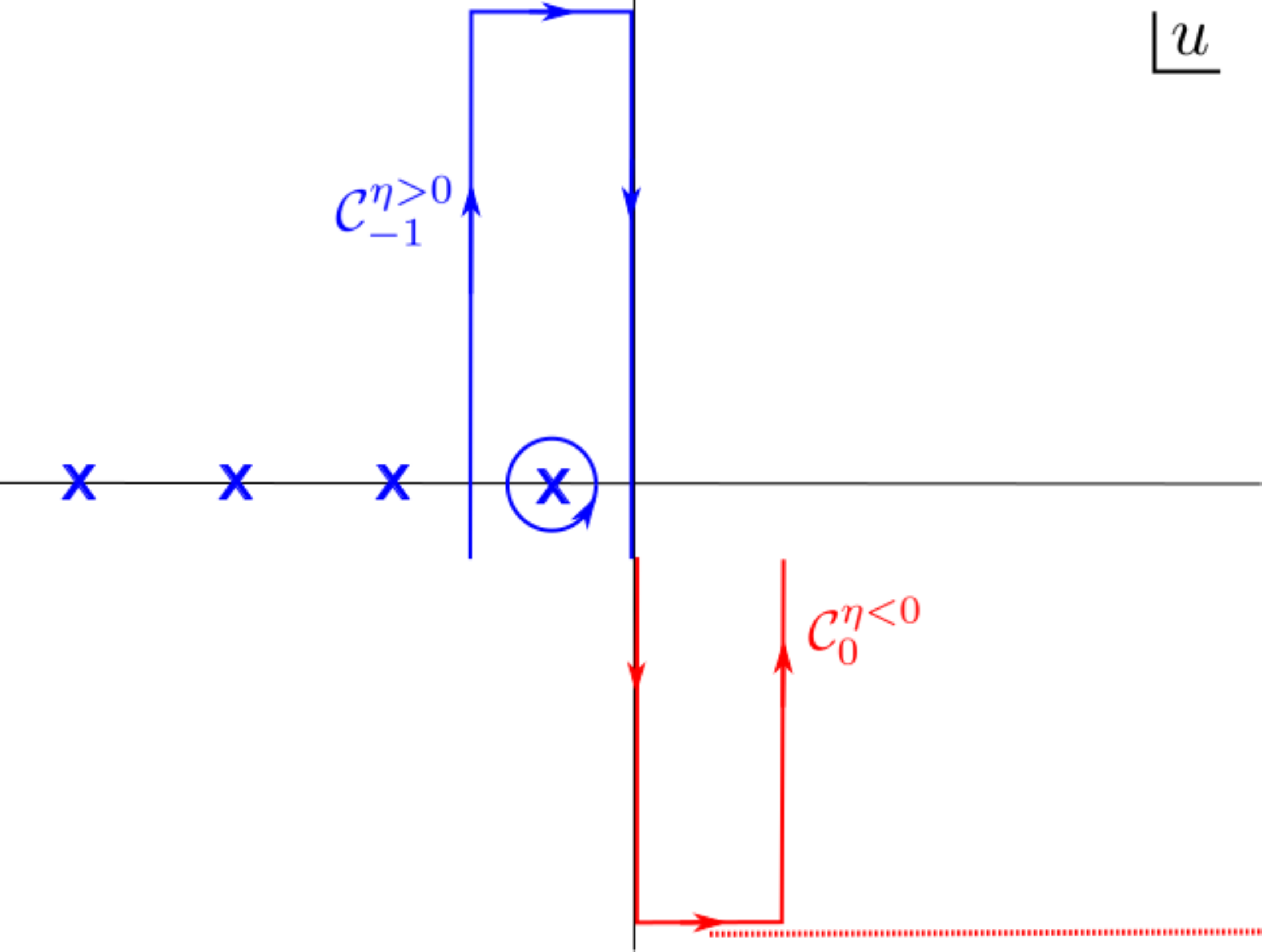}
\caption{JK contour $\CC_{-1}^\eta$ shown for $\eta>0$ in blue, and $\CC_{0}^\eta$ for $\eta<0$ in red.  Here we assume $\text{Im}(\tau)<0$, which implies that the contribution at $\text{Im}(u) \rightarrow \infty$ is included in the $\eta>0$ contour.}\label{fig:ex1jk}
\end{center}
\end{figure}

\subsection*{Example}

In Figure \ref{fig:ex1jk}, we illustrate the JK contour for the $U(1)_{-1/2}$ theory with a charge one chiral multiplet.  Note that:

\be \partial_u \cW = \tau - \frac{1}{2 \pi i} \log(1- e^{2 \pi i (u+\nu)}) \ee
Then one finds:
\be \text{Im}(\partial_u \cW) \rightarrow \left\{\begin{array}{ll} 
- \infty \;\;\; &\quad  u \rightarrow -\nu - n \\
\infty \;\;\; &\quad \text{Im}(u) \rightarrow -\infty \\
\text{Im}(\tau) \;\;\; &\quad \text{Im}(u) \rightarrow \infty \end{array} \right. 
\ee
in the respective limits.
Thus the $\eta>0$ contour surrounds the pole at $u=-\nu-n$, while the $\eta<0$ contour surrounds the ``pole at infinity'' due to the charged monopole $T_+$.  In addition, along the vertical boundaries of $\hat{\frak{M}}_\m$ at $\text{Re}(u) \in \Z$, and along the horizontal boundary at $\text{Im}(u) \rightarrow \infty$, a portion of the contour is selected depending on the sign of $\text{Im}(\partial_u \cW)$; the figure illustrates the behavior for $\text{Im}(\tau)<0$.

\subsubsection{The higher-rank case}
In the higher rank case, we conjecture the existence of a similar formula:
\be \label{JK formula higher rank}
Z_{\Mgp}= {(-1)^{\bf r}\ov |W_\GG|}\sum_{\m \in \Z_p^{\bf r}} \int_{\CC^\eta} d^{\bf r} u \;  \CF(u)^p\, \pif_a(u)^{\m_a}\, \,\pif_\alpha(u)^{\n_\alpha}   \,e^{2 \pi i (g-1) \Omega(u)}\, H(u)^g~,
\ee
with $H(u) \equiv \det_{ab}\d_{u_a}\d_{u_b}\CW$. Here $\eta$ is a non-zero covector in $\frak{h}^*$, and $\CC^\eta$ is an appropriate middle-dimensional ``JK contour'' in $\frak{h}^\C \cong \C^{\bf r}$.  Equivalently, by the same argument as in \eqref{JK formula rank one sum over fluxes}, we may rewrite this as:

\be \label{JK formula higher rank unfolded}
Z_{\Mgp}= {(-1)^{\bf r}\ov |W_\GG|}\sum_{\m \in \Z^{\bf r}} \int_{\CC_0^\eta} d^{\bf r} u \;  \CF(u)^p\, \pif_a(u)^{\m_a}\, \,\pif_\alpha(u)^{\n_\alpha}   \,e^{2 \pi i (g-1) \Omega(u)}\, H(u)^g~,
\ee
where $\CC^\eta_0$ is contained in the region $0 \leq \text{Re}(u_a) \leq 1$.  We will comment on the precise form of these contours below.

%%%%%%%%%%%%%%%%%%%%%%%%%%%%%

\subsection{Rank-one theories}
Let us explore some of the properties of the partition function formula of the previous section in the case of theories with a rank-one gauge group (that is, $\Fg=\frak{u}(1)$ or $\frak{su}(2)$).  We will comment on generalization to the higher-rank case in the next subsection.

\subsubsection{The $\sigma$-contour}
\label{sec:sigmacontour}

For $p \neq 0$, it is possible to express the $\Mgp$ partition function as a non-compact integral:
 
\be \label{Zmgp CB formula rank one general} Z_{\Mgp}=\frac{-1}{|W_G|}\sum_{\m \in \Z_p}   \int_{\CC_\sigma} du \; \cF(u)^p\, \pif(u)^{\m} \,\pif_\alpha(u)^{\n_\alpha}  \,e^{(g-1)\Omega(u)}\, H(u)^{g}~. 
\ee 
Here, $\CC_\sigma$ is a non-compact contour connecting $\text{Im}(u) \rightarrow -\infty$ with $\text{Im}(u) \rightarrow \infty$.  In other words, it is roughly an integral over imaginary $u$, or equivalently, over real $\sigma$.  This directly relates the contour prescription presented here to the one used in earlier work on the round $S^3$ \cite{Kapustin:2009kz, Hama:2010av, Jafferis:2010un}, where such an integral over real $\sigma$ was obtained instead. Here we derive the precise form of the non-compact contour, $\CC_\sigma$, by relating it to the JK contour prescription. Note that, unlike the more naive contour along the imaginary $u$ axis, the contour $\CC_\sigma$ always leads to a converging integral.

\subsection*{A simplification in the sum over fluxes}
To proceed, we will need a general fact about the sum over fluxes, which holds for all $p$.
  As we can see from \eqref{hyperplane 1} and \eqref{N pole}, a chiral multiplet of gauge charge $Q$ and $R$-charge $r$ contributes poles to the integrand if and only if:
\be \label{chiral bound} 
Q (p \; \text{Re}(u) - \m) < -(p \;\text{Re}(\nu) - \n) +  (r-1)(g-1)~.  
\ee
where $\nu, \n$ are the flavor parameters.
We also have ``monopole contributions'' that arise in the limit $\text{Im}(u) \rightarrow \mp \infty$, where the integrand takes the form:
\be
 \exp \pi i \left(- p \, k_{\mp} u^2 + 2 k_{\mp} u \m - 2p \tau_{\mp} u + 2 \tau_{\mp} \m + 2 u \n_{\mp} + 2  k_{R\mp}  u (g-1) \right)~.
 \ee
Here $k_\mp$ and $k_{R\mp}$ are the effective CS levels in this limit, which depend on the charges of the chiral multiplets, while $\tau_\mp$ and $\n_\mp$ are the effective parameters for the topological symmetry $U(1)_T$, which depend on the flavor symmetry parameters. Recall from \eqref{induced charges T} that we have the monopole charges $Q_\pm = \pm k_\mp$ and $r_\pm = \pm k_{R \mp}$. We then find that the monopole singularity at ${\rm Im}(u) \rightarrow \mp \infty$ contribute a ``pole at infinity'' only when:
\be  
 Q_{\pm}(p \; \text{Re}(u) - \m)  < -   ( \pm p \; \text{Re}(\tau_{\mp}) \mp  \n_\mp)+r_\pm(g-1)~.
\ee
We see that the integrand only has singularities, associated to either the chiral multiplets or the monopole operators, provided that:
\be \label{bound rank one} 
Q_\alpha (p \;\text{Re}(u) - \m) < \delta_\alpha 
\ee
where the index $\alpha$ runs over both  the matter and monopole contributions, and $\delta_\alpha$ is some constant that depends on the flavor symmetry parameters and $R$-charges.  
Note that if $Q_\alpha=0$ for a monopole operator,\footnote{A charge-zero chiral multiplet has a contribution which is independent of the gauge parameter, and so does not enter this analysis.} 
this bound becomes independent of $u$ and $\m$, and depends only on the flavor symmetry parameters through $\delta_\alpha$.  We will return to this point below.

In general, the allowed choices for the $R$-charge and the flavor symmetry parameters may be restricted by superpotential terms and by the Weyl symmetry.  Suppose for the moment that we lift any such  restriction, and allow independent mass parameters for all chiral multiplets, and complexified FI parameters in the Cartan of the gauge group.  Then it is clear from \eqref{chiral bound} that we may choose $\delta_\alpha$ arbitrarily for each chiral multiplet.  Similarly, by shifting the bare $U(1)_T$ parameters, we may take $\delta_\pm$, the bounds for the monopole contributions, to be arbitrary.
Although the answer we obtain in this way may only be defined in a non-physical region of parameter space, it is typically possible to analytically continue it back to the physical region at the end of the computation.

Returning to the expression  \eqref{JK formula rank one sum over fluxes} for the $\Mgp$ partition function of a rank-one gauge theory, we find it useful to decompose the contour $\CC_0^\eta$ into two pieces:
\be
 \CC_0^\eta = \CC_0^{\eta, \, {\rm bulk}} + \CC_0^{\eta,\, {\rm boundary}}~,
\ee
where $ \CC_0^{\eta, \, {\rm bulk}}$ is the part of the contour surrounding the poles in the integrand due to the charged chiral multiplets and monopole operators.  The remainder,  $ \CC_0^{\eta,\, {\rm boundary}}$, can be further decomposed as:
\be \label{c boundary definition}
\CC_0^{\eta,\, {\rm boundary}}  = \CC_0^{\eta,\,\text{Re}(u)=0} + \CC_0^{\eta,\,\text{Re}(u)=1}
+ \delta_{Q_+,0} \;\CC_0^{\eta,\,\text{Im}(u)=-R}+ \delta_{Q_-,0}\; \CC_0^{\eta,\,\text{Im}(u)=R}~.
  \ee
Here the first two terms consist of the parts of the vertical lines $\text{Re}(u)=0,1$ contained in $\CC^\eta_0$, while  the third and fourth terms are only included if the corresponding monopole charge vanish, $Q_+=0$ and/or $Q_-=0$ (otherwise these pieces are included as part of  $\CC_0^{\eta, \, {\rm bulk}}$). 

It follows from \eqref{bound rank one} that, for sufficiently large negative $\m$, there are no poles in the integrand which are due to the positively-charged singularities.  Therefore, if we choose $\eta>0$ to compute the integral at large negative $\m$, the contour $\CC_0^{\eta >0, \, {\rm bulk}}$ gives a vanishing contribution, and the only contribution comes from the boundary contour.  Similarly, for sufficiently large positive $\m$, we may take $\eta<0$ and there will be no  contribution from $\CC_0^{\eta <0, \, {\rm bulk}}$.

For $p=0$, since the integrand is periodic under $u \rightarrow u+1$, the boundary contributions along $\text{Re}(u)=0$ and $\text{Re}(u)=1$ cancel each other.  If, in addition, the monopole charges $Q_\pm$ are non-zero, then the sum over $\m$ truncates to a finite sum.  More generally, if one or both of $Q_\pm$ vanishes, this truncation only occurs provided the flavor symmetry parameters are picked so that the corresponding $\delta_\alpha$ satisfies \eqref{bound rank one}. Otherwise, there will be in general be a contribution from infinitely many flux sectors.  A similar truncation property in the sum over topological sectors was observed in \cite{Gadde:2015wta} in the context of the $S^2 \times T^2$ partition function (where there are no subtleties related to monopole contributions).  

For $p\neq0$, on the other hand, the pieces along the vertical boundaries at $\text{Re}(u)=0$ and $1$ no longer cancel. Their contributions actually add up and give rise to the $\sigma$-contour. 

\subsection*{Deriving the $\sigma$-contour}

To proceed, let us first assume,  for simplicity, that we have  $\delta_\alpha- Q_\alpha p<0$ for all $\alpha$.  In this case, there is no contribution from positively-charged singularities for any $\m\geq  0$, and no contributions from negatively-charged singularities for any $\m \leq 0$.  Moreover, with this assumption, the bound \eqref{bound rank one} is also violated for any zero-charge monopole operator, and so there is no contribution from the third and fourth terms in \eqref{c boundary definition}.  Then, if we choose $\eta>0$ for $\m \leq 0$ and $\eta<0$ for $\m > 0$, there are no bulk contributions from $\CC_0^\eta$ for any $\m$, and we only need to deal with the boundary contributions.

We set $p>0$ for definiteness. Let us rewrite the expression \eqref{JK formula rank one sum over fluxes} as:
\bea \label{sigma contour step two} 
Z_{\Mgp} &=  \frac{1}{|W_G|} \sum_{\m \in \Z_p} \sum_{n \in \Z}  \oint_{\CC_0^{\eta}} du \; \CF(u-n)^p \,\pif(u)^\m\,  \t\CJ(u) \cr
&= \frac{1}{|W_G|} \sum_{\m \in \Z_p} \bigg(\sum_{n =1}^\infty\oint_{\CC_0^{\eta<0}}+  \sum_{n=-\infty}^{0} \oint_{\CC_0^{\eta>0}} \bigg) du \;  \CF(u-n)^p \,\pif(u)^\m\,  \t\CJ(u)~,
\eea
where we wrote the periodic and $\m$-independent part of the integrand as $\t\CJ(u)$ to avoid clutter. 
Then, as argued above, the bulk contour, along with the third and fourth terms in \eqref{c boundary definition},  give a vanishing contribution, and we are left with:
\bea 
\frac{1}{|W_G|}  \sum_{\m \in \Z_p} \bigg( \sum_{n =1}^\infty  \bigg(  \int_{\CC_0^{\eta<0,\,\text{Re}(u)=0} } + \int_{\CC_0^{\eta<0,\,\text{Re}(u)=1} } \bigg)   du \; \CF(u-n)^p \,\pif(u)^\m\,  \t\CJ(u)  \cr
+ \sum_{n =-\infty}^{0}  \bigg( \int_{\CC_0^{\eta>0,\,\text{Re}(u)=0} } +\int_{\CC_0^{\eta>0,\,\text{Re}(u)=1} } \bigg)   du \;\CF(u-n)^p \,\pif(u)^\m\,  \t\CJ(u)  \bigg)~. 
\eea
Consider the sum over $n \leq 0$. Since the contours along $\text{Re}(u)=0$ and $\text{Re}(u)=1$ have opposite orientations, we see that it is a telescoping sum, with contributions canceling between adjacent terms.  That is, if we place an lower cutoff at $-N$, we find:
\bea\label{sum negative N}
&\sum_{n =-N}^0 \bigg(  \int_{\CC_0^{\eta>0,\,\text{Re}(u)=0} } + \int_{\CC_0^{\eta>0,\,\text{Re}(u)=1} } \bigg)   du \; \CF(u-n)^p \,\pif(u)^\m\,  \t\CJ(u)  \\
&= \int_{\CC_0^{\eta>0,\,\text{Re}(u)=0} } du \; \CF(u)^p \,\pif(u)^\m\,  \t\CJ(u)  +  \int_{\CC_0^{\eta>0,\text{Re}(u)=1} } du \; \CF(u+N)^p \,\pif(u)^\m\,  \t\CJ(u)~. 
\eea
Using $\CF(u+N)^p =\CF(u)^p\, \pif(u)^{-p N}$ and the fact that $|\pif(u)|=e^{-2 \pi \text{Im}(\partial_u \cW)}>1$ for $\eta >0$, by definition of the $\CC_0^\eta$ contour, we see that the second term in \eqref{sum negative N} vanishes as $N \rightarrow \infty$, and so the sum converges to the first term.  Similarly, the sum over positive $n$ converges to:
\be 
\int_{\CC_0^{\eta<0,\, \text{Re}(u)=1} } du \; \CF(u-1)^p \,\pif(u)^\m\,  \t\CJ(u)  = \int_{\CC_0^{\eta<0,\, \text{Re}(u)=0} } du \;  \CF(u)^p \,\pif(u)^\m\,  \t\CJ(u)~.
\ee
We  then find:
\be 
Z_{\Mgp} =\frac{1}{|W_G|}   \sum_{\m \in \Z_p}  \bigg( \int_{\CC_0^{\eta> 0,\, \text{Re}(u)=0}}+\int_{\CC_0^{\eta<0,\, \text{Re}(u)=0} } \bigg) du \;\CF(u)^p \,\pif(u)^\m\,  \t\CJ(u)~.
\ee
These two pieces include the portion of $\text{Re}(u)=0$ with $\text{Im}(\partial_u \CW)>0$ and $\text{Im}(\partial_u \CW)<0$, respectively, thus spanning the entire imaginary axis (up to a measure zero subset). The respective orientations are such that $\CC_0^{\eta>0,\, \text{Re}(u)=0} +\CC_0^{\eta<0,\, \text{Re}(u)=0}$ is the contour along $\text{Re}(u)=0$ from $\text{Im}(u)= - \infty$ to $\text{Im}(u)= \infty$.
We finally obtain:
\be \label{Zmgp CB formula rank one with assumption} 
Z_{\Mgp}= \frac{-1}{|W_G|}   \sum_{\m \in \Z_p} \int_{\text{Re}(u)=0} du \; \cF(u)^p\, \pif(u)^{\m} \,\pif_\alpha(u)^{\n_\alpha}  \,e^{(g-1)\Omega(u)}\, H(u)^{g}~.
\ee
This reproduces the formula \eqref{Zmgp CB formula rank one general} with $\CC_\sigma$ equal to the imaginary axis. 

The above derivation relied on the assumption that $\delta_\alpha-  Q_\alpha p<0$,  $\forall \alpha$.  In the more general case, the argument above, and the resulting integration contour $\CC_\sigma$, needs to be modified slightly.  In general, we may find a finite set of fluxes, $\m$, which have a non-zero contribution from $ \CC_0^{\eta, \, {\rm bulk}}$.  In addition, there may be contributions from third and fourth terms in \eqref{c boundary definition}.  Both contributions will add additional pieces to  the $\sigma$-contour.

Alternatively, a simple way to arrive at the correct contour is to start from a region of parameter space where the assumption $\delta_\alpha- Q_\alpha p<0$ holds, in which case the contour is the imaginary axis, and then analytically continue to the region of interest.  As we continuously vary parameters to perform this analytic continuation, we must deform the integration contour so that no poles cross it.  In particular, noting that the initial contour separates all poles due to positively charged chirals from those due to negatively charged ones, this must also be true of the general contour $\CC_\sigma$.  

Note this conditions does not uniquely fix the contour $\CC_\sigma$, however, all choices which separate poles appropriately will give the same result by holomorphy.  For $p>1$, we may in principle choose $\CC_\sigma$ differently for the $p$ different terms in the sum in \eqref{Zmgp CB formula rank one with assumption}.  However, it is always possible to find a single $\CC_\sigma$ which separates the poles due to positively and negatively charged fields for all $\m \in \Z_p$, and we will always make this choice.

\subsection*{Example}
\begin{figure}[t]
\begin{center}
\includegraphics[width=\textwidth]{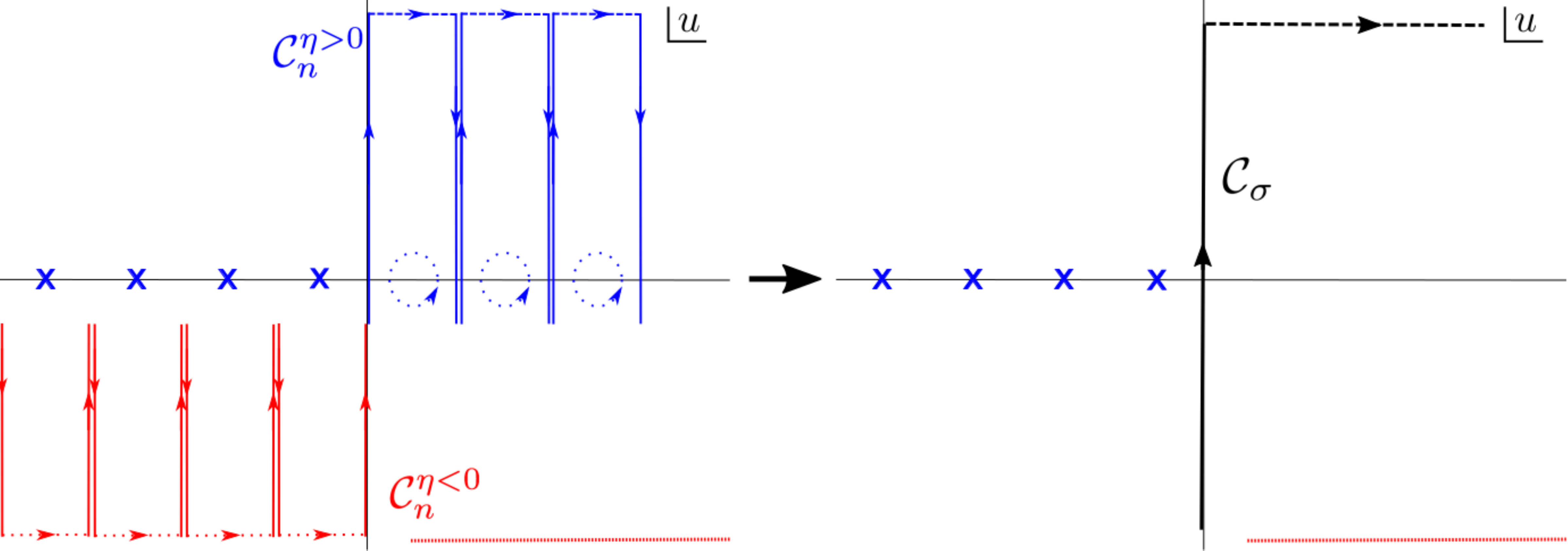}
\caption{Taking $\eta<0$ for $n<0$ and $\eta>0$ for $n \geq 0$, we see the contours do not enclose any poles, and so the bulk contributions vanish, leaving only the boundary contributions.  These sum to form the contour $\CC_\sigma$.  Here we have taken $\text{Im}(\tau)<0$.}\label{fig:ex1sigma}
\end{center}
\end{figure}

\begin{figure}[t]
\begin{center}
\includegraphics[width=\textwidth]{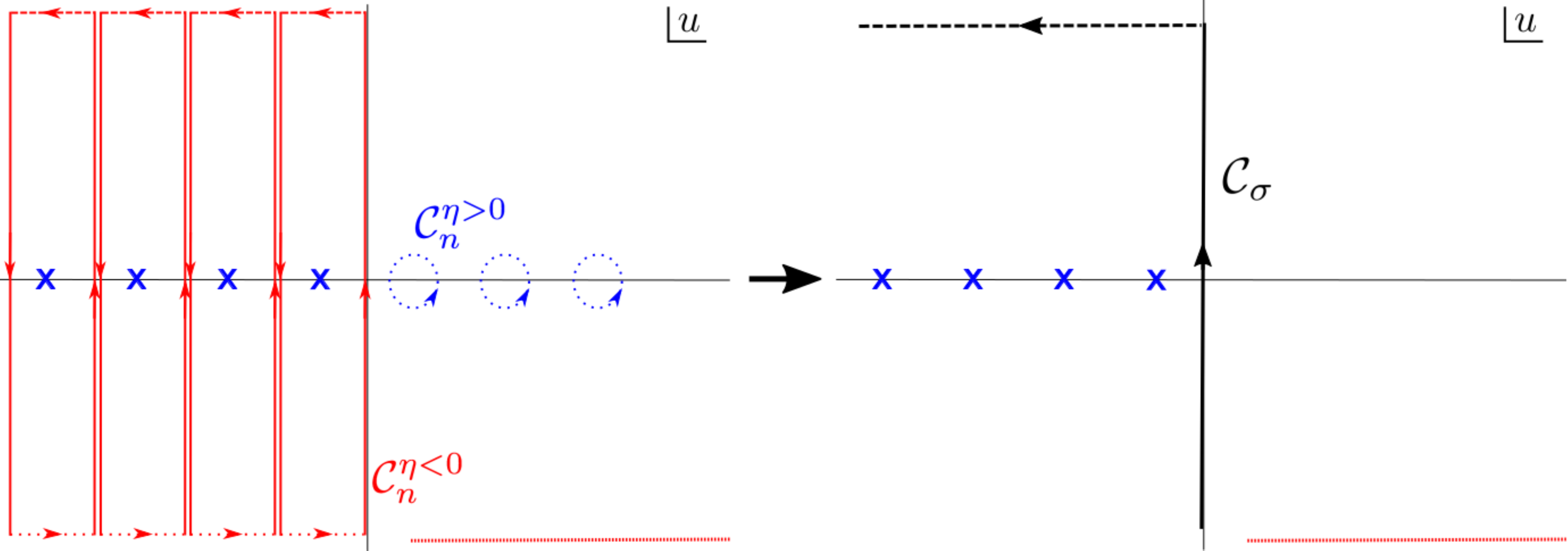}
\caption{Here we take $\text{Im}(\tau)>0$, and note the corresponding contours, $\CC_n^{\eta>0}$ and $\CC_n^{\eta<0}$.  Summing these as above to obtain the $\sigma$-contour, we see it now runs off to $\text{Re}(u)<0$ as $\text{Im}(u) \rightarrow \infty$.}\label{fig:ex1sigma2}
\end{center}
\end{figure}
Returning to our example of $U(1)_{-1/2}$ with a charge one chiral, note from Figure \ref{fig:ex1jk} that for $n \geq 0$, the contour $\CC_{n}^{\eta>0}$ does not enclose any poles of the integrand.   Similarly, for $n \leq 0$, the contribution to $\CC_{n}^{\eta<0}$ from the charged monopole, $T_+$, at $\text{Im}(u) \rightarrow -\infty$, vanishes.  Thus if we choose $\eta>0$ for $n\geq 0$ and $\eta<0$ for $n<0$, there are no contributions from the ``bulk'' part of the contour, shown in Figure \ref{fig:ex1sigma} as the dotted lines.  All that remains are the boundary contributions, which partially cancel between adjacent values of $n$, and leave the non-compact contour $\CC_\sigma$.

In more detail, the contour $\CC_\sigma$ pictured in Figure \ref{fig:ex1sigma} includes both the imaginary $u$ axis, as well as a horizontal piece at $\text{Im}(u) \rightarrow \infty$, which is the contribution from the uncharged monopole, $T_-$.  This piece extends towards positive $\text{Re}(u)$, which is a consequence of choosing $\text{Im}(\tau)<0$.  If we take $\text{Re}(\tau)<0$, which is equivalent to imposing $\delta_-<0$, then the contribution from this horizontal piece vanishes, and $\CC_\sigma$ is simply the imaginary $u$ axis.

More generally, we must include this piece of the contour to obtain a convergent integral.  Note that if we instead took $\text{Im}(\tau)>0$, we would obtain a different JK contour, as shown in Figure \ref{fig:ex1sigma2}.  This leads to a contour $\CC_\sigma$ which extends towards negative $\text{Re}(u)$.  To understand this behavior, note from \eqref{example large u} that in order for the integral to converge as $\text{Im}(u) \rightarrow \infty$ we must have $\text{Im}(\tau u)<0$.  One can check that this holds provided we take (for some $\delta>0$ depending on $\text{arg}\;\tau$):

\be \text{arg}\;u \in \left\{ \begin{array}{cc} 
(0,\delta) & \;\; \text{Im}(\tau)<0 \\
(\pi-\delta,\pi) & \;\; \text{Im}(\tau)>0 \\
 \end{array} \right. \;\;\;\; \text{as}\;\text{Im}(u) \rightarrow \infty\ee
These conditions are satisfied by the contours shown above.

\begin{figure}[t]
\begin{center}
\includegraphics[width=\textwidth]{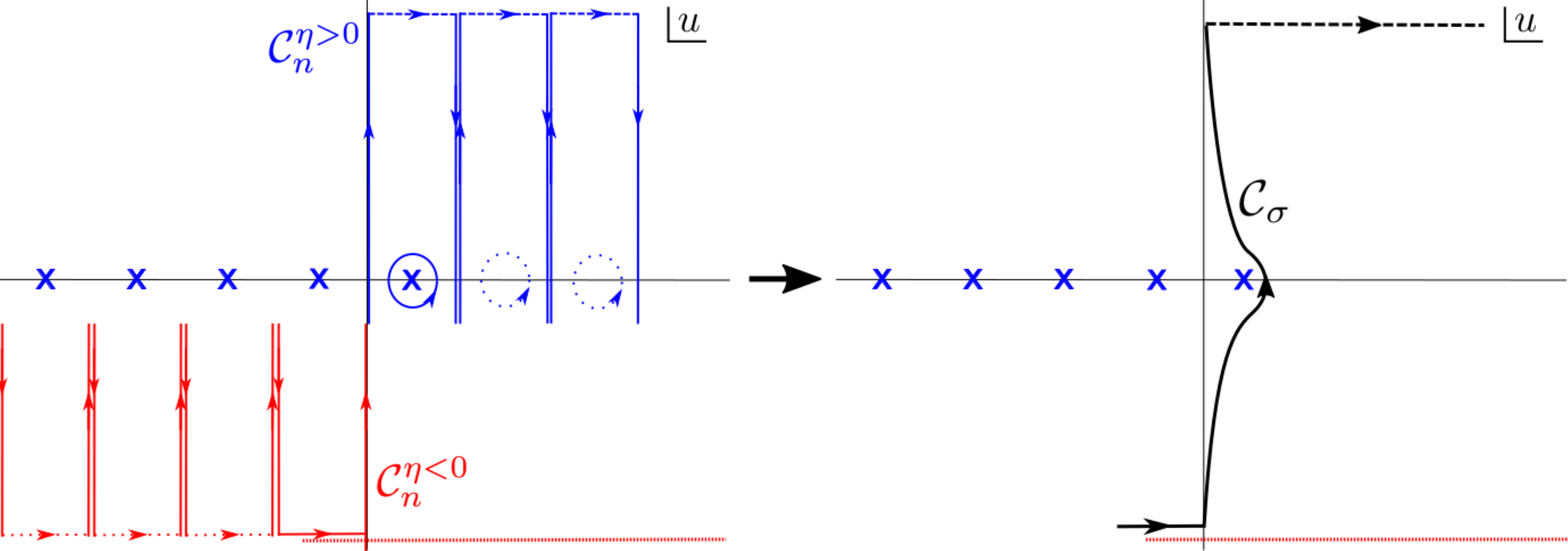}
\caption{For more general choices of parameters, there may be contributions from the bulk parts of $\CC^\eta_n$, indicated by solid lines.  Here the contour we find after summing all these pieces is homologous to the one shown at right, which separates the poles due to the positively and negatively charged fields}\label{fig:ex1sigma3}
\end{center}
\end{figure}
Finally, we note that if we vary the parameters $\nu$ and $\tau$, it may no longer be the case that all the bulk pieces of the contour vanish.  This is illustrated in Figure \ref{fig:ex1sigma3}.  In this case, we see that the contour we found above is supplemented by a finite number of additional pieces.  The resulting contour is homologous to one which separates the poles due to positively and negatively charged fields.

\subsubsection{Relation to the Bethe-vacua formula}

In this section, we have computed the partition function by supersymmetric localization, starting from the UV action.  In Section \ref{sec: BE formula}, we computed it instead using the low energy effective action, and found it was expressed as a sum over Bethe vacua.  In this subsection we relate these two prescriptions, and we argue that they give the same result. For completeness, we will present two arguments, relating the Bethe vacua formula to the JK contour in \eqref{JK formula rank one}, and then relating it directly to the $\sigma$-contour derived in the previous subsection.

\subsubsection*{Relation to JK-contour}

Let us first assume that the gauge group is $U(1)$.  We start again from the JK contour expression \eqref{JK formula rank one sum over fluxes}:
\be \label{BE relation start} 
Z_{\Mgp} = \sum_{\m \in \Z} \oint_{\CC_0^\eta}d u \;  \CJ(u) \, \pif(u)^\m~.
\ee
where we defined $\CJ(u)$ as in \eqref{def CJ}.
Note that, on the contour $\CC_0^\eta$, we have:
\be\label{condition on pif}
 |\pif(u)|<1 \quad {\rm if} \quad\eta <0~, \qquad \qquad
  |\pif(u)|>1 \quad {\rm if}\quad \eta >0~.
\ee
As before, we may choose to take $\eta <0$ for $\m \geq 0$ and $\eta >0$ for $\m <0$:
\be
Z_{\Mgp} = \sum_{\m =-\infty}^{-1} \oint_{\CC_0^{\eta>0}}d u \;  \CJ(u) \, \pif(u)^\m+\sum_{\m =0}^{\infty} \oint_{\CC_0^{\eta<0}}d u \;  \CJ(u) \, \pif(u)^\m~.
\ee
Then, both sums give converging geometric series due to \eqref{condition on pif}, and we can permute the sum and the integral. We then obtain:
\be\label{contour int inter BE contour}
Z_{\Mgp} = \left(- \oint_{\CC_0^{\eta>0}}+\oint_{\CC_0^{\eta<0}}\right)d u \; { \CJ(u)\ov 1- \pif(u)}= \oint_{\CC_{\rm BE}} d u \; { \CJ(u)\ov 1-\pif(u)}~,
\ee
where we defined the contour:
\be
\CC_{\rm BE} \equiv  \CC_0^{\eta<0} - \CC_0^{\eta>0}~.
\ee
This contour precisely  bounds the region $\hat{\frak{M}}$ remaining after all poles in the original integrand have been excised.  Therefore, by definition of $\hat{\frak{M}}$, the contour integral \eqref{contour int inter BE contour} does not pick any contributions from any of the poles of $\CJ(u)$. On the other hand $\CC_{\rm BE}$ is homologous to a contour that surrounds all the poles at the solutions to the Bethe equation $\pif(u)=1$ in an anti-clockwise manner. This directly leads to the Bethe-vacua formula:
\bea
\label{Bethe from JK}
&Z_{\Mgp} &=&\; \sum_{\h u | \Pi(\h u) = 1} 2 \pi i\, \text{Res}_{u= \h u}  {\CJ(u)\ov 1-\pif(u)}  \cr
&&=&\;- \sum_{\h u | \Pi(\h u) = 1} \CJ(\h u) H(\h u)^{-1} 
=\sum_{\h u | \Pi(\h u) = 1} \CF(\h u)^p\, \CH(\h u)^{g-1}\, \pif_\alpha(\h u)^{\n_\alpha}~. 
\eea
where we used the identity $\d_u \pif(u) = 2 \pi i H(u) \pif(u)$ in the second line.

Finally, if the gauge group is non-abelian (\ie, for $\GG=SU(2)$ or $SO(3)$), we should exclude $\h u=0$ from the potential Bethe solutions. (At $g=0$, we have a vanishing residue due to the vector multiplet contribution, while we should exclude the $\h u=0$ contribution by hand in general.~\footnote{One can also argue for it by introducing a non-gauge-invariant real-mass regulator for the $W$-bosons \protect\cite{Benini:2016hjo}.})
  The non-zero solutions come in Weyl-equivalent pairs, $\{ \pm \h u \}$, which give the same contribution, and so we may count each pair once, cancelling the Weyl symmetry factor $|W_\GG|=2$.  We are then counting precisely the Bethe solutions \eqref{S BE 3d}, and we reproduce in this way the Bethe-vacua expression \eqref{ZMgp main} for the $\Mgp$ partition function.

\subsection*{Example}
\begin{figure}[t]
\begin{center}
\includegraphics[width=0.9\textwidth]{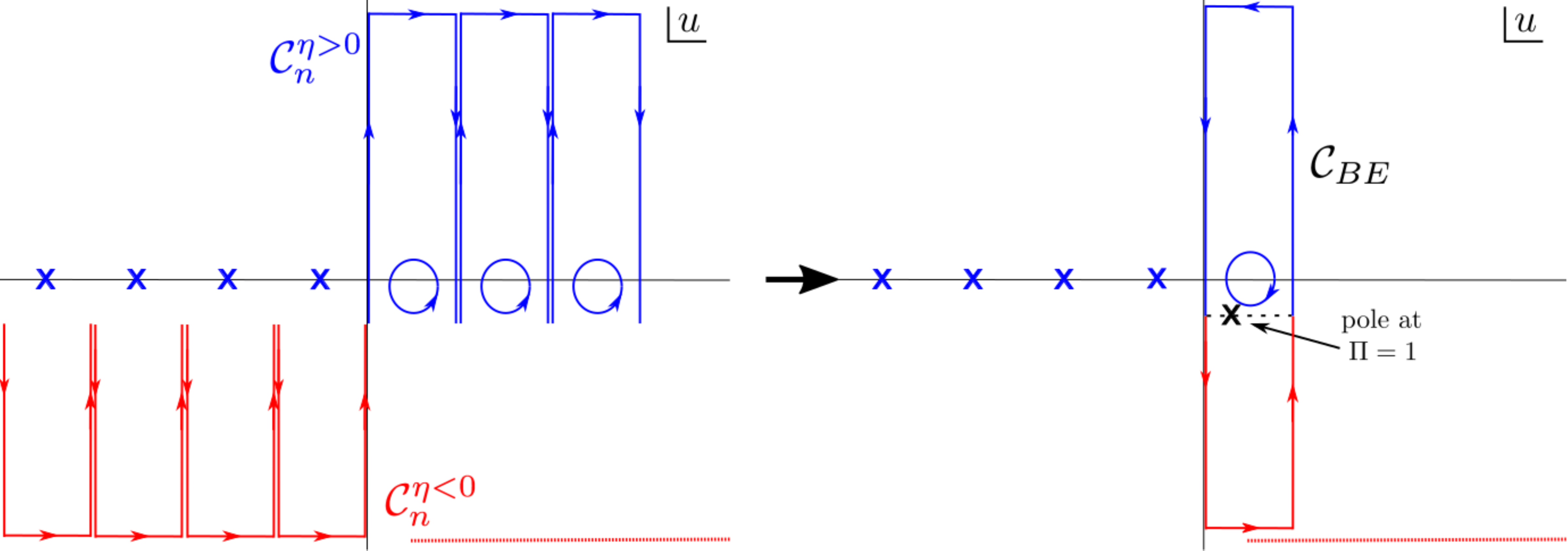}
\caption{Taking $\eta<0$ for $n<0$ and $\eta>0$ for $n \geq 0$, the geometric series converges and we find the integral of $ \frac{1}{1-\Pi(u)}\cF_\Phi(u+\nu) e^{-2 \pi i \tau u}$ on the region shown on the RHS.  This encloses a single pole at $\Pi=1$, corresponding to the Bethe vacuum.}\label{fig:ex1Bethe}
\end{center}
\end{figure}

Let us see how this argument works in the example we have been considering above.  If we start from the same contour as in the LHS of Figure \ref{fig:ex1sigma}, note that the geometric series in the sum over $\m\geq 0$, in the right half plane, converges on $\CC_\m^{\eta>0}$, since $|\Pi|<1$ there, while the sum over $\m<0$ converges along $\CC_\m^{\eta<0}$ since $|\Pi|^{-1}<1$ there.  Summing the geometric series, we find:

\be Z_{\Mgp} =- \int_{\CC_0^{\eta<0}-\CC_0^{\eta>0} } du \frac{1}{1-\Pi(u)}\cF_\Phi(u+\nu) e^{-2 \pi i \tau u} \ee
This is shown in Figure \ref{fig:ex1Bethe}.  This encloses the region $\hat{\frak{M}}_0$, and counts the residues from any poles in this region.  This includes only the single pole at $\Pi=1$, leading to the Bethe-vacua formula as above.

\subsubsection*{Relation to $\sigma$-contour}

Next, let us start from the $\sigma$-contour formula:
\be
Z_{\Mgp} ={1\ov |W_\GG|} \sum_{\m \in \Z_p} \int_{\CC_\sigma} d u\;\CF(u)^p \,\pif(u)^\m\,  \t\CJ(u)~.
\ee
As noted above, $\CC_\sigma$ is a contour which separates the poles due to the positively charged fields (including monopole singularities at infinity) from those due to the negatively charged ones. For $p>0$, the positively-charged singularities are to the left of $\CC_\sigma$.
By performing the finite sum over $\m$ and using the difference equation for $\CF(u)$, we find:
\bea\label{c prime rewrite}
Z_{\Mgp} &={1\ov |W_\GG|} \int_{\CC_\sigma} d u \;   \frac{\CF(u)^p-\CF(u-1)^p}{1-\Pi(u)}\t\CJ(u) \\
&= {1\ov |W_\GG|} \bigg( \int_{\CC_\sigma} -  \int_{\CC_\sigma-1}  \bigg)d u \; \frac{\CF(u)^p \t\CJ(u)}{1-\Pi(u)}
=  {1\ov |W_\GG|}  \oint_{\CC'} d u  \; \frac{\CF(u)^p \t\CJ(u)}{1-\Pi(u)}~,
\eea
where $\CC'$ is a contour which encloses all the poles of the integrand between $\CC_\sigma$ and $\CC_\sigma-1$. (Roughly speaking, it encloses the strip $-1\leq\text{Re}(u)\leq 0$ in anti-clockwise manner.)  There are two types of poles that may occur in that region: those from the original integrand, $\CF(u)^p \t\CJ(u)$, and those at solutions to $\pif(u)=1$.  

Let us first consider the poles from the original integrand.  The poles due to negatively-charged fields must lie to the right of $\CC_\sigma$, and so they cannot be enclosed by $\CC'$.  On the other hand, poles from positively-charged fields may lie inside $\CC'$, and suppose one lies at some $u_*$.  Then by assumption, there is no pole at $u_*+1$ for any $\m \in \Z_p$, and so $\CF(u_*+1)^p \Pi(u_*)^\m \t\CJ(u_*+1) = \CF(u_*)^p \t\CJ(u_*) \Pi(u_*)^{\m-p}$ is finite.  In particular, taking $\m=p-1$, we see $\CF(u_*)^p \t\CJ(u_*) \Pi(u_*)^{-1}$ is finite, and so $\frac{\CF(u_*)^p \t\CJ(u_*)}{1-\Pi(u_*)}$ is finite as well (here we recall $\Pi(u) \rightarrow \infty$ for poles due to positively charged fields).
 
Then the only poles in \eqref{c prime rewrite} lie at solutions to $\Pi(u)=1$.  The partition function is then given by:
\bea\label{BE from Csigma}
Z_{\Mgp} &= {1\ov |W_\GG|} \sum_{\h u| \Pi(\h u)=1} 2 \pi i\; \mbox{Res}_{u= \h u}  \frac{\CF(u)^p \t\CJ(u)}{1-\Pi(u)}  \\
&= {1\ov |W_\GG|} \sum_{\h u| \Pi(\h u)=1} \cF(\h u)^p \Pi_\alpha(\h u)^{\n_\alpha} \CH(\h u)^{g-1}~.
\eea 
For a non-abelian rank-one gauge group, the same comments as written after \eqref{Bethe from JK} apply.  The final formula  precisely agrees with \eqref{ZMgp main}.

%%%%%%%%%%%%%%%%%%%%%%%%
\subsection{Higher-rank theories}
In this section, we briefly discuss how some of the above considerations generalize in the case of higher-rank gauge theories.  
 
\subsection*{Higher rank ``JK contour'': a conjecture}

Here we present a natural conjecture for the contour $\CC^\eta$ appearing in \eqref{JK formula higher rank}, \ie:

\be \label{JK formula higher rank two}
Z_{\Mgp}= {(-1)^{\bf r}\ov |W_\GG|}\sum_{\m \in \Z_p^{\bf r}} \int_{\CC^\eta} d^{\bf r} u \;  \CF(u)^p\, \pif_a(u)^{\m_a}\, \,\pif_\alpha(u)^{\n_\alpha}   \,e^{2 \pi i (g-1) \Omega(u)}\, H(u)^g~.
\ee
Although we do not derive it directly from a localization argument, it passes several consistency checks, as we outline below and in Appendix \ref{sec:HRJK}.  We leave its careful derivation to future work.

First we recall the usual JK residue prescription \cite{JK1995,Benini:2013xpa}.  Generically, the integrand in \eqref{JK formula higher rank two} may have non-trivial residues at the intersection of $r$ complex-codimension-one singular hyperplanes associated to chiral multiplets.~\footnote{For special choices of parameters, there may be ``non-regular'' singularities where more than $r$ hyperplanes intersect, and the JK prescription in these cases is more complicated.  However, moving slightly away from such a point in parameter space we may typically resolve this into regular singularities, where we may apply the procedure above, and then analytically continue back to the point of interest. (This is true, in particular, in all the examples we will consider below.)}  The JK prescription, which depends on a choice of covector, $\eta \in \frak{h}^*$, determines which of these residues one should count.  Namely, if the corresponding chiral multiplets have charges $Q_\alpha^a$, $\alpha=1,...,r$, we count this residue (with an appropriate sign) if and only if:
\be \eta \in \text{Cone}_+(Q_\alpha) \ee
where the RHS is the positive cone of the $Q_\alpha$ in $i\frak{h}^*$, spanned by positive real multiples of the charge vectors $Q_\alpha$.  The final answer, obtained by summing over all such residues, is independent of the choice of $\eta$.

In our case, in order to properly deal with the ``singularities at infinity'' due to monopoles, it will be convenient to define an explicit contour, $\CC^\eta$, also labeled by a covector $\eta \in \frak{h}^*$, which will turn out to be closely related to this prescription.  Let us define:~\footnote{Here we choose the overall orientation on $\CC^\eta$ so that the top form $\wedge_{a=1}^r d \,\text{Re}(\partial_{u_a} \cW)$ is positive.  For example, in the rank one case, near a singularity due to a charge $Q$ chiral this form goes as $2 \pi Q d\theta$, where $u =u_*+ \epsilon e^{i \theta}$, correctly reproducing the orientation discussed above.  For higher rank, one can check that this agrees with the sign convention of the usual JK residue prescription.}

\be 
\label{C eta def} \CC^{\eta} = \{ u \in \frak{h}_\C \; | \; \text{Im}(\partial_{u_a} \cW) = - \eta^a , \;\; a=1,...,r \} 
\ee
To relate this to the usual JK residue prescription, recall that a chiral multiplet of charge $Q_\alpha$ may develop a pole when $Q_\alpha^a (p u_a - \m_a) = \beta_k$ for some set of parameters $\beta_k$.  In addition, there may be ``poles at infinity'' due to monopoles at large values of $\text{Im}(u_a)$. Then, in the vicinity of such a singularity, one can check that:
\be \label{dW dependence} \text{Im}(\partial_{u_a} \cW) \sim Q^a_\alpha \log \epsilon_\alpha \ee
where we have defined the small parameters:
\be \label{delta def} \epsilon_\alpha = \left\{ \begin{array}{cc} |Q_\alpha^a (p u_a - \m) - \beta_k| & \;\;\; \text{near a chiral singularity} \\ | e^{\pm 2 \pi i u_a} | & \;\;\; \text{near a monopole singularity} \end{array} \right. \ee
Now, let us rescale $\eta \rightarrow t \eta$ for large positive $t$, and consider some component of the contour $\CC^{t \eta}$.  Since $\text{Im}(\partial_{u_a} \cW)$ is parametrically large, this component must lie near some number, $k>0$, of singular hyperplanes.  Then, in the vicinity of these $k$ hyperplanes, we have:
\be \label{Im dW near r poles} \text{Im}(\partial_{u_a} \cW) \sim  \sum_{\alpha=1}^k Q^a_\alpha \log \epsilon_\alpha  \ee 
Since the $\log \epsilon_\alpha$ are negative near the intersection, the RHS is necessarily in the negative cone spanned by the $Q_\alpha$.  Then the only way it is possible to satisfy $\text{Im}(\partial_{u_a} \cW)=-t \eta^a$ is if $k=r$ (generically) and $\eta$ is in the positive cone of the $Q_\alpha^a$.   Explicitly, writing $\eta^a=\sum_\alpha c^\alpha Q_\alpha^a$, for $c^\alpha>0$, and taking $\epsilon_\alpha=e^{-t c^\alpha}$ we find a component of $C^\eta$ which wraps this intersection point.  Thus this residue is indeed counted by the integral over $C^\eta$, and so it counts the same residues as the usual JK prescription.  In particular it is independent of the choice of $\eta$.  This can also be seen directly by noting that changing $\eta$ continuously deforms the contour while not crossing any poles of the integrand, since these only occur when some $\text{Im}(\partial_{u_a} \cW)$ diverges.

We may also define the contour $\CC_0^\eta$ in \eqref{JK formula higher rank unfolded}.  This is contained inside the region $\frak{M}_0 = \{ u \in \frak{h}^\C \; | \; 0 \leq \text{Re}(u_a)\leq 1 \}$.  First, we define the portion of the contour in the interior of $\frak{M}_0$:
\be 
\CC_{0}^{\eta,\, {\rm bulk}} = \CC^\eta \cap \frak{M}_0~.
\ee
In addition to this bulk piece, the contour $\CC_0^\eta$ includes segments along the boundary of $\frak{M}_0$, at $\text{Re}(u_a)=0$ or $1$ for some $a$, similar to the rank one case.  We define:
\bea \label{C eta boundary def} 
&\CC^{\eta,\, {\rm boundary}}_{0} =   \cup_{a=1}^r \cup_{\omega \in \{0,1 \} } (-1)^{\omega} \Big\{ u \in \frak{M}_0 \; \Big| \;\text{Re}(u_a) = \omega\cr
&\quad\qquad\qquad\qquad\qquad \qquad  \text{and} \;  -\frac{\text{Im}(\partial_{u_b} \cW)}{\eta^b} = t~,\; b=1,...,r~, \; \; t\in [0,1] \Big\}~,
\eea
where $t$ runs over the interval $[0,1]$.
Note that this generically defines a dimension $r$ contour inside the boundary of $\frak{M}_0$.  The prefactor sets the relative orientation of these components, which is picked so that they match consistently with the interior components where they meet. Then we set:
\be
 \CC_0^\eta = \CC^{\eta,\,{ \rm bulk}}_{0} \cup \CC^{\eta,\, {\rm boundary}}_{0}~.
\ee
With this definition, one can check that the integral is invariant under continuously rescaling $\eta \rightarrow t \eta$, and more generally under any continuous change of $\eta$, as with the usual JK contour.  

As evidence that this is the correct contour for defining the $\Mgp$ partition function, in Appendix \ref{sec:HRJK} we present an argument relating the integral over this contour to the Bethe-vacua formula for the partition function in \eqref{ZMgp main}, generalizing the rank-one case.

\subsection*{The $\sigma$-contour}

We further conjecture that there exists an equivalent $\sigma$-contour, analogous to the one described in Section \ref{sec:sigmacontour}, of the form:
\be \label{Zmgp CB formula general rank} 
Z_{\Mgp}=\frac{(-1)^{\bf r}}{|W_G|}\sum_{\m \in \Z_p^{\bf r}}   \int_{\CC_\sigma} d^{\bf r}u \; \cF(u)^p\, \pif_a(u)^{\m_a} \,\pif_\alpha(u)^{\n_\alpha}  \,e^{(g-1)\Omega(u)}\, H(u)^{g}~,
\ee 
where $\CC_\sigma$ is a certain middle-dimensional non-compact contour connecting $\text{Im}(u_a) \rightarrow -\infty$ with $\text{Im}(u_a) \rightarrow \infty$.

To define $\CC_\sigma$, we first note that it is straightforward to  generalize the bound \eqref{bound rank one} to higher rank. One finds that the  chiral multiplets and monopole operators may only have poles provided that:
\be 
Q_\alpha^a (p \;\text{Re}(u_a) - \m_a) < \delta_\alpha~.
\ee
Let us assume for simplicity that we may pick flavor symmetry parameters and R-charges so that $\delta_\alpha- Q_\alpha p<0$ for all $\alpha$.  We then conjecture that we may take $\CC_\sigma$ to be the product of the imaginary $u_a$-axes for $a=1,...,r$.  This reproduces the prescription in \cite{Kapustin:2009kz, Hama:2010av, Jafferis:2010un}, which was obtained with a slightly different choice of localizing supercharge.  For more general parameters, one may derive the appropriate contour by analytic continuation. 

\subsection*{Relation of the $\sigma$-contour to the Bethe-vacua sum}

Here we show the equivalence of the conjectured $\sigma$-contour integral to the Bethe-vacua sum, \eqref{ZMgp main}.  As above, we assume that we may pick flavor symmetry parameters such that $\delta_\alpha- Q_\alpha p<0$ for all $\alpha$.  With such a simplification, the argument for the higher-rank case is a straightforward extension of that of the rank-one case.  We write:

\be\label{Zmgp CB formula general rank two}
Z_{\Mgp} =\frac{1}{|W_G|}\sum_{\m \in \Z_p^{\bf r}}   \int_{i \R^{\bf r}} d^{\bf r}u \; \CJ(u)\,  ) \prod_{a=1}^{\bf r}\pif_a(u)^{\m_a},
\ee
where $u=\{ u_a\}$ and $\m=\{\m_a\}$, where $a$ runs over a set of generators of the Cartan of the Lie algebra, $\Fh$.  As before, we first perform the sum over torsion fluxes:
\bea \label{holo sum higher rank}
Z_{\Mgp} &={1\ov |W_\GG|} \int_{i \R^{\bf r}} d^{\bf r} u \; \CJ(u) \prod_{a=1}^{\bf r} \frac{1-\pif_a(u)^p}{1-\pif_a(u)}\\
 &={1\ov |W_\GG|} \prod_a \bigg( \int_{i \R} - \int_{i \R-1} \bigg) d^{\bf r} u \; \frac{\CJ(u)}{\prod_a (1- \Pi_a(u))}~.
\eea
Here the contour is a product of the contour $i \R - (i \R-1)$ over each direction in the Cartan.  This encloses all the poles of the integrand in the region $-1 < \mbox{Re}(u_a) <0$, $a=1,...,{\bf r}$, and by our assumption above, these poles only arise at solutions to $\Pi_a(u)=1$, $a=1,...,{\bf r}$.  Thus we can write:

\bea \label{conjecture higher rank rewrite}
Z_{\Mgp} &={1\ov |W_\GG|} \sum_{\h u\,|\, \Pi_a(\h u)=1} (2 \pi i)^{\bf r}\, \mbox{Res}_{u = \h u} \frac{\CJ(u)^p }{\prod_a (1- \Pi_a(u))} \\
&= {1\ov |W_\GG|} \sum_{\h u\,|\, \pif_a(\h u)=0} \CF(\h u)^p\, \CH(\h u)^{g-1}\,\pif_\alpha(\h u)^{\n_\alpha}~.\eea
where we have used:
\be\label{dpdu higher rank}
\d_{u_a}  \Pi_b(u)= 2 \pi i \d_{u_a} \d_{u_b} \CW \equiv 2 \pi i H_{ab}
\ee
which contributes through a Jacobian factor of $(\det(\d_{u_a}  \Pi_b(u)))^{-1} = (2 \pi i)^{-{\bf r}} H(u)^{-1}$ to the residue.

Finally, in the case of non-abelian gauge groups, we should exclude those poles which are not acted freely by the Weyl group, while we count the remaining solutions up to the Weyl group action, canceling the symmetry factor.  In this way, we arrive at  the Bethe-vacua formula \eqref{ZMgp main}. This  completes the proof of the equivalence of the two prescriptions.

%%%%%%%%%%%%%%%%%%%%%%%%%%%%%%%%%%%
\section{The $S^3$ partition function and F-maximization}\label{sec: S3 and Fmax}

Consider the $S^3$ partition function for an $\CN=2$ supersymmetric gauge theory with integer $R$-charges $r_0$, and with generic flavor fugacities $\nu$ turned on.  The Bethe-vacua formula \eqref{ZMgp main} reads:
\be\label{ZS3R0}
Z_{S^3}(\nu) =  \sum_{\h u \in \CS_{\rm BE}} \CF(\h u,\nu) \CH(\h u,\nu)^{-1}
\ee
in this case.
As explained in Section  \ref{subsec: continuous R}, this result can be analytically continued to any allowed $R$-charge:
\be\label{R mixing real}
R= R_0 + Q^\alpha  t_\alpha~,\qquad\quad t_\alpha \in \R~,
\ee
simply by replacing $\nu_\alpha$ by $\nu_\alpha + t_\alpha$ in \eqref{ZS3R0}. (Here $t_\alpha \neq 0$ only for $\alpha$ corresponding to the free abelian subgroup of the flavor group.)

For any gauge theory that flows to an $\CN=2$ superconformal theory, we define a trial $F$-function as:
\be\label{Ftrial}
\t F_{S^3}(t) =- \log\left( \sum_{\h u \in \CS_{\rm BE}} \CF(\h u, t) \CH(\h u, t)^{-1}\right)~,
\ee
where we have set $\nu=t$ in \eqref{ZS3R0}.
The superconformal $R$-charge
\be
R= R_0 + Q^\alpha t_\alpha^\ast
\ee
{\it maximizes} the real part of $\t F_{S^3}(t)$ as a function of $t$ \cite{Jafferis:2010un,Closset:2012vg}. We then have:
\be\label{FS3 Fmaxed}
F_{S^3}= {\rm Re}\big[ \t F_{S^3}(t^\ast)\big]~.
\ee
In general, the right-hand-side of \eqref{FS3 Fmaxed} is only a local maximum, and we have to use our physical intuition to identify the correct superconformal $R$-charge. In practice, we choose $r_0=1$ as the integer $R$-charge for every elementary chiral multiplet, and we probe the $t_\alpha$ parameter space such that all the elementary $R$-charges  lie between $r=0$ and $1$. Given any $R$-charge that maximizes $F_{S^3}$, we should check that no gauge-invariant chiral operator violates the unitarity bound. A violation of the unitarity bound might signal the presence of free fields and accidental symmetries in the infrared---see {\it e.g.}  \cite{Morita:2011cs, Agarwal:2012wd, Safdi:2012re} for a discussion of such cases.

The formula \eqref{Ftrial} is an alternative to the matrix-model integral formula of  \cite{Kapustin:2009kz, Hama:2010av, Jafferis:2010un}. In the following, we demonstrate its utility by performing $F$-maximization in some simple theories. This provides highly non-trivial consistency checks of \eqref{ZS3R0}. This $F$-maximization method compares favorably to the usual method using the integral formula. The trial $F$ in \eqref{Ftrial} is given in terms of an explicit (albeit highly involved) function. In numerical studies, the usual integral method becomes more time-consuming as the rank of the gauge group increases, while in the present case the evaluation time depends principally on the number of Bethe vacua. We also avoid cumbersome issues of numerical integration, and the potential lack of convergence of the integral formula with a real $\sigma$ contour.

Note that $\t F_{S^3}$ is generally a complex function, whose imaginary part encodes parity-violating contact terms \cite{Closset:2012vp}. It is interesting to expand \eqref{Ftrial} around $t=t^\ast$. For instance, for a single $U(1)_F$ flavor symmetry mixing in \eqref{R mixing real}, we have:
\bea\label{FS3 expand}
&\t F_{S^3}(t^* + i m_F)&=&\; F_{S^3} + \pi i \left(\kappa_{RR} -{1\ov 12}\kappa_g\right)\cr
  &&& - 2 \pi \kappa_{RF}\,m_F+\half \left({\pi^2 \ov 2} \tau_{FF}-  2 \pi i \kappa_{FF}\right) m_F^2 + \cdots~,
\eea
with $m_F$ the real mass for $U(1)_F$, and the ellipsis denotes higher-order terms in $m_F$. The terms $\kappa_{RR}, \kappa_g, \kappa_{FR}, \kappa_{FF}$ are the contact terms discussed in Section \ref{subsec: regulated Zphi}, and $\tau_{FF}$ is the two-point function coefficient of the $U(1)_F$ conserved current \cite{Closset:2012vp}.

\subsection{The free chiral multiplet}\label{subsec: S3 free chiral}
Consider a chiral multiplet of $R$-charge $r\in \R$ coupled to a $U(1)_I$ vector multiplet with charge $Q\in \Z$ and real mass $\sigma$. The $S^3$ partition function reads:
\be\label{our ZS3 chiral}
Z_{S^3}^\Phi(\sigma, r) = \CF_\Phi(r-1+i Q\sigma)~,
\ee
with the function $ \CF_\Phi$ defined in \eqref{def FPhi of u}, and setting $u= i\sigma$ in order to compare with \cite{Jafferis:2011zi,Closset:2012vg}. We can easily check that:
\be
\t F_{S^3}^\Phi(\sigma, r)\equiv -\log\CF_\Phi(r-1+ i Q \sigma)
\ee
has a local maximum  $r=\half$, the superconformal $R$-charge of a free chiral multiplet, after we set $\sigma=0$. Expanding around $\sigma=0$ at $r=\half$, we obtain:
\be\label{FS3t expand}
\t F_{S^3}^\Phi(\sigma, r)= {\log{2}\ov 2}-{\pi i \ov 24}- {Q\ov 2} \pi \sigma+ {Q^2\ov 2} \left({\pi^2\ov 2} + \pi i\right)\sigma^2+\cdots~.
\ee
Comparing to \eqref{FS3 expand}, we read off:
\be
\kappa_{II}= - \half Q^2~, \qquad \kappa_{RI}= {1\ov 4} Q~, \qquad \kappa_{RR}- {1\ov 12}\kappa_g=-{1\ov 24}~.
\ee
This corresponds exactly to the $\kappa$ parameters \eqref{contact term ourreg} upon plugging in $r= \half$, providing another confirmation that our regularization of the chiral multiplet one-loop determinant indeed corresponds to those contact terms. We also see from \eqref{FS3t expand} that $F_{S^3}= \half \log{2}$ and $\tau_{FF}= Q^2$ for a free chiral, as should be the case in any regularization scheme.

We should also note that the chiral multiplet partition function \eqref{our ZS3 chiral} is related to the result $\t Z_{S^3}^\Phi$ of \cite{Hama:2010av, Jafferis:2010un} by:
\be
Z_{S^3}^\Phi(\sigma, r) = e^{{\pi i\ov 2} (r-1+i  Q \sigma)^2 -{\pi i \ov 12}}\, \t Z_{S^3}^\Phi(\sigma, r)~.
\ee
The discrepancy is simply because of our choice of a gauge-invariant but parity-violating regularization, leading to contact terms $\kappa\neq 0$ in \eqref{contact term ourreg}, while  the regularization scheme of  \cite{Hama:2010av, Jafferis:2010un} implicitly sets $\kappa=0$, which preserves parity but is inconsistent with gauge invariance.

\subsection{$U(N_c)_k$ theory with $N_f$ flavors} 
Consider a $U(N_c)$ theory at CS level $k>0$ with $N_f$ flavors (that is, $N_f$ pairs of fundamental and antifundamental chiral multiplets, with symmetric quantization). We will discuss this theory in more detail in Section \ref{sec:duality}. The abelian subgroup of the flavor group is $U(1)_A\times U(1)_T$, where $U(1)_T$ is the topological symmetry. We may assign a trial $R$-charge:
\be
r= 1+ t_A
\ee
to the chiral multiplets, where the only allowed mixing is with $U(1)_A$. (A  $\Z_2$ symmetry prevents any mixing with $U(1)_T$.)
%%%%%%%%%%%%%
\begin{table}
\begin{center}
\begin{tabular}{| c | cccccccc|}
 \hline 
 $k=0$   & $N_f = 1$ & $N_f = 2$ & $N_f = 3$ & $N_f = 4$ & $N_f = 5$ & $N_f = 6$ & $N_f = 7$&  $N_f=8$ \\
   \cline{1-9} $N_c = 1$  & \spc{$1/3$\\$.8724$ } & \spc{$.4085$ \\$1.934$ } &  \spc{$.4370$  \\$2.838$ }  & \spc{$.4519$\\ $3.679$} &  \spc{$.4611$\\$4.486$ } & \spc{$.4674$\\$5.272$ } & \spc{$\sim.47$\\$\sim6.0$ }& \spc{.4753\\ 6.805} \\
  $N_c = 2$ &  - & \spc{$1/4$\\$2.079$ }  & \spc{$.3417$\\$4.722$ } & \spc{$.3852$\\$6.875$ } &  \spc{$.4101$\\$8.817$ }& \spc{$.4263$\\ $10.64$ } & \spc{$.4375$\\ $12.38$} & \spc{$.4458$\\ $14.07$}\\
   $N_c = 3$ &  - & - & \spc{$(.2181)$\\$(4.162)$  } &\spc{$.3058$\\ 8.188$$} & \spc{$.3517$\\ $11.81$} & \spc{ $.3802$\\$15.03$} &\spc{$.3996$\\ $18.02$} &  \spc{$.4136$\\ $20.85$} \\ 
$N_c = 4$ &  - & - &- & \spc{$(.3333)$\\ $(6.334)$} & \spc{$.2809$\\  $12.19$} &\spc{$.3276$\\ $17.51$}& \spc{$.3574$\\$22.15$}&  \spc{$.3783$\\ $26.42$}  \\
  \hline
\end{tabular}
\vskip0.2cm
%%%%
\begin{tabular}{| c | cccccccc|}
 \hline 
 $k=1$ & $N_f = 1$ & $N_f = 2$ & $N_f = 3$ & $N_f = 4$ & $N_f = 5$ & $N_f = 6$& $N_f = 7$  & $N_f = 8$ \\
   \cline{1-9} $N_c = 1$  & \spc{$.3845$\\ $1.023$ } &\spc{$.04198$\\ $1.976$} & \spc{$.4407$\\$2.855$} & \spc{$.4535$\\$3.688$}&\spc{$.4619$\\$4.492$}&\spc{$.4678$\\$5.276$} &\spc{$\sim .47$\\$\sim6.0$} &\spc{$\sim .47$\\$\sim6.8$}   \\
 $N_c = 2$  &  \spc{$1/4$\\ $.3466$} &  \spc{$.3106$\\$2.635$}&  \spc{$.3591$\\$4.888$}&  \spc{$.3914$\\$6.944$}&  \spc{$.4129$\\$8.852$}&  \spc{$.4277$\\$10.66$}&  \spc{$.4383$\\$12.40$} & \spc{$\sim.446$\\$14.08$} \\  
$N_c = 3$  & - &   \spc{$1/4$\\$1.386$}&  \spc{$.2878$\\$4.939$}&  \spc{$.3278$\\$8.592$}&  \spc{$.3600$\\$11.98$}&  \spc{$.3839$\\$15.11$}&  \spc{$.4015$\\$18.07$}&  \spc{$.4147$\\$20.88$} \\  
$N_c = 4$  & - &  - & \spc{$1/4$\\ $3.119$}& \spc{$.2770$\\$7.928$} &\spc{$.3089$\\$13.03$}& \spc{$.3382$\\ $17.83$} & \spc{$.3621$\\ $22.31$} & \spc{$.3808$\\$26.51$}  \\
  \hline
\end{tabular}
\end{center}
\caption{Values of the superconformal $R$-charges $r$ and of $F_{S^3}$, respectively, for $U(N_c)$ SQCD with $N_f$ flavors, some low values of $N_c$ and $N_f$ and with CS level $k=0$ and $k=1$, determined by $F$-maximization.}
\label{tab:Fmax}
\end{table}
%%%%%
%%%%%%%%%%%%
The Bethe vacua correspond to all the choices of $N_c$ roots of the degree-$(N_f+k)$ polynomial:
\be
P(x)= (x y_A-1)^{N_f} - q (-x)^k (x-y_A)^{N_f}~,
\ee
where $y_A= e^{2\pi i \nu_A}$ and $q= e^{2\pi i \tau}$ are the fugacities for $U(1)_A$ and $U(1)_T$, respectively. 
The twisted superpotential of this theory reads:
\bea
&\CW(u, \nu_A) &=& {k+N_f\ov 2} \sum_{a=1}^{N_f} u_a(u_a+1)+ {N_c N_f\ov 2} \nu_A(\nu_A+1) \cr
&&&+ {N_f\ov (2 \pi i)^2}\sum_{a=1}^{N_c} \left(\dilog{\left(e^{2 \pi i (u_a+\nu_A)}\right)}+\dilog{\left(e^{2 \pi i (-u_a+\nu_A)}\right)}\right)~,
\eea
where we only turned on $\nu_A$, setting $\tau$ and all other mass parameters to zero. 
Let
\be
\h u_a =   \log(\h x_a)/(2 \pi i)
\ee
denote a Bethe vacua, where $\h x_a$ is a choice of $N_c$ roots of $P(x)$. The formula \eqref{Ftrial} gives us:
\be\label{Ftrial NcNf}
\t F_{S^3}(r) =- \log\left( \sum_{\h u \in \CS_{\rm BE}}{ \CF(\h u, \nu_A) \CH(\h u, \nu_A)^{-1}}\right)\Big|_{\nu_A= r-1}~,
\ee
with:
\bea
&\CF(u, \nu_A) &=&\; \exp{\Big(2 \pi i \left(\CW(u, \nu_A)-u_a\d_{u_a}\CW(u, \nu_A)-\nu_A\d_{\nu_A}\CW(u, \nu_A)\right)\Big)}~,\cr
&\CH(u, \nu_A) &=&\; (-1)^{\half N_c(N_c-1)} \prod_{\substack{a,b=1\\ a\neq b}}^{N_c} \left(1- e^{2 \pi i (u_a-u_b)}\right)\; \det_{a,b}\Big( \d_{u_a}\d_{u_b}\CW(u, \nu_A)\Big)~.
\eea
It is easy to maximize \eqref{Ftrial NcNf} using {\it Mathematica}, at least for $N_f+k$ small enough. We present some examples in the Table \ref{tab:Fmax}. They are in perfect agreement with results previously reported in the literature \cite{Willett:2011gp, Benini:2011mf, Safdi:2012re}. There are a few cases, denoted by  $\sim$, where our numerical evaluation of \eqref{Ftrial NcNf} was inaccurate near the $F$-maximizing value of $r$. In all other cases, we can easily reach a high precision for $r$ and $F_{S^3}$, although it becomes time-consuming as the number of Bethe vacua,
\be
|\CS_{\rm BE}| =\mat{N_f+k\\ N_c}~,
\ee
 increases. When $k=0$, there are cases where naive $F$-maximization leads to unphysical results, given in parenthesis in Table \ref{tab:Fmax}; the physical values were computed in \cite{Safdi:2012re}.

Note also that, in the special case $N_c= k+ N_f$ with $k>0$, we obtain $r= 1/4$ and $F_{S^3}= {N_f^2}\log{\sqrt2}$.  Indeed, the infrared theory consists of $N_f^2$ free mesons \cite{ Willett:2011gp}. This is the limiting case of the Giveon-Kutasov duality \cite{Giveon:2008zn}. For $N_c=N_f>1$ with $k=0$, we should have $r= 1/4$ and $F_{S^3}= {(N_f^2+2)}\log{\sqrt2}$, which can be obtained using the Aharony dual theory \cite{Aharony:1997gp, Safdi:2012re}.

%%%%%%%
\section{Matching $Z_{\Mgp}$ across supersymmetric dualities}
\label{sec:duality}

The Bethe-vacua formula \eqref{W corr 3d}  provides a simple way to study supersymmetric dualities. If two distinct three-dimensional $\CN=2$ gauge theories $\CT$ and $\CT_D$ are infrared dual, their supersymmetric partition functions and correlation functions must agree on any $\Mgp$. This implies:
\be\label{dual rel corr gen}
\langle W \rangle_{\Mgp}^{\CT}(\nu,\n) =  \langle W_D \rangle_{\Mgp}^{\CT_D}(\nu,\n)~,
\ee
where:
\bea\label{dual rel corr gen2}
& \langle W \rangle_{\Mgp}^{\CT}
&=& \sum_{\h u \in \CS_{\rm BE}} W(\h x)\, \CF(\h u,\nu)^p\, \CH(\h u,\nu)^{g-1} \,\pif_\alpha(\h u,\nu)^{\n_\alpha}~,\cr
& \langle W_D \rangle_{\Mgp}^{\CT_D} &=&  \sum_{\h u^D \in \CS_{\rm BE}^D} W_D(\h x^D)\, \CF_D(\h u^D,\nu)^p\, \CH_D(\h u^D,\nu)^{g-1} \, \pif_{\alpha D}(\h u^D,\nu)^{\n_\alpha}~,
\eea
in the dual theories.
Here $\nu_\alpha$, $\n_\alpha$ are the flavor fugacities and fluxes, respectively (the product over the index $\alpha$ is implied), $W$ is a loop operator in theory $\CT$, and $W_D$ is the loop operator it maps to under the duality.  For \eqref{dual rel corr gen} to hold for arbitrary $g,p,\n_\alpha$, and operator $W(x)$, it is necessary and sufficient that there exists a one-to-one ``duality map'' between the supersymmetric vacua: 
\be\label{duality map}
 \CD:\quad \CS_{BE} \rightarrow \CS_{BE}^D  :\quad \h u \mapsto   \CD(\h u)= \h u^D~,
 \ee
such that
\be\label{duality rel F H}
 \CF(\h u,\nu)=  \CF_D(\h u^D,\nu)~, \qquad 
 \CH(\h u,\nu)  = \CH_D(\h u^D,\nu)~,\qquad
\pif_\alpha(\h u,\nu) =\pif_{\alpha D}(\h u^D,\nu)~.
\ee
Note that, due to the difference equation \eqref{prop F shift}, the matching of the fibering operators, $\CF=\CF_D$, implies the matching of the flavor flux operators, $\pif_\alpha=\pif_{\alpha D}$. Finally, the map between Wilson loops follows from the duality map \eqref{duality map}. By definition, $W$ and $W_D$ are dual if and only if: 
\be
W(\h x) = W(\h x^D)~,
\ee
on every pair of dual vacua $\h x$ and $\h x^D$. Duality relations between Wilson loops in many infrared dualities were studied explicitly in \cite{Kapustin:2013hpk, Closset:2016arn}.

As argued in Section \ref{sec:onshell}, the fibering, flux, and handle-gluing operators evaluated at a Bethe vacuum can be obtained from the ``on-shell'' twisted superpotential and effective dilaton potential \eqref{Wonshell}. To prove the equivalence of the partition functions, it thus suffices to demonstrate that:
\be\label{equal onshell W}
\CW^l(\nu)= \CW_D^l(\nu)~, \qquad\quad \Omega^l(\nu)= \Omega_D^l(\nu)~, \qquad l=1, \cdots, |\CS_{\rm BE}|~,
\ee
modulo the integer-quantized branch cut ambiguities, \eqref{ambiguity W3d}.  In this section we will prove the equality \eqref{dual rel corr gen} for a number of non-trivial dualities by   checking \eqref{equal onshell W}, which then implies \eqref{duality rel F H}.  In most of the examples below,  the matching of the handle-gluing operators (and flux operators) was already checked in \cite{Closset:2016arn,Benini:2016hjo}, by considering theories on $\Sigma_g \times S^1$, so we will mostly focus on matching the fibering operators.~\footnote{In \protect\cite{Closset:2016arn,Benini:2016hjo}, the duality relations $\CH=\CH_D$ were checked up to a sign. We insist that there is no sign ambiguity once we treat the parity anomaly consistently (and include the correct signs in the classical CS terms, as reviewed in Appendix \protect\ref{app: spin CS}). We will see some examples of this below.}

\paragraph{Note on conventions.}  In the remainder of this section, we will use a rescaled twisted superpotential:
\be
\tCW = (2 \pi i)^2 \CW~.
\ee
Let us recall that a (regularized) chiral multiplet $\Phi $ and the $U(1)$ and gravitational Chern-Simons terms contribute:
\be
 \tCW_\Phi(x) = \dilog(x)~,\qquad\quad \tCW_{\rm CS}(x) = \frac{k}{2} \log x(\log x + 2 \pi i)- {\pi^2\ov 6} k_g
  \ee
  to $\tCW$, respectively. 
In the following, it will be important to keep track of our branch-cut conventions. We define the logarithm $\log{z}$ such that $-\pi <\mbox{Im}(\log z) \leq \pi$ ({\it i.e.} with a branch cut along the negative real axis), and we define the dilogarithm $\dilog(z)$  such that:
\be  \label{bcs} 
\partial_z \dilog(z) = - \frac{\log(1-z)}{z}~,
\ee
with a single branch cut along the real axis with ${\rm Re}(z)>1$.

%%%%%%%%%%%
\subsection{Two-term dilogarithm identities and abelian mirror symmetry}\label{subsec: dual1}
There are two elementary identities involving two dilogarithms:
\bea\label{2terms dilog id}
&\dilog(z^{-1}) + \dilog(z) = -{\pi^2\ov 6}- \half \log^2(-z)~,\cr
&\dilog(1-z) + \dilog(z)=  {\pi^2\ov 6} - \log(z) \log(1-z)~.
\eea
These identities correspond, at the level of the twisted superpotential, to the following properties of $3d$ $\CN=2$ theories.

\paragraph{Massive chiral multiplets.} We already pointed out in Section \ref{subsec: W3d} that the first identity in \eqref{2terms dilog id} corresponds to the fact that two chiral multiplets with a superpotential $W= \Phi_1\Phi_2$ is ``dual'' to an empty theory. More precisely, consider two chirals with $U(1)$ charges $\pm 1$ in the $U(1)_{-\half}$ quantization, of $R$-charges $r_1=r$, $r_2=2-r$, respectively. The low-energy theory corresponds to an empty theory with $U(1)$ CS level $k=-1$  and gravitational CS level $k_g=-2$. This corresponds to:
\be
\CW_{\Phi_1\Phi_2}= \dilog(x) + \dilog(x^{-1})  = - \frac12 \log x(\log x + 2 \pi i) + \frac{\pi^2}{3}~,
\ee
where we wrote the first line of \eqref{2terms dilog id} on the principal branch of the log.
This implies the identity
\be\label{iden FPhi}
\CF_\Phi(u) \CF_\Phi(-u) = e^{\pi i  u^2 - \frac{\pi i}{6} } 
\ee
for the fibering operators, which is independent of the branch cuts as a function of $u$---both sides of  \eqref{iden FPhi} are meromorphic functions on the $u$ plane.
The low energy-theory also has the gauge-$R$ and $RR$ CS levels $k_R= -r+1$ and $k_{RR}=-(r-1)^2$, respectively. The effective dilaton reads:
\be
\Omega_{\Phi_1\Phi_2} = -{r-1\ov 2 \pi i} \log(1-x)   -{1-r\ov 2 \pi i} \log(1-x^{-1})= -(r-1)u +\half (r-1)~,
\ee
which reproduces those CS levels, since:
\be
\CH_{\Phi_1\Phi_2} = e^{2 \pi i \Omega_{\Phi_1\Phi_2} }= (-1)^{k_{RR}}x^{k_{R}} ~.
\ee

\paragraph{The elementary mirror symmetry duality.} Let us consider a $U(1)_{\frac12}$ gauge theory with a single chiral multiplet of charge $1$ and $R$-charge $r$. The effective twisted superpotential and effective dilaton read:
\bea\label{W U1half}
&\CW(x,q) &=&\; \dilog(x)+\half \log x(\log x+2 \pi i)+\log q \log x- {\pi^2\ov 6}~,\cr
&\Omega(x, q) &=&\; -{r-1\ov 2 \pi i } \log(1-x)~,
\eea
where $q=e^{2 \pi i \tau}$ is the fugacity for the $U(1)_T$ topological symmetry. This corresponds to the non-zero bare contact terms $\kappa=\half$ (for the gauge symmetry), $\kappa_R=-\half (r-1)$ and $\kappa_{RR}=-\half (r-1)^2$, in addition to the FI term. The Bethe equation for this theory has a single solution:
\be
\frac{q x}{x-1} =1 \;\;\; \Rightarrow\;\;\; \h x = {1\ov 1-q}~.
\ee
Substituting back into \eqref{W U1half}, we find the on-shell twisted superpotential and dilaton potential:
\bea\label{onshell W expl U1half}
&\CW^{(1)}(q) = \dilog\left(1\ov 1-q\right)+\half  \log (1-q)\left(\log(1-q) -2 \pi i\right)-\log q \log (1-q)~, \cr
& \Omega^{(1)}(q)= {r\ov 2 \pi i }\log(1-q)- {r\ov 2 \pi i} \log{q} +{r\ov 2}~, 
\eea
where  $\Omega^{(1)}$ is defined as in \eqref{Wonshell}.
To obtain the fibering operator from \eqref{onshell W expl U1half}, we must choose the ``physical'' branch of the twisted superpotential as discussed in Section \ref{sec:onshell}.  This is the condition:
\be
 \frac{\partial \CW}{\partial \log x}(\h x) = 0~, 
\ee
which indeed holds for all $q$.

This theory is dual to a free chiral multiplet of charge $1$ under the $U(1)_T$ global symmetry, and $R$-charge $-r+1$.  This chiral multiplet can be identified with the gauge-invariant monopole operator $T_+$ in the original theory, whose induced charges can be computed from \eqref{induced charges T}.  This chiral multiplet is quantized with $\kappa_{TT}= - \half$, $\kappa_{TR}= -{r\ov 2}$ and $\kappa_{RR}= -\half (r-1)^2+r$ (mod $2$), so that the dual twisted superpotential is simply given by:
\be
\CW_D(q)= \dilog(q)~,
\ee
while the dual  effective dilaton is exactly the same as $\Omega^{(1)}(q)$ in \eqref{onshell W expl U1half}. The non-trivial identity of the on-shell twisted superpotentials:
\be
\CW_D(q)= \CW^{(1)}(q)~,
\ee
directly follows from \eqref{2terms dilog id}.  This duality relation implies:
\be\label{id CF basic mirror}
 \CF_\Phi(\h u) e^{- 2\pi i \tau \h u- \pi i \h u^2 +{\pi i \ov 12}} = \CF_\Phi(\tau)~,\qquad {\rm with}\quad 
 \h u= -{1\ov 2 \pi i}\log(1-q)~,
\ee
which is the identity $\CF^{(1)}(\tau)= \CF^{(1)}_D(\tau)$ between the on-shell fibering operators, seen as meromorphic functions of $\tau$. As a consistency check, one can easily check \eqref{id CF basic mirror} numerically.

\subsubsection{Gauging flavor symmetries and general abelian mirror symmetry}

From this basic duality, it is possible to construct a mirror dual description of a more general abelian gauge theory.  The idea is to start from several decoupled copies of this duality and gauge appropriate flavor symmetries on each side to obtain the desired theories \cite{Kapustin:1999ha}.  Here we illustrate this procedure in a simple case, constructing the mirror dual of $U(1)_{k=-\frac{N}{2}}$ with $N$ charge-one chiral multiplets.  We focus on the on-shell twisted superpotential for simplicity; the matching of effective dilaton can be shown similarly, and then that of the $\Mgp$ partition function follows from the general discussion in Section \ref{sec:onshell}.

To construct the original theory, we start with $N$ copies of the free chiral multiplet.  This theory has a $U(N)$ symmetry, which we decompose as $U(1) \times SU(N)$, with corresponding parameters $z$ for the overall $U(1)$ and $y_i$ with $\prod_i y_i =1$, for the $SU(N)$.%
~\footnote{More precisely, $U(N)\equiv (U(1) \times SU(N))/\Z_N$, and as a result, the flavor symmetry is $SU(N)/\Z_N$. 
}  Then the twisted superpotential of this theory, including a level $-\frac{1}{2}$ contact term for each chiral, is:
\be \label{U(1) N starting point} 
\tCW(z, y, q) = \sum_{i=1}^N \dilog(z y_i) + \log q \log z~.
\ee
For later convenience, we introduced a background vector multiplet with a BF coupling to the $U(1)$ flavor symmetry, with a corresponding  parameter $q$ .  We then gauge the $U(1)$ symmetry corresponding to $z$ by solving the Bethe equation:
\be 
\exp \left( \frac{\partial \tCW}{\partial \log z} \right) = z\prod_{i=1}^N (1-z y_i)^{-1}  = 1~.
\ee
This has $N$ solutions, which may be inserted into \eqref{U(1) N starting point} to find the on-shell twisted superpotential for the $N$ vacua.

To construct the mirror dual theory, we note that the $N$  free chiral multiplets we started with are dual to $N$ copies of the $U(1)_{\half}$ theory with one charged chiral multiplet. This has twisted superpotential:
\bea \label{U(1) N dual starting point}
& \tCW_D(x, z, y , q)& =&\; \sum_{i=1}^N \left( \dilog(x_i ) +\half \log x_i (\log x_i +2 \pi i) +  \log (z y_i) \log x_i \right) \cr
 &&&\;+  \log q \log z - \frac{N \pi^2}{6}~.
 \eea
Here $x_i$ are the parameters for the $U(1)^N$ gauge symmetry.  If we solve the Bethe equations for the $x_i$ and substitute the solutions, we obtain \eqref{U(1) N starting point}, and subsequently solving the Bethe equation for $z$ will give the same $N$ solutions as above.  It is more illuminating, however, to first solve the Bethe equation for $z$.  This gives:
\be \label{U(1) N dual BE}
 \frac{\partial \tCW_D}{\partial \log z}= \log q + \sum_{i=1}^n \log y_i  = 0~.
  \ee
To solve this, introduce the new variables $\t x_{i  \sim i+N}$ by:
\be 
\log x_i =-\frac{1}{N} \log q + \log \t x_i- \log  \t x_{i+1} 
 \ee
Then the twisted superpotential becomes:
\bea \label{U(1) N dual}
& \tCW_D(\t x, q, y)& =&\; \sum_{i=1}^N \bigg( \dilog(q^{-{1\ov N}} \t x_i \t x_{i+1}^{-1}) +\log y_i  (\log \t x_i-\log \t x_{i+1})  \cr
&&&\qquad   + \half \log(q^{-{1\ov N}} \t x_i \t x_{i+1}^{-1})(\log(q^{-{1\ov N}} \t x_i \t x_{i+1}^{-1})+2 \pi i)  \bigg) - \frac{N \pi^2}{6}~.
 \eea
This gives the twisted superpotential of a circular quiver with gauge group $U(1)^{N}/U(1)_{\rm diag}$, which is the mirror description of the original theory.  The $U(1)_T$ topological symmetry of the original theory maps to the $U(1)_{\rm diag}$ flavor symmetry of the dual. The $U(1)^{N-1}$ maximal torus of the  $SU(N)$ flavor symmetry of the original theory corresponds to the topological symmetries $U(1)_{T_i}$ of the quiver, while the full $SU(N)$ is expected to be realized in the infrared.  From \eqref{U(1) N dual}, one can solve the Bethe equations for the $N-1$ gauge variables $\t x_i$ to construct the on-shell twisted superpotential.  This operation must give the same result as if we perform the gauging in the opposite order. This demonstrates the matching of the on-shell twisted superpotential across this particular mirror symmetry.

%%%%%%%%%%%%%%%%%%%%%% 
\subsection{$N_f=1$ SQED/$XYZ$ model duality}\label{subsec: dual2}
Consider a $U(1)$ theory with two chiral multiplets $Q$, $\t Q$ of gauge charges $\pm$, respectively, and $R$-charge $r$. The theory has a $U(1)_A\times U(1)_T$ flavor symmetry, with charges summarized in Table \ref{tab:SQED1 charges}. We turn on the corresponding fugacities $y_A=e^{2\pi i \nu_A}$ and $q= e^{2\pi i \tau}$. The effective twisted superpotential  is given by:
\bea
&\tCW_{\rm SQED}(x, y_A, q)&=&\; \dilog(xy_A)+\dilog(x^{-1}y_A) + \log{q}\log{x} \cr
&&& \;+\half \log{x}(\log{x}+2\pi i) + \half\log{y_A}(\log{y_A}+2 \pi i)- {\pi^2\ov 3}~.
\eea
The CS terms in the second line appear because we choose the symmetric quantization for the flavor $Q, \t Q$, such that the bare contact terms vanish. Similarly, the effective dilaton reads: 
\be
\Omega_{\rm SQED}(x,y_A, q) = -{r-1\ov 2 \pi i}\log(1-x y_A)-{r-1\ov 2 \pi i}\log(1-x^{-1} y_A)+ {r-1\ov 2 \pi i}\log{y_A}~.
\ee
The Bethe equation has a single solution:
\be
\h x = \frac{qy_A-1}{q-y_A}~.
\ee

This theory is dual to the $XYZ$ model, which consists of three chiral multiplets $(X,Y,Z)= (M, T_+, T_-)$ with charges given in Table \ref{tab:SQED1 charges}, and a cubic superpotential $W= M T_+ T_-$.
%%%%%%%%%%%%%%%%%
\begin{table}[t]
\centering
\be\nn
\begin{array}{c|c|ccc}
    &  U(1)& U(1)_A & U(1)_T  & U(1)_R  \\
\hline
Q  & 1&1 &0 &r \\
\t Q       & -1& 1 &0 &r \\
\hline
M & 0 & 2 & 0 & 2r\\
T^+ &0&- 1 & 1& -r+1\\
T^- &0& -1 & -1& -r+1
\end{array}
\ee
\caption{Gauge and flavor charges for $N_f=1$ SQED, and for  its dual.}
\label{tab:SQED1 charges}
\end{table}
The twisted superpotential of that theory is:
\bea
&\tCW_{XYZ}(y_A, q) &=&\; \dilog(y_A^2)+ \dilog(qy^{-1})+ \dilog(q^{-1}y_A^{-1})\cr
&&&\; + \half \log{q}(\log{q}+2 \pi i) + {3\ov 2}\log{y_A}(\log{y_A}+2 \pi i)- {\pi^2\ov 2}~.
\eea
The CS levels are again chosen so that the bare contact terms vanish. Similarly, the effective dilaton is such that:
\be\label{HXYZ}
\CH_{XYZ}(y_A, q)= e^{2 \pi i \Omega_{XYZ}(y_A, q)}={ (1-q y^{-1}_A)^r (1-q^{-1} y_A^{-1})^r \ov(1-y_A^2)^{2r-1}}\,  y_A^{3r -1}~.
\ee
It is straightforward to check that the $N_f=1$ SQED handle-gluing operator exactly reproduces \eqref{HXYZ} on the Bethe vacuum: 
\be
\CH_{\rm SQED}(\h x, y_A, q)= \CH_{XYZ}(y_A, q)~.
\ee
On the other hand, we can check that the twisted superpotentials also match on-shell:
\be\label{W match SQED}
\tCW_{\rm SQED}(\h x, y_A, q)= \tCW_{XYZ}(y_A, q)
\ee
for a particular choice of branches. This relation follows from a well-known five-term relation for the dilogarithm, which can be written as:
\bea\label{5term identity}
&\dilog(w)+ \dilog(z) - \dilog(wz)+  \dilog\left({(1-z)w\ov w-1}\right)+ \dilog\left({(1-w)z\ov z-1}\right) \cr
&\qquad=- \half \log^2\left({1-w\ov 1-z}\right)~,
\eea
for a certain choice of branch.
By plugging $w=qy_A^{-1}$, $z=q^{-1}y_A^{-1}$ into \eqref{5term identity} and by using \eqref{2terms dilog id} several times, one can derive \eqref{W match SQED}. The fibering operators of the dual theories read:
\bea
&\CF_{\rm SQED}(u, \nu_A, \tau)&=&\; \CF_\Phi(u+\nu_A) \CF_\Phi(-u+\nu_A)e^{-\pi i (u^2+ \nu_A^2)+{\pi i\ov 6}}~,\cr
&\CF_{XYZ}(\nu_A, \tau) &=&\; \CF_\Phi(2 \nu_A)  \CF_\Phi(-\nu_A+ \tau)  \CF_\Phi(-\nu_A-\tau)e^{-\pi i (\tau^2 + 2 \nu_A^2) +{\pi i \ov 4}}~. 
\eea
The relation \eqref{W match SQED} implies a functional relation:
\be
\CF_{\rm SQED}(\h u(\nu_A, \tau), \nu_A, \tau)= \CF_{XYZ}(\nu_A, \tau)~.
\ee
One can also easily check this relation numerically.

%%%%%%%%%%%%%%%%%%%%%%%%%%
\subsection{Seiberg-like dualities}
Consider three-dimensional SQCD$[k, N_c, N_f, N_a]$, which consists of a $U(N_c)_k$ gauge theory at CS level $k$,~\footnote{In this language, the CS level $k$ may be half-integer, while $k+\half(N_f+N_a)$ must be integer.} coupled to $N_f$ fundamental and $N_a$ antifundamental chiral multiplets, denoted by $Q_i$ and $\t Q^j$, respectively. Without loss of generality, we consider $k\geq 0$ and $N_f \geq N_c$. This theory has a flavor group:
\be\label{flavor G SQCD}
SU(N_f)\times SU(N_a)\times U(1)_A \times U(1)_T~,
\ee
with charges summarized in Table \ref{tab:SQCD charges}. 
%%%%%%%%%%%%%%%%%
\begin{table}[t]
\centering
\be\nn
\begin{array}{c|c|ccccc}
    &  U(N_c)& SU(N_f) & SU(N_a)  & U(1)_A &  U(1)_T & U(1)_R  \\
\hline
Q_i        & \bm{N_c}& \bm{\overline{N_f}} & \bm{1}& 1   & 0   &r \\
\tilde{Q}^j   & \bm{\overline{N_c}}  &  \bm{1}& \bm{N_a}  & 1   & 0   &r \\
\end{array}
\ee
\caption{Charges of the chiral multiplets of 3d $\CN=2$ SQCD. 
}
\label{tab:SQCD charges}
\end{table}

Three-dimensional SQCD has an infrared-dual description whose precise form depends on the parameters $k$ and $N_f-N_a$ \cite{Aharony:1997gp, Giveon:2008zn, Benini:2011mf}. The dual theory has a gauge group $U(n_f-N_c)$, with
\be\label{def nf for sqcd}
n_f  \equiv \begin{cases}
k+ {N_f+ N_a\ov 2} &\qquad {\rm if} \qquad k \geq \half(N_f-N_a)~,\\
N_f &\qquad {\rm if} \qquad k \leq \half(N_f-N_a)~.
\end{cases}
\ee
The $U(n_f-N_c)$ vector multiplet is coupled to $N_a$ fundamental and $N_f$ antifundamental chiral multiplets, denoted by $q^i$ and $\t q_j$, respectively. It also contains  $N_f N_a$ gauge singlets ${M^j}_i$, and $d_C \leq 2$ additional singlets in special cases.~\footnote{ We have $d_C=0,1, 2$ if $k> \half(N_f-N_a)$, $k=\half(N_f-N_a)>0$ or $k=\half(N_f-N_a)=0$, respectively.}
 The gauge-singlets are coupled to the gauge sector through the usual Seiberg-dual superpotential.

All these dualities can be derived by massive deformations of the so-called Aharony duality \cite{Aharony:1997gp}, which is the case $k=0$, $N_f=N_a$. In the following, we discuss the equality of supersymmetric partition functions for Aharony-dual theories.  (We refer to \cite{Closset:2016arn}  for a more detailed review of Seiberg-like dualities.)

\subsubsection{Aharony duality}
\paragraph{Electric theory.} Consider a $U(N_c)$ vector multiplet coupled to $N_f$ pairs of fundamental and antifundamental chiral multiplets  $Q_i,\t Q^j$ of $R$-charge $r$. 
Let us introduce the parameters: 
\be
y_i= e^{2 \pi i \nu_i}~,\quad \t y_j= e^{2 \pi i \t\nu_j}~,\quad y_A= e^{2 \pi i \nu_A}~,\quad q=e^{2\pi i \tau}~, 
\ee
for the flavor group \eqref{flavor G SQCD}, such that:
\be
\sum_{i=1}^{N_f}\nu_i =- \sum_{j=1}^{N_f}\t \nu_j = -N_f \nu_A~. 
\ee
The effective twisted superpotential of this theory reads:
\bea\label{sqcd W NcNf}
&\tCW^{[N_c, N_f]}_{\rm SQCD}&=&\; \sum_{a=1}^{N_c}\Bigg( \log{q}\log{x_a}+ {N_f\ov 2}\log{x_a}(\log{x_a}+2\pi i) + \sum_{i=1}^{N_f}\dilog{(x_a y_i^{-1})}    \cr
&&&    +\sum_{j=1}^{N_f}\left(\dilog{(x_a^{-1}\t y_j)}  
+ \half\log{\t y_j}(\log{\t y_j}+2\pi i) -{\pi^2\ov 3} \right)\Bigg)~.
\eea
The integer CS terms in \eqref{sqcd W NcNf} are chosen such that most of the bare contact terms vanish. More precisely, we have:
\be\label{contact terms for aha dual}
\kappa=0~, \qquad \kappa_{AA}=\kappa_{AT}=\kappa_{TT}=\kappa_g=0~,\qquad \kappa_{SU(N_f)}= -\kappa_{SU(N_f)'}=-\half{N_c}~,
\ee
where $\kappa$ is the gauge contact term, and $ \kappa_{SU(N_f)}$, $ \kappa_{SU(N_f)'}$ are the $SU(N_f)\times SU(N_f)$ contact terms.
 Similarly, the effective dilaton is given by:
\bea
&\Omega^{[N_c, N_f]}_{\rm SQCD}&=&\;  \sum_{a=1}^{N_c}\Bigg( -{r-1\ov 2 \pi i} \sum_{i=1}^{N_f}\log(1-x_a y_i^{-1})-{r-1\ov 2 \pi i} \sum_{j=1}^{N_f}\log(1-x_a^{-1} \t y_j)\cr
&&&\;\qquad\quad +{r-1\ov 2 \pi i }\log y_A\Bigg) -{1\ov 2 \pi i}\sum_{\substack{a,b=1\\ a\neq b}}^{N_c}  \log(1-x_a x_b^{-1})~.
\eea
The Bethe equations of this theory,
\be
P(x_a)=0~,\quad \forall a~, \qquad \qquad x_a\neq x_b \quad {\rm if} \quad a\neq b~,
\ee
are given in terms of a single polynomial of degree $N_f$:
\be\label{Px Aha dual}
P(x) \equiv \prod_{i=1}^{N_f}(x-y_i) - qy_A^{-N_f} \prod_{j=1}^{N_f} (x- \t y_j)~.
\ee
The Weyl group is the symmetric group $S_{N_c}$ that permutes the $x_a$'s.
Therefore, a Bethe vacuum $\h x^{(l)}\equiv \{\h x_a^{(l)}\}_{a=1}^{N_c}$ consist of a choice of $N_c$ distinct roots of $P(x)$, and  there are
\be
|\CS_{\rm BE}|= \mat{N_f\cr N_c} 
\ee
distinct Bethe vacua. 

\paragraph{Magnetic theory.} The Aharony dual theory is a $U(N_f-N_c)$ theory with $N_f$ pairs of fundamental and antifundamentals $q_j$, $\t q^i$ of $R$-charge $1-r$, together with $N_f^2$ gauge singlets ${M^j}_i$ of $R$-charge $2r$, and two additional singlets $T_\pm$ of $R$-charge:
\be\label{rpm Aha dual}
r_{\pm}\equiv -N_f(r-1) - N_c+ 1~,
\ee
and a superpotential
\be
W= M^j_i \t q^i q_j + T_+ t_- + T_- t_+~.
\ee
The gauge singlets are identified with the ``mesons'' ${M^j}_i= \t Q^j Q_i$ and with the monopole operators $T_\pm$ in the $U(N_c)$ theory. All the charges are given in Table  \ref{tab: Aharony duality charges}.
%%%%%%%%%%%%%%%
\begin{table}[t]
\centering
\be\nn
\begin{array}{c|c|ccccc}
    & U(N_f-N_c)& SU(N_f) & SU(N_f)  & U(1)_A &  U(1)_T & U(1)_R  \\
\hline
q_j   &\bm{N_f-N}&    \bm{1} & \bm{\overline{N_f}} & -1   & 0   &1-r \\
\t q^i      & \bm{\overline{N_f-N_c}}& \bm{N_f}    & \bm{1}    & -1   & 0   &1-r \\
{M ^j}_i      & \bm{1} &\bm{\overline{N_f}}& \bm{N_f} &  2   & 0   &2 r \\
T_+       & \bm{1} & \bm{1} & \bm{1} & -N_f   & 1   & -N_f(r-1) -N_c +1\\
T_-     & \bm{1} & \bm{1} & \bm{1} & -N_f   & -1   & -N_f(r-1) -N_c +1
\end{array}
\ee
\caption{Chiral multiplet charges in the Aharony dual theory. }
\label{tab: Aharony duality charges}
\end{table}
%%%%%%%%%%%%%%%%%%

The twisted superpotential reads:
\be\label{W SQCD Aha dual}
\tCW_D = \tCW_{D, {\rm gauge}}+  \tCW_{D, {\rm singlet}}~,
\ee
with:
\bea\label{W SQCD Aha dual2}
& \tCW_{D,{\rm gauge}} &=&\; \sum_{\ba=1}^{N_f-N_c}\Bigg(- \log{q}\log{x_\ba}+ {N_f\ov 2}\log{x_\ba}(\log{x_\ba}+2\pi i)+ \sum_{j=1}^{N_f}\dilog{(x_\ba \t y_j^{-1})}  \cr
&&&\;\;\qquad +\sum_{i=1}^{N_f}\Big(\dilog{(x_\ba^{-1} y_i)}  +\half \log{y_i}(\log{y_i}+2 \pi i)-{\pi^2\ov 3} \Big) \Bigg)~, 
\eea
\bea
& \tCW_{D, {\rm singlet}}&=&\; \sum_{i=1}^{N_f}\sum_{j=1}^{N_f} \left(\dilog(y_i^{-1}\t y_j)+ \half \log{\t y_j}(\log{\t y_j }+ 2 \pi i)-{\pi^2\ov 6}  \right) \cr
&&&\; \;+ \dilog(q y_A^{-N_f})  + \dilog(q^{-1}y_A^{-N_f})+\half \log{q}(\log{q}+2 \pi i) \cr
&&&\; \; + N_f^2 \log{y_A}(\log{y_A}+2 \pi i)- {\pi^2\ov 3}~.
\eea
In \eqref{W SQCD Aha dual}, we identified the dual FI parameter $\tau_D$ with minus the $U(N_c)$ FI parameter, $\tau_D= -\tau$, and the bare contact terms are the same as in \eqref{contact terms for aha dual}. Similarly, the dual effective dilaton reads:
\be
\Omega_D= \Omega_{D,{\rm gauge}}+  \Omega_{D,{\rm singlet}}
\ee
with:
\bea
&\Omega_{D,{\rm gauge}} &=&\;  \sum_{\ba=1}^{N_f-N_c}\Bigg( {r\ov 2 \pi i} \sum_{j=1}^{N_f}\log(1-x_\ba \t y_j^{-1})+{r\ov 2 \pi i} \sum_{i=1}^{N_f}\log(1-x_\ba^{-1}  y_i)\cr
&&&\;\qquad\quad +{r\ov 2 \pi i }\log y_A\Bigg) -{1\ov 2 \pi i}\sum_{\substack{\ba,\b b=1\\ \ba\neq \b b}}^{N_f-N_c}  \log(1-x_\ba x_{\b b}^{-1})~,
\eea
\bea
& \Omega_{D,{\rm singlet}}&=&\; \sum_{i=1}^{N_f}\sum_{j=1}^{N_f}\left(-{2r-1\ov 2 \pi i}  \log(1- y_i^{-1}\t y_j)  + {2 r-1\ov 2 \pi i}\log{y_A}\right)\cr
&&&\; \; - {r_+-1\ov 2 \pi i} \log(1- q y_A^{-N_f})- {r_--1\ov 2 \pi i} \log(1- q^{-1} y_A^{-N_f})\cr
&&&\; \; +{N_f(N_f(r-1)+ N_c)\ov 2 \pi i}\log{y_A}+ \half (N_f-N_c)~,
\eea
with $r_{\pm}$ given by \eqref{rpm Aha dual}.  The dual Bethe equations read:
\be
P(x_\ba)=0~,\quad \forall \ba~, \qquad \qquad x_\ba\neq x_{\b b} \quad {\rm if} \quad \ba\neq \b b~,
\ee
in terms of the same polynomial \eqref{Px Aha dual} as the original theory. Let $\{\h x_\alpha\}_{\alpha=1}^{N_f}$ denote the roots of $P(x)$.  The duality map \eqref{duality map} is simply:
\be
\h x^{(l)}= \{\h x_a\} \mapsto  \h x^{(l)}_D = \{\h x_\ba\} =\{\h x_a\}^c~,\qquad \qquad \{\h x_\alpha\}_{\alpha=1}^{N_f} = \h x^{(l)} \cup \h x^{(l)}_D~.
\ee
That is, given $\h x$ choice of $N_c$ roots of $P(x)$, the dual vacuum $\h x_D$ in the Aharony-dual theory is given by the $N_f-N_c$ complement set of roots. Note that the duality of Section \ref{subsec: dual2} is also the special case $N_f=N_c=1$ of Aharony duality.

\paragraph{Matching the fibering operators.}
To match the fibering operators across the duality, we need to prove that:
\be\label{rel W WD aha}
\tCW^{[N_c, N_f]}_{\rm SQCD}(\h x)= \tCW_D(\h x_D)~,
\ee
for a particular choice of branch, where the dependence on the many flavor parameters is left implicit. Using the first relation in \eqref{2terms dilog id}, one can show that \eqref{rel W WD aha} is equivalent to:
\be\label{id W aha}
\tCW^{[N_f, N_f]}_{\rm SQCD}(\h x) =\tCW_{D, {\rm singlet}}~,
\ee
where $\h x$ in \eqref{id W aha} is given by the $N_f$ roots  $P(x)$ in $\{\h x_\alpha\}$ of \eqref{Px Aha dual}. Note that \eqref{id W aha} is independent of $N_c$. This relation corresponds to a known  multi-variable generalization of the five-term dilogarithm identity, which was studied thoroughly in the mathematical literature  \cite{Lewin1991, Kirillov:1994en}. 
This implies the duality relations:
\be
\CF(\h x) = \CF_D(\h x_D)~,
\ee
which are independent of branch cut ambiguities. In particular, the relation \eqref{id W aha} implies:
\bea\label{rel CF full aha}
&\prod_{\alpha=1}^{N_f} \left[ \prod_{i=1}^{N_f}\CF_\Phi(\h u_\alpha - \nu_i) \prod_{j=1}^{N_f} \left[\CF_\Phi(-\h u_\alpha + \t\nu_j) e^{-\pi i \h u_\alpha^2 -\pi i \t\nu_j^2 + {\pi i\ov 6}} \right] e^{-2 \pi i \tau \h u_\alpha} \right]\cr
&\qquad \quad= \prod_{i=1}^{N_f}  \prod_{j=1}^{N_f} \left[ \CF_\Phi(- \nu_i + \t\nu_j)\; e^{-\pi i \t\nu_j + {\pi i \ov 12}}\right]
 \cr
 &\qquad \quad\qquad \times\CF_\Phi(\tau- N_f\nu_A) \; \CF_\Phi(-\tau- N_f\nu_A) \;e^{-\pi i \tau^2- 2 \pi i N_f\nu_A^2 +{\pi i \ov 6}}~,
\eea
where $\h x_\alpha= e^{2 \pi i \h u_\alpha}$ are the roots of $P(x)$.
One can also check \eqref{rel CF full aha} numerically.

\paragraph{Matching the handle-gluing operators.} To complete the proof of the equality \eqref{dual rel corr gen} for Aharony duality, we must also prove that:
\be\label{rel H Hd Aha}
\CH(\h x) = \CH_D(\h x_D)~,
\ee
for any pair of dual vacua. The handle-gluing operators are rational functions of the $x_a$ and $x_\ba$ variables, and the relation \eqref{rel H Hd Aha} can be proven using rather straightforward algebraic manipulations, as explained in \cite{Closset:2016arn}.

%%%%%%%%%%%%%%%%%%%%%%%
\subsubsection{Decoupling limits and Seiberg-like dualities}
All other dualities for SQCD$[k, N_c, N_f, N_a]$ can be derived from Aharony duality by real mass deformation. We refer to Appendix C of \cite{Closset:2016arn} for a detailed review. 

To obtain SQCD$[k, N_c, N_f, N_a]$, we consider a particular massive deformation of  SQCD$[0, N_c, n_f, n_f]$, with $n_f$ defined in \eqref{def nf for sqcd}. We can take the decoupling limit at the level of the effective twisted superpotential and effective dilaton. The number of Bethe vacua,
\be
|\CS_{\rm BE}|= \mat{n_f\cr N_c}~,
\ee
stays constant upon deformation. We can then study the identities \eqref{equal onshell W} as we take a decoupling limit. Typically, both sides of a duality relation diverge as the mass goes to infinity, but with an identical coefficient on both sides. Therefore, we can cancel the divergences and deduce the identity for the IR theory from its UV parent  \cite{Benini:2011mf}. In the following, we demonstrate this behavior at the level of the Bethe equations.

Let us define $k_c\equiv \half (N_f- N_a)$. Consider first the case $k\geq k_c$. This can be obtained from SQCD$[0, N_c, n_f, n_f]$ by integrating out $k-k_c$ fundamental chiral multiplets $Q_\alpha$ with positive real mass and $k+k_c$ antifundamental chiral multiplets $\t Q^\beta$ with positive real mass, while the remaining $N_f$ fundamental chiral multiplets $Q_i$ and $N_a$ antifundamental chiral multiplets $Q_j$ remain light.
Let us denote by $m_0\rightarrow \infty$ the real mass parameter that we send to infinity, and by $y_0\rightarrow 0$ the corresponding fugacity. The gauge and flavor fugacities must be rescaled according to:
\bea
&x^{-1}\,y_i &\rightarrow&\quad x^{-1}\,y_i~, &\qquad\qquad & x\, \t y_j^{-1}&\rightarrow&\quad x\, \t y_j^{-1}~, \cr
& x^{-1}y_\alpha &\rightarrow&\quad y_0\, x^{-1} \,y_A^{-1}~,  &\qquad \qquad& x\, \t y_\beta^{-1} &\rightarrow &\quad y_0\, x \, y_A^{-1}~,\cr
&q&\rightarrow&\quad y_0^{k_c} q~,
\eea
which also implies:
\be
x^{n_f}\rightarrow y_0^{-k_c} \, x^{n_f}~, \qquad \quad  y_A^{n_f} \rightarrow y_0^{k}\, y_A^{n_f}~.
\ee
The case $k_c \geq k$ can be obtained similarly. We start from SQCD$[0, N_c, N_f, N_f]$ and we integrate out $k_c+ k$ antifundamental multiplets $\t Q^\beta$ with positive real mass and $k_c-k$ antifundamental multiplets $\t Q^\gamma$ with negative real mass. The relevant scaling is:
\bea
&x^{-1}\,y_i &\rightarrow&\quad x^{-1}\,y_i~, &\qquad\qquad & x\, \t y_j^{-1}&\rightarrow&\quad x\, \t y_j^{-1}~, \cr
&  x\, \t y_\beta^{-1}&\rightarrow&\quad y_0\,  x\,  y_A^{-1}~, ~,  &\qquad \qquad& x\, \t y_\gamma^{-1} &\rightarrow &\quad y_0^{-1}\, x \, y_A^{-1}~,\cr
&q&\rightarrow&\quad y_0^{k_c} q~,
\eea
and
\be
x^{N_f}\rightarrow y_0^{-k} \, x^{N_f}~, \qquad \quad  y_A^{N_f} \rightarrow y_0^{k}\, y_A^{N_f}~.
\ee
It is easy to apply this scaling to any of the various operators that enter the supersymmetric partition function. By considering the  limit $y_0 \rightarrow 0$ at the level of the Bethe equations, we obtain new Bethe equations $P(x_a)=0$ in terms of the polynomial:
\be
P(x) = \prod_{i=1}^{N_f}(x-y_i) - (-1)^{k-k_c} q y_A^{Q_+^A} x^{k+k_c} \prod_{j=1}^{N_a} (x- \t y_j)~,
\ee
where we defined:
\be
Q_+^A =  \begin{cases}
-N_f&\qquad {\rm if} \qquad k \geq  k_c~,\\
-k - \half(N_f+N_a) &\qquad {\rm if} \qquad k \leq  k_c~.
\end{cases}
\ee
We may similarly study this decoupling limit at the level of the twisted superpotential. It is obvious from the general properties of $\CW$, and in particular from the limits  \eqref{W limit},  that we reproduce in this way the correct low energy theories, including all the correct gauge and flavor Chern-Simons levels.

\subsection{The ``duality appetizer''}
As our last example, we consider the ``duality appetizer'' of \cite{Jafferis:2011ns}.  It relates the following theories: Theory A is an $SU(2)$ gauge theory with CS level $k=1$, coupled to a single adjoint chiral multiplet $\Phi$.  Theory $B$ is  a free chiral multiplet, $Z$, together with a decoupled  $U(1)_{k=2}$ topological sector.
The operator $Tr \Phi^2$ in theory $A$ is mapped to $Z$ in theory $B$.  Correspondingly, there is a single $U(1)_F$ flavor symmetry which acts on $\Phi$ with charge $1$, and on $Z$ with charge $2$.~\footnote{The $U(1)_{2}$ sector in Theory $B$ also has a topological symmetry $U(1)_T$, which we ignore. In the partition function, we just set the $U(1)_2$ FI parameter to zero.}

The handle-gluing operators across the duality were matched in \cite{Benini:2015noa}. Let us show that the fibering operators match as well.  The effective twisted superpotential of theory A is given by:
\be\label{DAW} 
\tCW_A(x, y) = \dilog(x^2 y) + \dilog(y) + \dilog(x^{-2} y) + \frac{1}{2} \log{y}(\log{y} +2 \pi i) + \log^2{x}~.
\ee 
The corresponding Bethe equation can be written as:
\be
 (x^2-1) \left( (x+x^{-1})^2-(1+y^{-1})^2\right) = 0~.
 \ee
The solutions $x=\pm 1$ correspond to fixed points of the Weyl group action, $x \rightarrow x^{-1}$, and  are thus discarded.  The remaining four solutions come in two Weyl pairs, with:
\be \label{DABEsol} 
\h x+\h x^{-1} = \pm (1+y^{-1} ) 
\ee
which correspond to the two physical vacua of this theory.  
Let us define $ \h x=\alpha$ to be one of the first solution in \eqref{DABEsol}, so that $\h x=-\alpha$ gives the other  solution.  Then $y^{-1} = \alpha+\alpha^{-1}-1$. Plugging this relation into the effective twisted superpotential \eqref{DAW}, we find that the on-shell twisted superpotential for the two vacua are:
\bea\label{DAW onshell}
 &\tCW_A^{(\pm)}(\alpha) &=&\;  \dilog\left(\frac{\alpha^2}{\alpha+\alpha^{-1}-1}\right) +\dilog\left(\frac{\alpha^{-2}}{\alpha+\alpha^{-1}-1}\right) \cr
&&&\; + \dilog\left(\frac{1}{\alpha+\alpha^{-1}-1}\right) + \log (\pm \alpha) (\log (\pm \alpha) + 2 \pi i)\cr
&&&\;  + \frac{1}{2} \log (\alpha+\alpha^{-1}-1)(\log( \alpha+\alpha^{-1}-1) -2 \pi i) ~.
\eea
Note that, up to a change of branch:
\be\label{rel vacua thA}
 \tCW_A^{(-)} =  \tCW_A^{(+)} - \pi^2~.
 \ee
In other words, the only difference between the on-shell twisted superpotentials in the two vacua can be attributed to a relative gravitational CS term.

Turning to theory B,  the contribution from the scalar $Z$ is:
\be 
\tCW_Z = \dilog(y^2) + \log^2{ y} = \dilog\left(\frac{1}{(\alpha+\alpha^{-1}-1)^2}\right) + \log^2\left(\frac{1}{\alpha+\alpha^{-1}-1}\right)~,
\ee
where we set the $U(1)_F$ CS term such that the $\kappa_{FF}$ bare contact term vanishes. 
For the $U(1)_2$ sector at zero FI parameter, the two vacua contribute only gravitational CS terms $k_g=0$ and $k_g=6$.~\footnote{The twisted superpotential for the $U(1)_2$ sector is $\tCW_{\rm top}= \log^2 x$ and the Bethe equation $x^2=1$ has two solutions $x=\pm 1$, leading to $\tCW_{\rm top}=0, -\pi^2$ on-shell.}  This is precisely the difference \eqref{rel vacua thA} between the two vacua in Theory $A$.  Thus it remains only to check the matching of the twisted superpotential of one of the vacua.  The precise statement, including a relative gravitational CS term $\Delta k_g= -2$, is:
\be\label{rel WA WB app}
 \tCW_A^{(-)}(\alpha) = \tCW_B(\alpha) \equiv   \tCW_Z(\alpha)  + \frac{1}{3} \pi^2~.
\ee
This follows from the identity:
\bea  
&\dilog\left(\frac{\alpha^2}{\alpha+\alpha^{-1}-1}\right) +\dilog\left(\frac{\alpha^{-2}}{\alpha+\alpha^{-1}-1}\right) + \dilog\left(\frac{1}{\alpha+\alpha^{-1}-1}\right) \cr
&- \frac{1}{2} \log(\alpha+\alpha^{-1}-1)(\log(\alpha+\alpha^{-1}-1)+2 \pi i) + \log \alpha (\log \alpha - 2 \pi i) \cr
& = \dilog\left(\frac{1}{(\alpha+\alpha^{-1}-1)^2}\right) + \frac{1}{3} \pi^2~.
\eea
As with all identities involving dilogarithms evaluated at rational functions of a single variable, this can be derived by repeated applications of the five-term identity.  The relation \eqref{rel WA WB app} implies the matching of the dual fibering operators.

%%%%%%%%%%%%%%%%%%%%%%%%%%%%%%%%%%%%%%%%%%%%%%%%
\section*{Acknowledgements} We would like to thank Ofer Aharony, Benjamin Assel, Matthias Blau, Guido Festuccia, Sergei Gukov, Victor Mikhaylov, Daniel Park, Shlomo Razamat and Itamar Yaakov 
for interesting discussions and comments.
This research was supported in part by Perimeter Institute for
Theoretical Physics. Research at Perimeter Institute is supported by the Government of Canada through Industry Canada and by the Province of Ontario through the Ministry of Economic Development \& Innovation. 
The work of HK was made possible through the support of a grant from the John
Templeton Foundation. The opinions expressed in this publication are those of the
author and do not necessarily reflect the views of the John Templeton Foundation.  BW was supported in part by the National
Science Foundation under Grant No. NSF PHY11-25915. CC and HK gratefully acknowledges support from the Simons Center for Geometry and Physics, Stony Brook University at which some of the research for this paper was performed.

%%%%%%%%%%%%%%%%%%%%%%%%%%%%%%%%%%%%%%%%%%%%%%%%

%%%%%%%%%%%%%
\appendix
\section{The $\Mgp$ geometry}\label{app: geom}
In this Appendix, we briefly summarize our geometric conventions and we provide some additional details about the geometry and topology of $\Mgp$. We also briefly discuss torsion line bundles over $\Mgp$ with $p\neq 0$.

\subsection{The $\Mgp$ geometry}
We follow the geometry conventions of \cite{Closset:2016arn}, which closely follows \cite{Closset:2012ru,Closset:2013vra}. Let us consider the three-manifold $\Mgp$, a $U(1)$ principal bundle over the Riemann surface, $\Sigma_g$,  with metric:
\be\label{metric Mgp appendix}
ds^2(\Mgp)= \beta^2 \big(d\psi +  \CC(z,\bz) \big)^2 + 2 g_{z\bz}(z, \bz) dz d\bz = (e^0)^2+ e^1 e^{\b1}~.
\ee
The coordinates are $(x^\mu)=(\psi, z, \bz)$, with $\psi \in [0, 2 \pi)$ an angular coordinate along the $S^1$ fiber, and the $z, \bz$ local coordinates on the base $\Sigma_g$. 
The two-dimensional metric $2 g_{z\bz}$ is a complete Hermitian metric on $\Sigma_g$, written in a local patch. The quantity $\CC$ is a $U(1)$ connection over $\Sigma_g$ with first Chern number $p$:
\be
{1\ov 2 \pi} \int_{\Sigma_g} d\CC=p~.
\ee
The complex frame $(e^a)=(e^0, e^1, e^{\b1})$ is defined in \eqref{canonical frame}. The frame indices $a=0, 1, \b1$  are lowered using $\delta_{ab}$ with $\delta_{00}=1$ and $\delta_{1\b 1}= \half$. The orientation is such that $\epsilon^{0 1 \b 1}= -2 i$ and the $\gamma$-matrices are:
\be
\left\{{(\gamma^a)_\alpha}^\beta\right\} = \left\{\gamma^0, \gamma^1, \gamma^{\b1}\right\}= \left\{ \mat{1 & 0 \cr 0 &-1}~,\; \mat{0& -2\cr 0&0}~,\; \mat{0& 0\cr -2&0}  \right\}~. 
\ee
The metric \eqref{metric Mgp appendix} has an Killing vector $K = {1\ov \beta} \d_\psi$. 
Let us also define the one-form:
\be\label{def eta app}
\eta_\mu dx^\mu= \beta (d \psi + \CC)~,
\ee
which satisfies $K_\mu \eta^\mu=1$, and the tensor:~\footnote{This notation is slightly redundant, since $\eta_\mu= K_\mu$ with our particular choice of metric, but we find it convenient to use $\eta$ for the one-form defining the THF \protect\cite{Closset:2012ru, Closset:2013vra}.}
\be\label{def phi app}
{\Phi_\mu}^\nu= - {\epsilon_\mu}^{\nu\rho}\eta_\mu~.
\ee
We have:
\be
\eta^\mu \eta_\mu=1~, \qquad  {\Phi_\mu}^\nu {\Phi_\nu}^\rho =- {\delta^\mu}_\nu+ \eta^\mu \eta_\rho~.
\ee
The objects $\eta$ and $\Phi$ define a metric-compatible {\it transversely holomorphic foliation} (THF) on $\Mgp$. This means that there exists adapted coordinates $\psi, z, \bz$  such that the transition functions between patches are of the form 
\be\label{change of coord adpt}
\psi'= \psi -\lambda(z, \bz)~,\qquad z'= f(z)~,
\ee
with $\lambda$ real and $f(z)$ a holomorphic function of $z$  \cite{Closset:2012ru}. We are considering a particular THF on $\Mgp$ such that the foliation $\eta$ is also an $S^1$ fibration, and the leaves of the THF are the $S^1$ fibers.
Note that,  under a change of coordinates \eqref{change of coord adpt}, we also have the gauge transformation:
\be
\CC'= \CC+ d \lambda~,
\ee
so that $\eta$ is a well-defined one-form.

A THF is a natural three-dimensional analog of a complex structure. In the present case, the THF is simply the uplift of the complex structure on $\Sigma_g$. 
Let us define the projection operators:
\bea\label{projectors}
& {{\rm P_0}^\mu}_\nu= \eta^\mu \eta_\nu~, \cr
& {\Pi^\mu}_\nu =\half \left({\delta^\mu}_\nu - i {\Phi^\mu}_\nu- \eta^\mu \eta_\nu\right)~, \cr
&{ \t\Pi^\mu}_{\phantom{\mu}\nu} = \half \left({\delta^\mu}_\nu +i {\Phi^\mu}_\nu- \eta^\mu \eta_\nu\right)~,
\eea
which satisfy ${\rm P_0}+ \Pi + \t\Pi= \mathbbm{ 1}$. They allow us to decompose any one-form $\alpha$ into vertical, holomorphic and (horizontal) anti-holomorphic components, respectively:
\be\label{dec alpha}
\alpha = \alpha_0 \eta + \alpha_z dz +\alpha_\bz d\bz~.
\ee
In particular, a holomorphic one-form, $\omega \in \Lambda^{1,0}\Mgp$, is such that:
\be
  \omega_\mu {\Pi^\mu}_\nu= \omega_\nu~.
\ee
Its single component $\omega_z$ transforms as $\omega_{z'}'=(\d_z f(z))^{-1}\omega_z$ under a change of adapted coordinates. By definition,  $\omega_z$  is a section of an {\it holomorphic line bundle} over $\Mgp$ \cite{Closset:2013vra}.~\footnote{This is known as an $h$-foliated bundle in the mathematical literature \protect\cite{gomez-mont1980}.} 
We call that particular holomorphic line bundle the {\it canonical bundle}, denoted by $\CK$:
\be
\omega_z \in \Gamma[\CK]~.
\ee
$\CK$ is the pull-back of the canonical line bundle on $\Sigma_g$, and its first Chern class is given by \eqref{c1K}.
Similarly, a holomorphic vector $X\in T^{1,0}\Mgp$ satisfies $ {\Pi^\mu}_\nu X^\nu = X^\mu$, and is given by
\be
X= X^z(\d_z - \CC_z \d_\psi)
\ee
in local coordinates. In the main text, we mainly use the frame basis, so that $\omega= \omega_1 e^1$ and $X= X^1 \d_1$, with $\d_1= e_1^\mu \d_\mu$.

Note that the Levi-Civita connection $\nabla$ does not commute with $\eta_\mu$, and therefore does not preserve the decomposition \eqref{dec alpha}. We define a metric- and THF-compatible connection $\h \nabla$, such that
\be
\h \nabla_\mu g_{\nu\rho}=0 ~, \qquad \h \nabla_\mu \eta_\nu=0~.
\ee
It is given by:
\be
{\h \Gamma^\mu}_{\phantom{\nu}\mu\rho}= { \Gamma^\nu}_{\mu\rho}
+  {K^\nu}_{\mu\rho}~, \qquad K_{\nu\mu\rho}=  i H \left(\eta_\nu \Phi_{\mu\rho}- \eta_\rho \Phi_{\mu\nu}+ \eta_\mu \Phi_{\nu\rho}  \right)~,
\ee
with ${ \Gamma^\nu}_{\phantom{\nu}\mu\rho}$ the Christoffel symbols. Here $ {K^\nu}_{\phantom{\nu}\mu\rho}$ is the contorsion tensor. The adapted spin connection is:
\be
\h\omega_{\mu\nu\rho}=  \omega_{\mu\nu\rho}- K_{\nu\mu\rho}~.
\ee
We will denote the adapted covariant derivative, acting on any field, simply by $D_\mu$.~\footnote{$D_\mu$ will also denote the $R$- and gauge-covariant derivative acting on charged fields.}
It commutes with the projectors \eqref{projectors} and it is therefore compatible with the decomposition into vertical, holomorphic and anti-holomorphic component, which we used extensively in Section \ref{sec: 3}.
Note that we have:
 \be
 D_\mu \psi = (\d_\mu - {i\ov 4}\h\omega_{\mu ab}\epsilon^{abc}\gamma_c)\psi
 \ee
on a Dirac fermion $\psi$, and similarly on fields of any definite three-dimensional spin.
The adapted connection has torsion:
\be
{T^{\nu}}_{\mu\rho}={K^{\nu}}_{\mu\rho}-{K^{\nu}}_{\rho\nu}= 2 i H \eta^\nu \Phi_{\mu\rho}~.
\ee
In particular,  we have:
\be\label{expl comm scalar}
[D_\mu , D_\nu] \varphi = - 2 i H \Phi_{\mu\nu} \eta^\rho D_\rho \varphi~.
\ee
when acting on a scalar field $\varphi$.

We can also check that the Lie derivative and the adapted covariant derivative are equal along  the Killing vector $K^\mu$ :
\be
\CL_K = K^\mu D_\mu~,
\ee
for fields of any spin.
This is useful in order to check that the supersymmetry transformations of Section \ref{sec: 3} realize the supersymmetry algebra \eqref{susy algebra 3d}. The following identities are also useful:
\be
d\eta =  2 p \beta \; d{\vol}(\Sigma_g)~,\qquad\quad
\eta\wedge d\eta= 2 p \beta \; d{\vol}(\Mgp)~,
\ee
where $d{\vol}(\Sigma_g)$ and $d{\vol}(\Mgp)$ are the volume forms on $\Sigma_g$ and $\Mgp$, respectively. We normalized the volumes to $\vol(\Sigma_g)=\pi$ and $\vol(\Mgp)= 2 \pi^2 \beta$.  Note that the volume form $d{\vol}(\Sigma_g)$ is exact unless $p=0$.

\subsection{Cohomology and homology of $\Mgp$}
Some useful homological properties of $\Mgp$ are described in \cite{Blau:2006gh}, to which we refer for further details.
Let us assume that $p \neq 0$. By the Gysin sequence, we have the cohomology groups:
 \bea\label{cohomology Mgp}
& H^1(\Mgp, \Z) \cong H^1(\Sigma_g, \Z) \cong \Z^{2g}~, \cr
 & H^2(\Mgp, \Z) \cong H^1(\Sigma_g, \Z) \oplus \Z_p \cong \Z^{2g}\oplus \Z_p~, 
 \eea
where the torsion subgroup $\Z_p$ is given by:
\be
{\rm coker}\left( c_1: H^0(\Sigma_g, \Z)\rightarrow  H^2(\Sigma_g, \Z)\right) \cong \Z_p~,
\ee
with $c_1$ the first Chern class of the $U(1)$ principal bundle over $\Sigma_g$.
The homology of $\Mgp$ follows from \eqref{cohomology Mgp} by Poincar\'e duality:
\be
 H_1(\Mgp, \Z) \cong \Z^{2 g} \oplus \Z_p~, \qquad \qquad  H_2(\Mgp, \Z) \cong \Z^{2 g}~.
\ee

One can also define a Dolbeault-like cohomology \cite{gomez-mont1980, Closset:2013vra} of the transversely holomorphic foliation, which carries interesting information. For instance, infinitesimal deformations of a holomorphic line bundle $L$ are valued in:
\be\label{h01 M}
H^{0,1}(\Mgp,\C) \supset \C~.
\ee
We did not compute $H^{0,1}(\Mgp,\C)$ from first principles.~\footnote{In \protect\cite{Closset:2013vra}, it was shown explicitly that $H^{0,1}(\CM_{0,1},\C)\cong \C$ for the three-sphere. 
} For our purposes, it is sufficient to note that the one-form $\eta$ is a $(0,1)$-form such that
\be
\t\d \eta=0~, \qquad \eta \neq \t\d(\cdots)~,
\ee
in the notation of \cite{Closset:2013vra}---see equation (5.15) in that reference. Therefore, $\eta$ generates the one-dimensional subgroup of $H^{0,1}(\Mgp)$ indicated on the right-hand side of \eqref{h01 M}. 

Deformations of holomorphic line bundles sit in $H^{0,1}(\Mgp)$, therefore any holomorphic line bundle $L$  has at  least a one-parameter family of deformations. The corresponding line bundle modulus is denoted by $u$ or $\nu$ in the main text. In general, we can have other deformations of the bundle, corresponding (roughly speaking) to flat connections along the $\Sigma_g$ base. However, those additional deformations are $Q$-exact in the supersymmetric field theory \cite{Closset:2013vra}.

\subsection{Flat connection of a torsion line bundle}
Let us review some elementary facts about flat connections for torsion bundles over $\Mgp$ ($p\neq 0$).
We focus  on the case $g=0$ case---the Lens space $L(p,p-1)$---where we can write explicit formulas. The $S^2$ base can be covered by two coordinate patches. The $z$ coordinate
\be
z= \tan{\theta\ov 2} e^{i \phi}
\ee
covers the northern patch of the sphere, and the $z'={1\ov z}$
coordinate covers the southern patch. With the standard round metric \eqref{round S3 metric}, we have the change of coordinates 
\be\label{change coord Lens space}
\psi'=\psi -{i p\ov 2} \log\left(z\ov \bz\right)= \psi + p \phi~, \qquad\qquad z'={1\ov z}~, 
\ee
between the north and southern patches (each patch has topology $D^2 \times S^1$, with $D^2$ the open disk).
Consider a flat connection $a$ for a non-trivial bundle $L$. On the northern patch, we take:
\be
a= a_\psi d\psi~,
\ee
with $a_\psi$ some constant to be determined. For the holonomy $\exp\left(- i \int_\gamma a\right)$ of the fiber to be well-defined, we must also have
\be
a'= a_\psi  d\psi'
\ee 
on the southern patch. The two descriptions are related by $a'= a+ d\lambda$ for some gauge parameter $\lambda$, and comparing to \eqref{change coord Lens space} we see that:
\be
 d\lambda = a_\psi p \, d\phi~.
 \ee
We must have $a_\psi p \in \Z$ for this transition function to be well-defined on the overlap, leading to:
\be\label{apsi explicit}
a_\psi = -{\m\ov p}~, \qquad \m \in \Z~.
\ee
This corresponds to a first Chern class:
\be
c_1(L)= \m\in \Z_p
\ee
for the corresponding line bundle.~\footnote{To check the sign, note that the relation $a^{S}= a^{N}- \m d\phi$ is precisely the relation between the southern and northern patch connections of a flux $\m$ Dirac monopole on $S^2$.} The relation \eqref{apsi explicit} is also valid for $g>0$.

%%%%%%
\section{Supersymmetry on $\Mgp$}\label{app: 3d Atwist}
In this Appendix, we provide additional details about the supersymmetric background of Section \ref{sec: 3}. Curved-space supersymmetry for $\CN=2$ supersymmetric theories with an $\CR$-multiplet is governed by the generalized Killing spinor equations \cite{Klare:2012gn, Closset:2012ru}:
\bea\label{KSE}
&(\nabla_\mu -i A_\mu^{(R)})\zeta= - \half  H \gamma_\mu \zeta+ {i\ov 2}V_\mu\zeta - \half \epsilon_{\mu\nu\rho}V^\nu \gamma^\rho\zeta~, \cr
&(\nabla_\mu +i A_\mu^{(R)})\t\zeta= - \half  H \gamma_\mu \t\zeta - {i\ov 2}V_\mu\t\zeta + \half \epsilon_{\mu\nu\rho}V^\nu \gamma^\rho\t\zeta~.
\eea
A supersymmetric background on a compact three-manifold $\CM_3$ with Riemannian metric $g_{\mu\nu}$ consist of background values for the $\CN=2$ ``new-minimal'' supergravity fields:
\be
g_{\mu\nu}~,  \qquad H~, \qquad V_\mu~,  \qquad A_\mu^{(R)}~,
\ee
that preserve certain Killing spinors $\zeta$, $\t\zeta$.~\footnote{Note that $A_\mu^{(R)}=A_\mu -{3\ov 2}V_\mu$ in the notation of \protect\cite{Closset:2012ru}.}

Consider $\CM_3=\Mgp$ with the metric \eqref{metric Mgp appendix}. 
Given the THF \eqref{def eta app}-\eqref{def phi app} and the Killing vector $K^\mu= \eta^\mu$, the general solution  to \eqref{KSE} preserving one $\zeta$ and one $\t \zeta$ reads \cite{Closset:2012ru}:
\bea\label{gen sol KSE}
& H = {i\ov 2}\epsilon^{\mu\nu\rho} \eta_\mu\d_\nu \eta_\rho + i \kappa~,\cr
& V_\mu = - \epsilon_{\mu\nu\rho}\d^\nu \eta^\rho- \kappa \eta_\mu~,\cr
& A_\mu^{(R)}= \CA_\mu^{(R)} + \half \epsilon_{\mu\nu \rho} \d^\nu\eta^\rho+ \d_\mu s~,
\eea
with $\CA^{(R)}$ given by:
\be\label{def CAR ii}
\CA^{(R)}_\mu= {1\ov 4} {\Phi_\mu}^\nu \d_\nu \log\sqrt{g}
\ee
in the adapted coordinates $\psi, z, \bz$. The function $\kappa$ in \eqref{gen sol KSE} satisfies $K^\mu \d_\mu \kappa=0$ and is otherwise arbitrary. It couples to the real central charge $Z$ of the three-dimensional $\CN=2$ theory \cite{Closset:2013vra}. In this work, we choose:
\be
\kappa=0~.
\ee
This leads to a simple relation between our $\Mgp$ background and the A-twist background on $\Sigma_g$. While small deformations by $\kappa$ do not affect supersymmetric observables, one could consider a ``large'' deformation such that $\kappa$ introduces a flux for the central charge $Z$ \cite{Closset:2013vra}. This would lead to a Dirac quantization condition for real mass and FI parameters. We do not consider such backgrounds.

Note that $S^3$ background of \cite{Kapustin:2009kz, Hama:2010av, Jafferis:2010un} corresponds to $\kappa \neq 0$ such that $V_\mu=0$. This does not affect the $S^3$ partition function, however, because there is no possible central charge flux on $S^3$. Therefore, our results for $\CM_{0,1}\cong S^3$ must be in agreement with  \cite{Kapustin:2009kz, Hama:2010av, Jafferis:2010un}, as we indeed find to be the case.

Setting $\kappa=0$ in \eqref{gen sol KSE} gives us the background fields
\be
H= i p \beta~,\qquad\quad
  V_\mu= - 2 p \beta\, \eta_\mu~, \qquad\quad
A_\mu^{(R)} = \CA^{(R)}_\mu + p \beta \eta_\mu+ \d_\mu s~,
\ee
Since $V_\mu= 2 i H \eta_\mu$, we find it convenient to use the background field $H$ and $\CA_\mu^{(R)}$ only, as in  \eqref{H and CAR}. The Killing spinor equations \eqref{KSE} can be simplified by using the adapted connection $\h\nabla$, as discussed in Section \ref{sec: 3}. We obtain \eqref{KSE twist}, which is simply:
\be\label{KSE twist app}
D_\mu\zeta=0~, \qquad \qquad
D_\mu\t\zeta=0~,
\ee
in terms of the covariant derivative $D_\mu$ define above, including the $U(1)_R$ gauge field $\CA_\mu^{(R)}$.

%%%%%%%%%%%%%%%%%%
\subsection{A-twisted field variables}\label{app: twisted fields}
Using the Killing spinors $\zeta$, $\t\zeta$, we may build the one-forms:
\be
p_\mu = \zeta\gamma_\mu \zeta~, \qquad \quad \t p_\mu = \t\zeta\gamma_\mu\t \zeta~,
\ee
of $R$-charge $2$ and $-2$, respectively. We have:
\be\label{def p pb ii}
p_\mu dx^\mu =p_{\b1} e^{\b 1} = -e^{2 i s} e^{\b1}~, \qquad \qquad \t p_\mu dx^\mu =\t p_1 e^1= e^{-2 i s} e^{1}~.
\ee
In particular, $\t p_1$ is a nowhere-vanishing section of $\CK\otimes (L^{(R)})^{-2}$, where $L^{(R)}$ is the $R$-symmetry line bundle. This implies that $\CK \cong (L^{(R)})^2$ up to a topologically trivial line bundle.

After decomposing any field into vertical and horizontal components, like in \eqref{dec alpha}, we may assign two-dimensional spins in the frame basis, as explained above \eqref{def twisted spin}.
 For instance, $p_1$ has $2d$ spin $s_0=1$ and $R$-charge $-2$, and it has vanishing A-twisted spin \eqref{def twisted spin}.
We find it convenient to use field variables adapted to the A-twist, exactly like in \cite{Closset:2015rna,Closset:2016arn}. Let us briefly review the definitions:

\paragraph{$A$-twisted chiral multiplet.}
The twisted fields in the chiral and antichiral multiplets are related to the flat-space fields of \cite{Closset:2012ru} by:
\bea\label{Atwistvar}
&\CA = (\t p_1)^{r\over 2}\, \phi~, 
&\qquad&\t\CA = (p_{\b1})^{r\over 2} \t\phi~, \cr
& \CB =\sqrt2  (\t p_1)^{r\over 2} \zeta\psi~,\quad 
&& \t\CB =- \sqrt2  ( p_{\b1})^{r\over 2}  \t\zeta\t\psi~, \cr
&\CC =-{1\over \sqrt2} (\t p_1)^{r\over 2} p_{\b1} \, \t\zeta\psi~, \qquad
& &\t\CC ={1\over \sqrt2} (p_{\b1})^{r\over 2}\t p_1 \, \zeta \t\psi~, \cr
&\CF = (\t p_1)^{r\over 2} p_{\b1}\, F~, 
&&\t\CF = (p_{\b1})^{r\over 2}\t p_1 \,\t F~. \cr
\eea
Here $p_{\b 1}$ and $\t p_1$ are the sections of $\b\CK \otimes L^2$ and $\CK \otimes L^2$ as defined in \eqref{def p pb ii}.  By constructions, all the A-twisted fields have $R$-charge zero and two-dimensional spin \eqref{def twisted spin}. In particular, $\CA, \CB$ have twisted spin ${r\ov 2}$ and $\CC, \CF$ have twisted spin ${r-2\ov 2}$.

\paragraph{$A$-twisted vector multiplet.} The gauginos $\lambda$ and $\t \lambda$ in the vector multiplet of  \cite{Closset:2012ru}  are related to the A-twisted fields \eqref{gauginiA} by:
\be
\Lambda_\mu \equiv \t\zeta\gamma_\mu\lambda~, \qquad \qquad
\t\Lambda_\mu \equiv -\zeta\gamma_\mu\t\lambda~.
\ee
The gaugino supersymmetry variation can be written as:
\be
\delta\Lambda_\mu = i \eta_\mu (D- \sigma H) + i \left({\delta_\mu}^\nu + i {\Phi_\mu}^\nu\right) \Big(\d_\nu \sigma + \half {\epsilon_\nu}^{\lambda\rho}f_{\lambda\rho}\Big)~, \qquad\quad \t\delta \Lambda_\mu=0~,
\ee
and similarly for $\t\Lambda_\mu$.

%%%%%
%%%%%%%%%%%%
\section{Spin-structure dependence of the $U(1)$ Chern-Simons action}\label{app: spin CS}
Consider a $U(1)$ connection $a=a_\mu dx^\mu$ on an (oriented) three-manifold $\CM_3$. Whenever $a_\mu$ is a connection on a topologically non-trivial bundle, the CS action is defined by:
\be\label{def CS action 4d}
S_{\rm CS}= {i k\ov 4\pi}\int_{\CM_3} a \wedge f \equiv  {i k\ov 4\pi}\int_{\CN_4} f \wedge  f \; \mod 2\pi i~,
\ee
with $f=da$ and $k\in \Z$ the CS level. Here $\CN_4$ is a four-manifold with boundary $\d\CN_4= \CM_3$, and the three-dimensional connection is extended to the connection of a line bundle over $\CN_4$. An important subtlety is that the CS action depends on the spin structure of $\CM_3$ if $k$ is odd. In that case, $\CN_4$ must also be a spin manifold, whose spin structure restricts to the spin structure specified on the boundary. This can introduce an explicit dependence of the CS action \eqref{def CS action 4d} on the choice of spin structure \cite{Dijkgraaf:1989pz}. This point was emphasized recently  {\it e.g.} in \cite{Seiberg:2016gmd}.~\footnote{See also \protect\cite{Mikhaylov:2015nsa}. We thank Victor Mikhaylov for very illuminating discussions on the matter.}

First of all, note that the definition \eqref{def CS action 4d} is independent of the choice of $\CN_4$. If we consider two different choices of four-manifolds (with spin structures) $\CN_4$ and $\CN_4'$, we have:
\be\label{SN4N4p}
S_{\rm CS}[\CN_4]- S_{\rm CS}[\CN_4']=  {i k\ov 4\pi}\int_{\CM_4} f \wedge  f = \pi i k \, q(f)~,
\ee 
 where the compact four-manifold $\CM_4$ is the union of $\CN_4$ and $\CN_4'$ (with reversed orientation) glued along $\CM_3$. Here $q(f)$ is a topological invariant of the $U(1)$ line bundle  on $\CM_4$, which is always integer. If we specify a spin structure on $\CM_3$, then $\CM_4$ is also a spin manifold and $q(f)$ is an {\it even} integer \cite{Dijkgraaf:1989pz}; therefore the definition \eqref{def CS action 4d} makes sense for any integer $k$. 

Now, consider two distinct spin structures on the same three-manifold $\CM_3$, which we denote by $\CM_3^\pm$, and consider some choice of bounding spin four-manifolds $\CN_4^\pm$. The difference between the CS actions on $\CM_3^+$ and $\CM_3^-$ is again given by \eqref{SN4N4p}, but the compact four-manifold $\CM_4$ is not spin in general, since the spin structures on $\CN_4^+$ and $\CN_4^-$ are not compatible on the $\CM_3$ boundary. Therefore, the CS actions on $\CM_3^\pm$ might differ by some integer multiple of $\pi i$; in other words, the exponentiated action $e^{-S_{\rm CS}}$ might include a {\it sign} that depends on the choice of spin structure on $\CM_3$. 

We are particularly interested in the three manifolds $\CM_3^\pm \cong \Sigma_g \times S^1_\pm$, where the spin structures correspond to either the periodic ($+$) or anti-periodic ($-$) boundary condition for fermions along the $S^1$ (and with some given spin structure on $\Sigma_g$). In order to preserve supersymmetry, we choose the periodic spin structure, $\CM_{g, 0}\cong \Sigma_g \times S^1_+$. Consider a $U(1)$ line bundle with first Chern number $\m\in \Z$ and a flat connection $a_\psi$ along the $S^1$. We can easily see that:
\be
e^{-S_{CS (-)}}= e^{-2 \pi i k \,a_\psi \m}~,
\ee
for the anti-periodic spin structure,
because we can extend $\CM_3^-\cong \Sigma_g\times S^1_-$ to $\CN_4^- \cong \Sigma_g \times D^2$, with $D^2$ a disk with $S^1_-$ as its boundary.~\footnote{To see this, we can first set $a_\psi=0$, in which case $S_{CS (-)}=0$ because $a$ extends to a flat gauge field along $D^2$. The deformation by a flat connection corresponds to a shift by a well-defined one-form $\delta a$, and the shift in the CS action is given by the three-dimensional formula $\int \delta a \wedge da$.} On the other hand, one can show that \cite{Mikhaylov:2015nsa}:~\footnote{
In the general case, we have:
$$
S_{CS (+)} - S_{CS (-)} = {i k\ov 2} \int_{\CM_3} f \cup x~,$$
with $x\in H^1(\CM_3, \Z_2)$ encodes the change of spin structure between $\CM_3^+$ and $\CM_3^-$  \protect\cite{Mikhaylov:2015nsa}.
}
\be
S_{CS (+)} - S_{CS (-)} = \pi i k \m~, 
\ee
and therefore:
\be
e^{-S_{CS (+)}}= (-1)^{k \m} e^{-2 \pi i k \,a_\psi \m}~.
\ee
This is the correct result on the supersymmetry-preserving $\Sigma_g\times S^1$ background; the sign was previously missed by \cite{Benini:2015noa, Benini:2016hjo, Closset:2016arn}. The case $T^2 \times S^1$ is discussed explicitly in \cite{Seiberg:2016gmd}.

Note that a closely related sign $(-1)^{k\m}$ appears in  \eqref{CS torsion and sign} from the CS action on any $\Mgp$ with  $p\neq 0$, with $\m\in \Z_p$. In that case, the sign is necessary for the CS action to be invariant under large gauge transformations, $\m \sim \m +p$.

Incidentally, similar signs seem to be important for other supersymmetric backgrounds, in particular for the 3d superconformal index \cite{Kim:2009wb, Imamura:2011su, Kapustin:2011jm} (the ``untwisted'' $S^2 \times S^1$) and the Lens space partition function \cite{Benini:2011nc, Alday:2012au}. In those cases, {\it ad hoc} signs were introduced {\it e.g.} in \cite{Dimofte:2011py, Imamura:2013qxa, Nieri:2015yia} in the sum over topological sectors. It would be interesting to review those results accordingly.

%%%%%

\section{Localization and JK contours}\label{app: loc}

In this Appendix, we sketch the derivation  of the final formula \eqref{JK formula rank one}  for $p\neq 0$, using supersymmetric localization. The derivation is very similar to the $p=0$ discussed in Appendix B of \cite{Closset:2016arn}, apart from subtleties regarding the contribution from infinity on the $u$-plane. We focus on the case where the gauge group has rank one.

The generalization to any higher-rank gauge groups should follow from the previous works---see  \cite{Benini:2013xpa,Hori:2014tda} and \cite{Benini:2015noa,Closset:2015rna, Benini:2016hjo, Closset:2016arn}, except for complications due to the contributions from the ``boundaries'' at infinity, which we did not study rigorously. We presented a conjecture in the main text, and we provide some additional evidence for it below (see section  \ref{sec:HRJK}).

\subsection{Localization for $\GG=U(1)$}
\label{app:JK derivation}
In Section \ref{sub: Vector localization}, we saw that the SYM action \eqref{S YM full} admits the following scalar and one-form zero modes in the vector multiplet:
\be
\CV_0 = (u~,\, \t u~,\, \Lambda_0~,\, \t\Lambda_0~,\, \h D)~, \qquad\quad \CV_I = (\alpha_I~,\, \t \alpha_I~,\, \Lambda_I~,\, \t\Lambda_I)~, \quad I=1, \cdots, g~.
\ee
For  $p\neq 0$, the variable $u=i\beta (\sigma+ia_0)$ is valued in $\mathbb{C}$. In this ``Coulomb branch'' background, the partition function can be written as::
\be\label{integral singular}
Z_{g,p} = \lim_{e^2\rightarrow 0} \sum_{\m \in \mathbb{Z}_p} \int \prod_{I=1}^g d\CV_I \int_\Gamma d\h D \int_{\fM} \frac{d u d\t u}{\beta} \int d \Lambda_0 d\t\Lambda_0
~\CZ_{\m} (\CV_0, \CV_I)\ ,
\ee
where $\fM$ denotes the complex $u$-plane and $\CZ_\m$ is the contribution from the one-loop determinant and the classical action contribution at flux sector $\m$. We also have defined the measure
\be
d \CV_I = \frac{1}{\beta \vol (\Sigma_g)} d\alpha_I d\t\alpha_I d\Lambda_Id\t\Lambda_I\ .
\ee
The normalization of the path integral is chosen for convenience. In the end, we fix the overall normalization by comparing our result against known results \cite{Closset:2016arn}.
For future convenience, we perform the change of variable $\t u\rightarrow \t u'$ and $\t \Lambda_0 \rightarrow \t\Lambda'_0$ according to
\be
\t u = \t u'/k^2\ ,\t \Lambda_0 = \t \Lambda'_0/k^2~,
\ee
with $k$ some small real  parameter (not to be confused with a CS level).

Note that the contribution from the one-loop determinant and classical action $\CZ_{\m}$ has singularities at $\{u|u=u_*\}$ in $\fM$ where the chiral multiplets become massless. There also exists a potential singularity associated to the boundary at infinity. For singularities in the bulk, we first define the $\epsilon$-neighborhood $\Delta_\epsilon$ of these singularities as $(u-u_*)(\t u'-\t u'_* )\leq \epsilon^2$. Then, if we take the limit $\epsilon\rightarrow 0$ and $e\rightarrow 0$ in a way that $\epsilon$ is sufficiently smaller than $e$, we can show that the integral gets contribution only from the region $\fM\backslash\Delta_\epsilon$. We will discuss the contribution from infinity below.

The integration over the scalar gaugino zero-modes $\Lambda_0, \t \Lambda'_0$ can be performed by using the following relation from the residual supersymmetry of zero-modes:
\be
\delta \CZ_\m = \left(-2i\beta \t\Lambda'_0\partial_{\t u'}-\h D\partial_{\Lambda_0}+i\t\Lambda_I \partial_{\t \alpha_I}\right) \CZ_\m = 0\ ,
\ee
which implies
\be
\left.\partial_{\Lambda_0}\partial_{\t\Lambda'_0}\CZ_{\m}\right|_{\Lambda_0=\t\Lambda'_0=0}
=\left.\frac{1}{\h D} \left(2i\beta \partial_{\tilde u'}+i\t\Lambda_I\partial_{\t \alpha_I}\partial_{\t \Lambda'_0}\right) \CZ_\m \right|_{\Lambda_0=\t\Lambda'_0=0}\ .
\ee
Since there are no singularities in a compact domain of $\alpha_I$ as long as $\epsilon>0$, the second term which involves the total derivative $\partial_{\t \alpha_I}$ does not contribute to the path integral. 
We have
\be\label{zeromodeformula}
Z_{g,p} = \lim_{\epsilon, e\rightarrow 0} \sum_{\m \in \mathbb{Z}_p} \int \prod_{I=1}^g d\CV_I \int_\Gamma \frac{d\h D}{\h D} \int_{\fM\backslash\Delta_\epsilon} du d\tilde u' \left.\partial_{\tilde u'} \CZ_{\m}\right|_{\Lambda_0=\t\Lambda'_0=0}\ .
\ee
Here $\Delta_\epsilon$ also implicitly includes an excised region ``at infinity''. The $u$-plane integral  in \eqref{zeromodeformula} then reduces to a sum of contour integrals over all the components of the boundary $\d \Delta_\epsilon$.

To evaluate $\CZ_\m$ for non-zero $\h D$, we first expand the fields into the Fourier modes along the $S^1$ fiber:
\be
\phi = \sum_{n\in \mathbb{Z}} \phi_n (z,\bar z) e^{in \psi}\ .
\ee 
Let us  define the two-dimensional variables (until \eqref{return to rank one} we consider the generalization to higher rank, as it is essentially the same argument):
\be
Q\sigma_n = \frac{1}{i\beta} (Q^a u_a+ n)\ ,~~~Q\t \sigma'_n = -\frac{1}{i\beta}(Q^a \t  u'_a/k^2+n)\ .
\ee
At fixed $n$, we can have the spectrum $\{\lambda_n\}$ of the twisted Laplacian $D_1 D_{\b 1}$ on the two-dimensional base $\Sigma_g$, with:
\be
-4 D_1 D_{\b 1} \phi_n (z,\bar z)= \lambda_n \phi_n (z, \bar z)\ .
\ee
Recall that the scalar $\phi_n$ is valued in the line bundle \eqref{line bundle chiral}.
The contribution from the full chiral multiplet is given by:
\be\label{D nonzero chiral one-loop}
\left. Z^{\Phi} \right|_{\Lambda_0=\t\Lambda'_0=0}= \left. Z^{\Phi}_{\text{zero}}  Z^{\Phi}_{\text{massive}} \right|_{\Lambda_0=\t\Lambda'_0=0}\ ,
\ee
where the first factor is the contribution from the zero modes of $D_{\b 1}$ (or $D_1$):
\be\label{Zzero Phi}
Z_{\text{zero}}^{\Phi} = \prod_{\n \in \mathbb{Z}} (Q\sigma_n)^{n_C}\left(\frac{Q\t\sigma'_n}{Q\t\sigma'_n Q \sigma_n+iQ\h D}\right)^{n_B}\ .
\ee
Here $n_B$ and $n_C$ are the number of $\CB_n$ and $\CC_n$ zero-modes, respectively, with $n_B-n_C= p n+  Q\m + (g-1) (r-1)$ by the Riemann-Roch theorem.
Evaluating \eqref{Zzero Phi} at $\h D=0$ gives the 1-loop determinant \eqref{Zphi gen}. Since the non-zero modes pair among themselves, they do not contribute in the limit $\h D=0$. When $\h D\neq 0$, the contribution can be written as:
\bea
&Z_{\text{massive}}^{\Phi}&=&\; \prod_{\n \in \mathbb{Z}}\prod_{\lambda_n} \left[\frac{\lambda_n + Q \t \sigma'_n Q \sigma_n}{\lambda_n + Q^2 \t\sigma'_n\sigma_n+i Q\h D}\right]\cr
&&&\;\qquad \times \left(1-2i\frac{(Q\t\sigma'_n)(Q\t \Lambda_{\b 1})(Q\Lambda_1)}{(\lambda_n + Q^2 \t \sigma'_n\sigma_n)(\lambda_n+ Q \t \sigma'_n Q\sigma_n+iQ\h D)}\right)~.
\eea
The zero modes have singularities at finite $u=u_i^*$ where the hypermultiplets with charge $Q_i$ become massless, as well as at the infinite boundary $|u|\rightarrow \infty$. Each singualrity defines a ``hyperplane" $H_i$ or $H_\infty$ respectively.

First of all, let us consider the contour for the $\h D$-integral in \eqref{zeromodeformula}. In order to ensure convergence of the integral, we define the contour $\Gamma$ by $\mathbb{R}+i\delta$, where $\delta$ is a real number that satisfies $|\delta|<|Q_i\epsilon^2/k^2|$ for all singular hyperplane $H_i$'s.  The sign of $\delta$ is determined by the condition $\eta(\delta)>0$, where $\eta \in i\frak h^* \cong \R$ is a covector that we choose. The final answer does not depends on the choice of $\eta$.

Let us first integrate out the one-form zero modes $\Lambda_I, \t\Lambda_I$ and the flat connections on the base $\Sigma_g$.  This procedure is the same as in the $p=0$ case. We take the limit $k\ll \epsilon$ so that the summation over the modes that couples to $\Lambda_I, \t\Lambda_I$ can be simplified: For this, we note that
\bea
\log\CZ^{\Phi}|_{\substack{s\text{-th order in }\h D^a\\ \t\Lambda_{\b 1}=\Lambda_{1}=0}} &= \sum_{\lambda_n, n} \left(\frac{-iQ^a}{\lambda_n + Q^2\sigma_n\t\sigma_n'}\right)^s \\
&= \sum_{n\in \mathbb{Z}}\frac{(-iQ^a)^s}{\Gamma(s)} \int_{0}^{\infty}dt~ t^{s-1}\left(\sum_{\lambda_n}e^{-t\lambda_n}\right)e^{-t Q^2\t\sigma'_n\sigma_n}\ .
\eea
In the small $k$ limit (large $\t\sigma'_n$ limit), only the small $t$ expansion of the heat kernel,
\be
\sum_{\lambda_n} e^{-t\lambda}=\frac{1}{4\pi t}\sum_{l=0}^\infty a_lt^l~,
\ee
 contributes for fixed $n$. Performing the $t$ integral, we obtain
\be
\log\CZ^{\Phi}|_{\substack{s\text{-th order in }\h D^a\\ \t\Lambda_{\b 1}=\Lambda_{1}=0}} = \sum_n\left[\frac{a_0 (-i Q)^s}{4\pi (s-1)(Q^2\t\sigma'_n\sigma_n)^{s-1}}+ \frac{a_1  (-i Q)^s}{4\pi s (Q^2\t\sigma'_n\sigma_n)^s}+\cdots\right]\ .
\ee
Therefore, in the limit $k\rightarrow 0$, we are left with the following $\Lambda_1$, $\t\Lambda_{\b 1}$ and $\h D$ dependence:
\be
\cZ^{\Phi} = \exp\left[i\vol(\Sigma_g)\text{Im}(\partial_{u^a} \CW_{\Phi})\h D^a-i \beta \vol(\Sigma_g) \tilde \Lambda^a_{\bar 1}\Lambda^b_{1}H^{\Phi}_{ab}\right]\ ,
\ee
where:~\footnote{Note that the regularization used in \protect\cite{Closset:2016arn} is different from that we are using here. See Section \protect\ref{subsec: regulated Zphi}.}
\be
H^\Phi_{ab} =  Q^aQ^b\left(\frac{x^Q}{1-x^Q}\right)~.
\ee
This can be written as:
\be
H^\Phi_{ab} = \d_{u_a}\d_{u_b} \CW_\Phi~, \qquad \CW_\Phi = \frac{1}{(2\pi i)^2}\dilog (x^Q)~,
\ee 
in terms of the contribution of the chiral multiplet to the twisted superpotential. The dependence of the the classical action on the gauginos is also consistent with this expression:
\be
e^{-S_{CS}} = x^{k\m} \exp[i\vol(\Sigma_g)\text{Im}(\partial_{u^a} \CW_{\rm CS})\h D^a- i \beta \vol(\Sigma_g) k^{ab}\t\Lambda_{\b 1}^a\Lambda_1^b]~,
\ee
from the classical action. (Here we gave the formulas for the higher-rank case as well.)
Note that the one-loop and classical contributions do not depend on the flat connections $\alpha_I, \t\alpha_I$. We have
\be
\int \prod_{I=1}^g d\CV_I~\CZ_{\m}|_{\Lambda_0 = \t\Lambda_0=0} = H(u)^g \CZ_{\m}|_{\Lambda_0 = \t\Lambda_0=\Lambda_1 = \t\Lambda_{\b 1}=0}
\ee
after integrating over the gaugino zero-modes $\Lambda_1, \t \Lambda_{\b1}$, where we have defined:

\be H(g) \; = \; \det_{a,b} H_{ab} \; = \; \det_{a,b} \d_{u_a}\d_{u_b} \CW~. \ee
The final formula can be written as
\be
I = \sum_iI_{i}~,
\ee
where the summation runs over all the bulk singularities:
\bea \label{return to rank one}
&I_{i,\text{bulk}}=\cr
&\lim_{\substack{e\rightarrow 0\\R\rightarrow \infty}} \sum_{\m \in \mathbb{Z}_p}\int_{\Gamma}\frac{d\h D}{\h D}  \oint_{\partial \Delta_\epsilon^{Q_i}}du~ \cZ_\m(u) H(u)^g\exp\left[-\frac{\beta \vol(\Sigma_g)}{2e^2}\h D^2+i \vol(\Sigma_g)\text{Im}(\partial_u\CW) \h D\right]~,
\eea
and the contribution from the monopole singularities 
\bea\label{D_integral_infty}
&I_{\infty}=\cr
&\;\lim_{\substack{e\rightarrow 0\\R\rightarrow \infty}} \sum_{\m \in \mathbb{Z}_p}\int_{\Gamma}\frac{d\h D}{\h D} \oint_{\CC_R} du~ \cZ_\m(u) H(u)^g\exp\left[-\frac{\beta \vol(\Sigma_g)}{2e^2}\h D^2+i \vol(\Sigma_g)\text{Im}(\partial_u\CW) \h D\right]~.
\eea
%%%%
\begin{figure}[t]
\begin{center}
\includegraphics[width=0.9\textwidth]{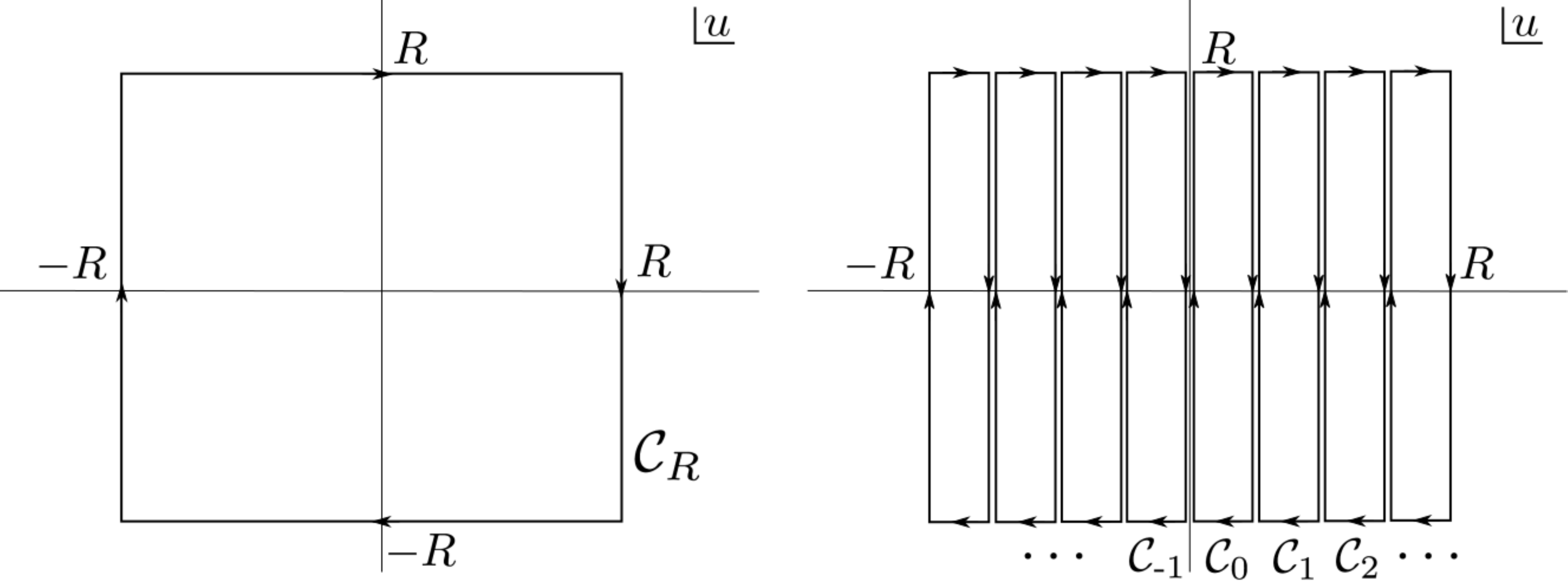}
\caption{We define the $u$-contour at infinity by $\lim_{R\rightarrow \infty}\CC_{R}$, where $\CC_R$ is given on the left. This contour can be decomposed into  $\lim_{R\rightarrow \infty}\CC_R=\sum_{k\in \mathbb{Z}}\CC_k$, a sum over an infinite number of contours around the strips of unit width,  as shown on  the right.}
\label{contour}
\end{center}
\end{figure}
%%%
We define the contour at infinity $\CC_R$ as shown on the left in Figure \ref{contour}. One can show that, once we take the limit $R\rightarrow \infty$ sufficiently faster than $e\rightarrow 0$ so that $e^N R\rightarrow \infty$, the integral \eqref{integral singular} does not get any contribution from the region outside of the contour $C_R$. (To show this, we generally need to turn on the regulator, as in \cite{Benini:2015noa}.) 
The easiest way to perform these integrals is to use the topological property of the theory on $\Sigma_g$. Using the fact that the final answer does not depend on the $\vol(\Sigma_g)$, we rescale it with a positive real $\mu\ll 1$:
\be
\h D \rightarrow \mu \h D\ ,~~\vol(\Sigma_g)\rightarrow \mu^{-1}\vol(\Sigma_g)\ .
\ee
Let us take the limit $\mu\rightarrow 0$, keeping $e$ finite. We have
\be
I_i = \lim_{\substack{e\rightarrow 0\\R\rightarrow \infty}} \sum_{\m \in \mathbb{Z}_p}\int_{\Gamma}\frac{d\h D}{\h D}  \oint_{\partial \Delta_\epsilon^{Q_i},C_R}du~ \cZ_\m(u) H(u)^g\exp\left[i \vol(\Sigma_g)\text{Im}(\partial_u\CW) \h D\right]\ .
\ee
Let us first assume $\text{Im}(\partial_u\CW)>0$.  Then we may close the contour in the upper-half $\h D$ plane.  If $\eta>0$, this contour surrounds no poles, so the result of the integral is zero, while for $\eta<0$, it surrounds the pole at $\h D=0$, and picks up the residue there.  Similarly, for $\text{Im}(\partial_u\CW)<0$, we close the contour in the lower half-plane, and get a contribution only when $\eta>0$.  Thus we find:
\bea
 &\int_{\Gamma}\frac{d\h D}{\h D} ~ \cZ_\m(u) H(u)^g\exp\left[i \vol(\Sigma_g)\text{Im}(\partial_u\CW) \h D\right]\ \cr
 &\quad= \left\{  \begin{array}{cc} 2\pi i\;\cZ_\m(u) H(u)^g &\qquad \text{if  }\; \sign (\text{Im}(\partial_u\CW)) = -\sign \eta~, \\  & \\ 0 &\qquad \text{otherwise}~. \end{array} \right.  
 \eea
Let us choose $\eta>0$ below for definiteness. For the bulk singularities, this rule picks up the residues from the poles with positive charges, as in the $p=0$ case:
\be
I_{i,\text{bulk}}(\eta>0)= (2\pi i)^2
\begin{cases}
 \sum_{\m \in \mathbb{Z}_p}
 \underset{u=u_i^*}{\text{Res}}~ \CZ_\m(u)~ H(u)^g  &\qquad {\rm if} \qquad Q_i>0~,\\
0 &\qquad {\rm if} \qquad Q_i<0~.
\end{cases}
\ee
As discussed in Section \ref{sec: Loc}, using the invariance under
\be\label{large_gauge}
u\rightarrow u+1~,\qquad \m\rightarrow\m+p~,
\ee
we can further massage this expression into
\be
I_{i,\text{bulk}}(\eta>0)=(2\pi i)^2 \begin{cases}
\sum_{\m \in \mathbb{Z}}
 \underset{\substack{u= u_i^*\\ 0\leq {\rm Re}(u) <1}}{\text{Res}} \CZ_\m(u)~ H(u)^g &\qquad {\rm if} \qquad Q_i>0~,\\
0 &\qquad {\rm if} \qquad Q_i<0~.
\end{cases}
\ee
The contribution from the ``poles at infinity'' does not directly follows from the $p=0$ case, since the geometry of the boundary at infinity  is different. We have
\be\label{D_integral_infty}
I_{\infty}=\lim_{\substack{e\rightarrow 0\\R\rightarrow \infty}} \sum_{\m \in \mathbb{Z}_p} \oint_{\CC_R^\eta} du~2\pi i\; \cZ_\m(u) H(u)^g~,
\ee
where the contour $\CC_R$ has been replaced by:
\be \CC_R^{\eta} = \{ u \in \CC_R \; \; | \; \; \sign (\text{Im}(\partial_u\CW)) = -\sign \eta \; \}\ .\ee

The contour $\CC_R$ consists of horizontal segments and vertical segments.  For the horizontal segments, where $\text{Im}(u) = \pm R \rightarrow \pm \infty$, one finds:
\be \text{Im}(\partial_u\CW)  \sim Q_{\mp}(\text{Im}(u) )~,
\ee
where $Q_\pm$ are the monopole charges defined in \eqref{induced charges T}.  Thus the contributions along the horizontal segments are included in the final contour, $C_R^\eta$, precisely when the sign of the monopole charge is the same as $\eta$, just as in the $p=0$ case.  If $Q_\mp=0$, the sign of $\text{Im}(\partial_u\CW)$ is determined by the flavor symmetry parameters, as discussed in an example in Section \ref{sec:jkrankone}.

Next, consider the vertical segments of $\CC_R$, where $\text{Re}(u) = \pm R \rightarrow \pm \infty$.  Recall that:
\be \cZ(u+n) = \Pi(u)^{-pn} \cZ(u)~,\ee
and therefore:
\be |\cZ(u+n)|= |\Pi(u)|^{-pn} |\cZ(u)| = e^{2 \pi p n \text{Im}(\partial_u \cW)} |\cZ(u)|~. 
\ee
Suppose first that we take $\eta>0$.  Then see that, for $\text{Re}(u) \rightarrow \infty$, the portions of the vertical line we are including, those with $\text{Im}(\partial_u \cW)<0$, have vanishing contribution.  Thus we can ignore the vertical line on the right.  However, the vertical line on the left, at $\text{Re}(u) \rightarrow -\infty$, has a large contribution, and can not be ignored.  Similarly, for $\eta<0$, we must include the vertical line on the right. 

It is also useful to write this expression in a way which clarifies the connection to the $p=0$ case.  Starting from the contour $\CC_R$, we decompose it into a sum over an infinite number of contours $\CC_n$,
\be
\lim_{R\rightarrow \infty} \CC_R= \sum_{n\in \mathbb{Z}} \CC_n\ ,
\ee 
as depicted in the right of Figure \ref{contour}. Here $\CC_n$ is a contour that goes around the boundary of an infinitely long strip $i\mathbb{R} \times [n,n+1]$.  Similarly, we decompose:
\be
\lim_{R\rightarrow \infty} \CC^\eta_R= \sum_{n\in \mathbb{Z}} \CC^\eta_n\ ,
\ee 
where $\CC_n^\eta$ is the portion of $\CC_n$ with $\sign \text{Im}(\partial_u \cW(u)) = -\sign \eta$. Noting that the \eqref{D_integral_infty} is invariant under the \eqref{large_gauge}, 
we can rewrite it as:
\bea
\label{sum rewrite}
I_{\infty} &= \sum_{\m \in \mathbb{Z}_p} \sum_{n\in\mathbb{Z}} \oint_{\CC^\eta_n} du~ 2\pi i\;\cZ_\m(u) H(u)^g ,\\
&=\sum_{\m \in \mathbb{Z}_p} \sum_{n\in\mathbb{Z}} \oint_{\CC^\eta_0} du~  2\pi i\;\cZ_{\m-np}(u) H(u)^g \\
&=\sum_{\m \in \mathbb{Z}} \oint_{\CC^\eta_0} du~  2\pi i\;\cZ_\m(u) H(u)^g \ .
\eea
In this formulation, the domain of the integration and of the sum is the same as in $p=0$ case, but with a different integrand. (Note that, for $p=0$, the integrand is periodic, and so the contributions along the vertical lines in $\CC_0^\eta$ cancel.)

\subsection{Relation of the JK contour to the Bethe-vacua formula}
\label{sec:HRJK}

Here we derive the Bethe-vacua formula \eqref{ZMgp main} for the $\Mgp$ partition function from the conjectured ``JK contour'' integral \eqref{JK formula higher rank unfolded} for higher-rank gauge groups. 
This provides some non-trivial evidence for the conjectured contour.

The  higher-dimensional contour integral \eqref{JK formula higher rank unfolded} takes the form:
\be \label{app:JK formula higher rank unfolded} Z_{\Mgp}= {1\ov |W_\GG|}\sum_{\m \in \Z^{\bf r}} \int_{\CC_0^\eta} d^{\bf r} u \;  \CJ(u)\, \pif_a(u)^{\m_a}~.
\ee
Consider the sum over all $\m \in \Z^r$. Let us split the sum as:
\be \sum_{\m \in \Z^{\bf r}} = \prod_{a=1}^{\bf r} \bigg( \sum_{\m_a=-\infty}^{0} + \sum_{\m_a=1}^\infty \bigg) = \prod_{a=1}^{\bf r}\sum_{\delta^a \in \{\pm \}} \sum_{\m_a^{\delta^a}=0}^\infty \ee
where we define $\m_a^{\pm} \in \Z_{\geq 0}$ by:
\be 
\m_a = \left\{ \begin{array}{cc} \m_a^+ +1 &\quad {\rm if }\quad \m_a \geq 1~,
 \\ -\m_a^-  &\quad {\rm if }\quad \m_a \leq 0~. \end{array} \right. 
 \ee
We then rewrite \eqref{app:JK formula higher rank unfolded} as:
\be \label{delta sum} Z_{\Mgp}= {1\ov |W_\GG|}  \prod_{a=1}^{\bf r}\sum_{\delta^a \in \{\pm \}} \sum_{\m_a^{\delta^a}=0}^\infty\int_{ \CC_0^{\eta^\delta}} d^r u \, \CJ(u) \pif_a(u)^{\m_a}~.
\ee
Here we have chosen a different contour, denoted $\CC_0^{\eta^\delta}$, for each choice of signs $\delta^a$, labeled by the covector:
\be\label{eta delta def}
 {(\eta^\delta)}^a=-\delta^a |\eta^a|, \quad \;\;\;a=1,\cdots ,r~.
\ee
Fix some $\delta^a$, and consider the sum over $\m_a$:
\be 
 \prod_{a=1}^{\bf r} \sum_{\m_a^{\delta^a}=0}^\infty\int_{ \CC_0^{\eta^\delta}} d^{\bf r} u\, \CJ(u) \Pi_a(u)^{\m_a}~.
\ee
Note that along $\CC_0^{\eta^\delta}$, we have, for each $a$:
\be \label{Pi bound} |\Pi_a|^{\delta^a} = e^{-2 \pi \delta^a \text{Im}(\partial_a \cW)} = e^{ -2 \pi |\text{Im}(\partial_a \cW)|} < 1 \ee
where we used the fact that $\text{sign}(\text{Im}(\partial_a \cW)) = \delta^a$ due to \eqref{eta delta def}. Thus the sum over $\m_a^{\delta^a}$ is a convergent geometric series. Using:
\be \sum_{\m_a^{\delta^a}=0}^\infty \Pi_a^{\m_a} = \delta^a \frac{\Pi_a}{1-\Pi_a}~,
\ee 
we find:
\be 
 \prod_{a=1}^{\bf r} \sum_{\m_a^{\delta^a}=0}^\infty\int_{ \CC_0^{\eta^\delta}} d^{\bf r} u\, \CJ(u) \Pi_a(u)^{\m_a} = \bigg(\prod_a\delta^a \bigg) \int_{ \CC_0^{\eta^\delta}} d^{\bf r} u\, \CJ(u) \prod_{a=1}^{\bf r} \frac{\Pi_a}{1-{\Pi_a}}~.
 \ee
Summing over all choices of $\delta^a$, we have:

\bea \label{HRJK bethe int} Z_{\Mgp} &= {1\ov |W_\GG|} \sum_{\delta^a \in \{\pm \}} \bigg(\prod_a \delta^a\bigg) \int_{ \CC_0^{\eta^\delta}} d^{\bf r}u \, \CJ(u) \prod_a \frac{\Pi_a}{1-{\Pi_a}} \\
&= {1\ov |W_\GG|}  \int_{\tilde{\CC}_0^\eta} d^{\bf r} u\,  \CJ(u) \prod_a \frac{\Pi_a}{1-{\Pi_a}}\eea
where we have defined:

\be \label{c tilde def} \tilde{\CC}_0^\eta=\sum_{\delta^a \in \{\pm \}} \bigg(\prod_a \delta^a\bigg)  \CC_0^{\eta^\delta}  \ee
Now we use the independence of the answer on $\eta$, and consider taking $\eta \rightarrow \epsilon \eta$ for $0<\epsilon\ll1$.  Note that, in this limit, the boundary contributions become negligible, since the interval in \eqref{C eta boundary def} includes a vanishingly small range of values of $\text{Im}(\partial_a \cW)$, and so $\CC_0^\eta \approx \CC^{\eta,\, {\rm bulk}}_{0}$ .  Thus, what remains are the contours at:
\be
 \text{Im}(\partial_{u_a} \cW) = \pm \epsilon \eta^a~.
\ee
Given their relative orientations, from \eqref{c tilde def},  the contributions from these nearby contours cancel everywhere except in the neighborhood of the singularities in \eqref{HRJK bethe int} at $\Pi_a=1$. The integral then captures the residues at the simultaneous solutions to:
\be 
\Pi_a=1, \;\;\; a=1, \cdots, {\bf r}~, 
\ee
\ie, the solutions to the Bethe equations.  Then, by an argument identical to the one leading to \eqref{conjecture higher rank rewrite}, summing the residues leads to the formula \eqref{ZMgp main}, given by a sum over Bethe vacua.

%%%%%
%%%%%%%%%%%%
\section{Gauging flavor symmetries using the on-shell $\CW$ and $\Omega$}\label{app: gauging}

In this appendix, we describe in more detail the procedure of gauging global symmetries at the level of the $\Mgp$ partition function.  As discussed in Section \ref{sec:onshell}, this is achieved most easily by working with the on-shell twisted superpotential and effective dilaton:

\be \label{onshell W O} \CW^l(\nu), \;\;\; \Omega^l(\nu),\quad \;\;\; l=1,...,|\CS_{BE}| \ee
Namely, suppose one is given these objects for a three dimensional $\cN=2$ theory $\CT$, but one does not have any other information about the theory (\eg, a Lagrangian description).  Then, as explained in Section \ref{sec:onshell}, we may nevertheless use these to construct the $\Mgp$ partition function $Z_{\Mgp}[\CT]$.  Moreover, for any theory $\hat{\CT}$ obtained from $\CT$ by gauging flavor symmetries, we can use this data to construct the on-shell superpotential and effective dilaton for $\hat{\CT}$, and therefore also its $\Mgp$ partition function.

Let us recall here how this gauging operation works, and elaborate on some of the details.  Suppose the theory $\hat{\CT}$ is obtained by gauging the flavor symmetry associated to some subset $S$ of the flavor symmetry generators, whose parameters we relabel $\nu_i \rightarrow v_{\h a}$, ${\h a}=1,...,|S|$.\footnote{Here for simplicity we assume the group we are gauging is abelian; it is straightforward to extend the argument to the non-abelian case.} Then we claim that we should simply write the Bethe equation for $v_{\h a}$, in terms of $\CW^l$, \ie:
\be \label{uhat bethe equation} \Pi^l_{\h a}=\exp \bigg(2 \pi i \frac{\partial \CW^l}{\partial v_{\h a} }\bigg) = 1, \;\;\; \h a \in S, \ee
These equations should be solved for each $l$, and may have zero, one, or several solutions for each $l$.  The vacua of the new gauge theory is the union of these solutions for all $l$.  That is, the set $\hat{\CS}_{BE}$ of Bethe vacua of $\hat{\CT}$ is:
\bea
&\hat{\CS}_{BE} = \bigcup_{l \in \CS_{BE}}  \hat{\CS}_{BE}^l~, \cr
& \label{hat S_BE decomp} \hat{\CS}_{BE}^l \equiv  \Big\{ \; \h v_{\h a}^{l,j} \;\; | \;\; \exp \bigg( 2 \pi i \frac{\partial \CW^l}{\partial v_{\h a}}\bigg) \bigg|_{v_{\h a}= \h v_{\h a}^{l,j} }= 1 \Big\}~.\eea

An important consistency check of this procedure is the following.  Suppose we are told the on-shell objects in \eqref{onshell W O} actually come from a gauge theory.  That is, they were obtained by starting with a twisted superpotential, $\CW(u, v,\nu)$, which is a function of some gauge variables, $u_a$, $a=1,...,\rk$, and solving the Bethe equations:
\be
\CS_{BE} = \Big\{ u_a^l \;|\; \frac{\partial \CW(u^l, v,\nu)}{\partial u_a} = 0, \;\;\; a= 1,...,\rk  \Big\}~,
 \ee
 such that:
 \be \label{on shell W from Lagrangian} 
 \CW^l(v,\nu) =\CW(\h u^l(v, \nu), v,\nu)~.
 \ee
Then, another way to obtain the theory $\hat{\CT}$ is to gauge all the variables $u_a, v_{\hat a}$ at once.  In that case, we would find the Bethe equations:
\be \label{joint BE}
 \exp\bigg(2 \pi i \frac{\partial \CW}{\partial u_a}  \bigg) = 1,  \qquad\;\;\; \exp \bigg(2 \pi i \frac{\partial \CW}{\partial v_{\hat a}} \bigg) = 1
  \ee
which we should solve simultaneously as a function of the $u_a$ and $v_{\hat a}$.   This procedure must lead to the same Bethe vacua and $\Mgp$ partition function as we obtained by starting from \eqref{onshell W O}.

First, to see that we get the same set of vacua, note that if we solve the first set of equations in \eqref{joint BE} as a function of $u_a$, and for fixed $\hat{u}_a$, we find the solutions $u_a^l$ in \eqref{on shell W from Lagrangian}.  Next, we can plug in the $u_a^l$, where we consider all choices of $l$, and find the $\hat{u}_{\hat a}$ which solve the second set of equations in \eqref{joint BE}, that is:

\be 1 = \exp\bigg( 2 \pi i \frac{\partial \CW}{\partial v_{\hat a}}\bigg) \bigg|_{u_a=u_a^l} \ee
However, from \eqref{on shell fiber and flux}, the RHS is equal to $\Pi^l_{\hat a}$, and so this is equivalent to solving \eqref{uhat bethe equation}.

It remains to check that the various ingredients in the $\cM_{g,p}$ partition function that we obtain by the two methods agree.  For the fibering and flux operators, this follows straightforwardly from  \eqref{on shell fiber and flux}.  For the handle gluing operator, if we gauge $u_a$ and $v_{\hat a}$ simultaneously, we find, for $\hat{l} \in \hat{\CS}_{BE}$:

\be \label{HGO method one}
\CH^{\hat{l}}(\nu) = e^{\Omega(\h u^{\hat l},\h v^{\hat l},\nu)}\det \left( \begin{array}{cc} \displaystyle \frac{\partial \CW}{\partial u_a \partial u_b} & \displaystyle \frac{\partial \CW}{\partial u_a \partial v_{\hat b}} \\ \displaystyle \frac{\partial \CW}{\partial v_{\hat a}\partial u_b}  & \displaystyle \frac{\partial \CW}{\partial v_{\hat a}\partial v_{\hat b}}  \end{array} \right)  \bigg|_{u=\h u^{\hat l}, v= \h v^{\hat l}}
 \ee
On the other hand, starting from $\CW^l$ and $\Omega^l$ and solving the Bethe equations for the $v_{\hat a}$, we would find, in the notation of \eqref{hat S_BE decomp}:
\bea \label{HGO method two} 
&\cH^{l,j}(\nu) &=&\; e^{\Omega^l(v,\nu)}\det \frac{\partial \CW^l}{\partial v_{\hat a}\partial v_{\hat b}}   \bigg|_{v_{\hat a}= v^{l,j}_{\hat a}} \cr
&& =&\;   \bigg( \bigg( e^{\Omega(u,v,\nu)} \det \frac{\partial \CW}{\partial u_a \partial u_b}  \bigg) \bigg|_{u=u^l} \det \frac{\partial \CW^l}{\partial v_{\hat a}\partial v_{\hat b}}  \bigg) \bigg|_{v= v^{l,j}} 
\eea
At first sight, the expressions \eqref{HGO method one} and \eqref{HGO method two}  look quite different.  To see that they agree, let us first introduce the notation:
\be \left( \begin{array}{cc} \displaystyle \frac{\partial \CW}{\partial u_a \partial u_b} & \displaystyle \frac{\partial \CW}{\partial u_a \partial v_{\hat b}} \\ \displaystyle \frac{\partial \CW}{\partial v_{\hat a}\partial u_b}  & \displaystyle \frac{\partial \CW}{\partial v_{\hat a}\partial v_{\hat b}}  \end{array} \right)  \equiv \left( \begin{array}{cc} A_{ab} & B_{a \hat{b}} \\ C_{\hat{a} b} & D_{\hat{a} \hat{b}} \end{array} \right) \ee
Then the determinant in \eqref{HGO method one} can be written:
\be \label{method one rewrite}
\det \left( \begin{array}{cc} A_{ab} & B_{a \hat{b}} \\ C_{\hat{a} b} & D_{\hat{a} \hat{b}} \end{array} \right)  =  \det_{ab} A_{ab} \det_{\h a \h b} (D_{\hat{a} \hat{b}} - C_{\hat{a} b} A^{-1}_{ab} B_{a \hat{b}} )  \ee
Next, we rewrite the matrix appearing in second factor in \eqref{HGO method two} as:
\be \label{method two step one} \frac{\partial \CW^l}{\partial v_{\hat a}\partial v_{\hat b}}  =  
\frac{\partial \CW}{\partial v_{\hat a}\partial v_{\hat b}}  + \frac{\partial u^l_a}{\partial v_{\hat a}} \frac{\partial \CW}{\partial u_a\partial v_{\hat b}} 
= D_{\hat{a} \hat{b}} +  \frac{\partial u^l_a}{\partial v_{\hat a}} B_{a \hat{b}}~.
\ee
We differentiate \eqref{physical branch condition} to find:
$$ 0 = \frac{\partial}{\partial v_{\hat b}} \frac{\partial \CW}{\partial u_a}(u_a^l(\nu_i),\nu_i) = \frac{\partial^2 \CW}{\partial u_a \partial v_{\h b}} + \frac{\partial^2 \CW}{\partial u_a \partial u_b} \frac{\partial u^l_b}{\partial v_{\hat b}}  = C_{a \hat{b}} +\frac{\partial u^l_b}{\partial v_{\hat b}} A_{ab} $$
\be \Rightarrow  \frac{\partial u^l_b}{\partial v_{\hat b}}  = - A^{-1}_{ab} C_{a \hat{b}} \ee
Plugging this into \eqref{method two step one} gives:
\be \frac{\partial \CW^l}{\partial v_{\hat a}\partial v_{\hat b}}  = D_{\hat{a} \hat{b}} - C_{a \hat{b}} A^{-1}_{ab} B_{a \hat{b}}\ee
and so the determinants in \eqref{HGO method two} can be written as:
\be \det A_{ab} \det (D_{\hat{a} \hat{b}} - C_{\hat{a} b} A^{-1}_{ab} B_{a \hat{b}} )  \ee
agreeing with \eqref{method one rewrite} (after identifying the solution corresponding to $\hat{l}$ with that corresponding to the pair $(l,j)$), and completing the proof that \eqref{HGO method one} agrees with \eqref{HGO method two}.

\bibliographystyle{utphys}
\bibliography{bib3d}{}

\end{document}